\documentclass[showpacs,aps,nofootinbib]{revtex4-1}

\usepackage{amsfonts}
\usepackage{amsmath}
\usepackage{color}
\usepackage{graphicx}
\usepackage{dcolumn}
\usepackage{bm}
\usepackage{slashed}

\begin{document}

\def\bb    #1{\hbox{\boldmath${#1}$}}

\title{Magnetic States at Short Distances}
\author{Horace W. Crater$^{1*}$
\footnote[0]{${}^*$Email address: hcrater@utsi.edu}
 and Cheuk-Yin Wong$^{2,3\dagger}$
\footnote[0]{${}^\dagger$Email address: wongc@ornl.gov}
}
\affiliation{${}^1$The University of Tennessee Space Institute, Tullahoma, Tennessee
37388 \\
${}^2$Department of Physics and Astronomy, University of Tennessee,
Knoxville, TN 37996 \\
${}^3$Physics Division, Oak Ridge National Laboratory, Oak Ridge, TN 37831 }
\date{\today }

\begin{abstract}
The magnetic interactions between a fermion and an antifermion of opposite
electric or color charges in the $^{1}S_{0}^{-+}$ and $^{3}P_{0}^{++}$
states with $J=0$ are very attractive and singular near the origin and may
allow the formation of new bound and resonance states at short distances. In
the two body Dirac equations formulated in constraint dynamics, the
short-distance attraction for these states for point particles leads to a
quasipotential that behaves near the origin as $-\alpha ^{2}/r^{2}$, where $
\alpha $ is the coupling constant. Representing this quasipotential at short
distances as $\lambda (\lambda +1)/r^{2}$ with $\lambda =(-1+\sqrt{1-4\alpha
^{2}})/2$, both $^{1}S_{0}^{-+}$ and $^{3}P_{0}^{++}$ states admit two types
of eigenstates with drastically different behaviors for the radial wave
function $u=r\psi $. One type of states, with $u$ growing as $r^{\lambda +1}$
at small $r$, will be called usual states. The other type of states with $u$
growing as $r^{-\lambda }$ will be called peculiar states. Both of the usual
and peculiar eigenstates have admissible behaviors at short distances.
Remarkably, the solutions for both sets of $^{1}S_{0}$ states can be written
out analytically. The usual bound $^{1}S_{0}$ states possess attributes the
same as those one usually encounters in QED and QCD, with bound state
energies explicitly agreeing with the standard perturbative results through
order $\alpha ^{4}$. In contrast, the peculiar bound $^{1}S_{0}$ states, yet
to be observed, not only have different behaviors at the origin, but also
distinctly different bound state properties (and scattering phase shifts).
For the peculiar $^{1}S_{0}$ ground state of fermion-antifermion pair with
fermion rest mass $m$, the root-mean-square radius is approximately $1/m$,
binding energy is approximately $(2-\sqrt{2})m$, and rest mass approximately 
$\sqrt{2}m$. On the other hand, the $(n+1)$${}^{1}S_{0}$ peculiar state with
principal quantum number $(n+1)$ is nearly degenerate in energy and
approximately equal in size with the $n$$^{1}S_{0}$ usual states. For the $
{}^{3}P_{0}$ states, the usual solutions lead to the standard bound state
energies and no resonance, but resonances have been found for the peculiar
states whose energies depend on the description of the internal structure of
the charges, the mass of the constituent, and the coupling constant. The
existence of both usual and peculiar eigenstates in the same system leads to
the non-self-adjoint
property of the mass operator
and two non-orthogonal complete sets. As both sets of states are physically
admissible, the mass operator can be made self-adjoint with a single
complete set of admissible states by introducing a new peculiarity quantum
number and 
an enlarged Hilbert space that contains
 both the usual and peculiar states
in different peculiarity sectors. Whether or not these
newly-uncovered quantum-mechanically acceptable peculiar $^{1}S_{0}$ bound
states and $^{3}P_{0}$ resonances for point fermion-antifermion systems
correspond to physical states remains to be further investigated.
\end{abstract}

\maketitle
\affiliation{${}^1$The University of Tennessee Space Institute, Tullahoma, Tennessee
37388 \\
${}^2$Department of Physics and Astronomy, University of Tennessee,
Knoxville, TN 37996 \\
${}^3$Physics Division, Oak Ridge National Laboratory, Oak Ridge, TN 37831 }
\affiliation{The University of Tennessee Space Institute, Tullahoma, Tennessee 37388 \\
Department of Physics and Astronomy, University of Tennessee, Knoxville, TN
37996 \\
Physics Division, Oak Ridge National Laboratory, Oak Ridge, TN 37831 }

\section{INTRODUCTION}

It is well known that for some combinations of the spin configurations and
orbital motion the magnetic interaction can be strongly attractive and
singular\footnote{
A potential is quantum mechanically singular if it is more attractive than $
-1/4r^{2}$ at the origin in the context of ($-d^{2}/dr^{2}$ $-1/4r^{2})$.
See \cite{Case}.} at short distances \cite{Bar77,Bar81,Won86}. We can
illustrate this by a classical example as shown schematically in Fig. 1(a)
where a positive charge $q^{+} $ is making a circular orbit about a fixed
negative charge $q^{-}$ whose spin ${\ \bb s} (q^{-})$ is pointing in a
direction opposite to the orbital angular momentum of $q^{+}$ \cite{Won86}.
In the external field problem, (e.g., Fermi's treatment of hyperfine
structure), the charged particle $q^-$ at rest with a magnetic moment $\bb
\mu (q^{-})$ generates a vector potential ${\bb A}={\bb\mu }(q^{-})\times {
\bb  r}/r^{3}$ which acts on the other particle, $q^+$. Such a
\textquotedblleft magnetic" interaction can be very attractive when the
spins and the orbital angular momentum are oppositely aligned, as shown in
the configuration of ($q^{+}q^-$) in Fig.\ 1, where the vector potential ${
\bb A}$, arising from the $q^{-}$ magnetic dipole moment ${\bb\mu }(q^{-})$,
is parallel to the $q^{+}$ orbital momentum ${\bb p}$. The interaction $(-
\bb p\cdot \bb A)$ from $q^-$ acting on $q^{+}$ is attractive and is
proportional to $[\bb L (q^+) \cdot \bb  s(q^-)]/r^{3}$ that is quite
singular in nature. At short distances it may overwhelm the centrifugal
barrier that is proportional to $1/r^{2}$. Similarly, the interaction from $
q^+$ acting on $q^{-}$ will be likewise attractive and singular if the spin
of the ${\bb  s}(q+)$ is parallel to the electron spin ${\ \bb s}(q-)$ and
pointing in the same direction, resulting in the total spin of the $
q^{+}q^{-}$ system aligning opposite to the orbital angular momentum, as in
the ${}^{3}P_{0}^{++}$ state with $S=1,L=1$, $J=0$, $P=+1$, and $C=+1$.

The ${}^{3}P_{0}^{++}$ state is not the only state with a strong magnetic
interaction. One can envisage classically another spin configuration, the $
^{1}S_{0}^{-+}$ state, that also has attractive and singular magnetic
interactions. As illustrated schematically in Fig.\ 1(b), a fermion $q^{-}$
with an electric or color charge interacts with an antifermion $q^{+}$ of
opposite electric or color charge with spins ${\ \bb s}(q^{-})$ and ${\ \bb s
}(q^{+})$ pointing in a opposite directions in the $^{1}S_{0}^{-+}$ state
configuration. With the spins opposite to each other, the magnetic moments
of $q^{-}$ and $q^{+}$ are parallel to each other. The interaction between
the magnetic moments is \cite{Jac62} $H_{\mathrm{int}}=-({8\pi }/{3}){\bb\mu 
}_{q^{-}}\cdot {\bb\mu }_{q^{+}}\delta (\bb r),$ which is attractive and
singular at short distances. The strong and singular magnetic interaction
may overcome other repulsive interactions and may allow the formation of
bound states of the fermion and antifermion system at short distances. For
brevity of notation, the quantum numbers $P$ and $C$ in and $^{1}S_{0}^{-+}$
and ${}^{3}P_{0}^{++}$ will be understood.

\begin{figure}[h]
\includegraphics[scale=0.65]{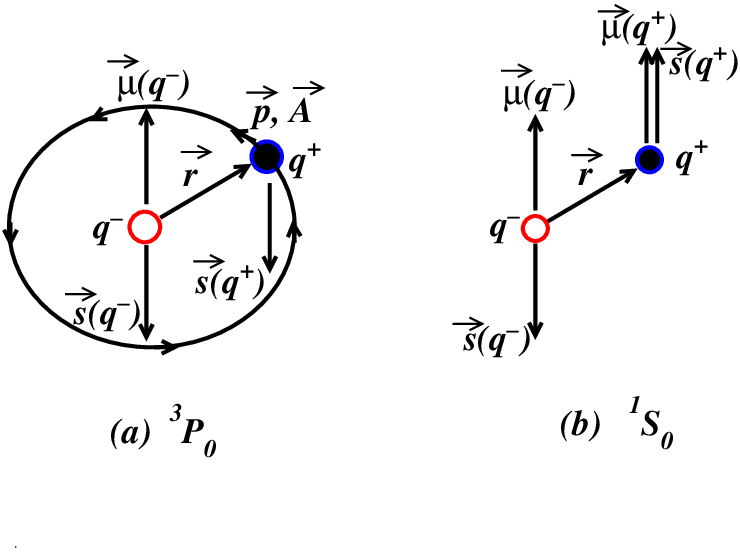} \vspace*{-1.0cm} 
\caption{ (a) The schematic picture of the $^3 P_0$ state spin configuration
and the orbital motion of a negative charge $q^-$ and a positive charge $q^+$
that can lead to a strong magnetic attraction at short distances. Here, 
${\bb \mu}(q^\pm)$ is the magnetic moment of the charge $q^\pm$
arising from its spin $\bb s(q^\pm)$. (b) The schematic picture of the spin
configurations of $q^-$ and $q^+$ in the $^1 S_0$ state. }
\end{figure}

Previously, one of us (CYW), in collaboration with R. L. Becker, studied the 
$(e^{+}e^{-})$ system using the Kemmer-Fermi-Yang equation \cite{Kem37} with
interactions consisting of the Coulomb interaction and the vector (magnetic)
interaction, ${\bb A}_{i}={\bb\mu }_{j}\times ({\bb r}_{i}-{\bb r}_{j})/|{
\bb r}_{i}-{\bb r}_{j}|^{3}$, in connection with a possible scalar $
^{3}P_{0} $ magnetic resonance \cite{Won86}. The interest was to investigate
whether there could be a resonance at the mass of 1.579 MeV that might
explain the anomalous positron peak in heavy-ion collisions near the Coulomb
barrier \cite{Sch83}. The experimental evidence for the anomalous positron
peak later turned out to be negative when greater statistics were
accumulated \cite{Ahm95}. Nevertheless, it remains of interest to study the
behavior of the two-body system at short distances and see how the
attractive magnetic interaction in the ${}^{3}P_{0}$ state may reveal itself
in some observable properties.

While the use of the Kemmer-Fermi-Yang equation with a two-body magnetic
interaction is useful to motivate an approximate description \cite{Won86}, a
consistent relativistic description of the two-body interaction at short
distances can be found in the relativistic two body Dirac equations (TBDE)
formulated in Dirac's constraint dynamics \cite{dirac,cnstr,cra82,cww}.
These relativistic two body Dirac equations give a good description to the
entire meson mass spectrum (excluding most flavor-mixed mesons) with
constituent world-scalar and vector potentials depending on just two or
three invariant functions, in previous relativistic quark-model calculations \cite{crater2,tmlk,unusual}.

The application of the TBDE equations to two-body bound and resonance states
in quantum electrodynamics has intrinsic merits. In Ref.\ \cite{bckr}, the
properties of these TBDE equations that made them work so well for the
relativistic quark model were investigated by solving them nonperturbatively
(i.e.\ analytically or numerically) in quantum electrodynamics (QED), where
order $\alpha ^{4}$ perturbative solutions are well known. The two coupled
Dirac equations in the constraint formalism depend on Lorentz-covariant
potentials between the two constituents and act on a 16-component wave
function. An exact Pauli reduction led to a second-order relativistic 
Schr\"{o}dinger-like equation for a reduced four-component wave function 
with an
effective interaction containing all the dependencies on spin, orbital
angular momentum, and tensor operators. We were able to solve the TBDE
nonperturbatively (analytically or numerically) as well as perturbatively
because the spin dependent short-distance components of the effective
interaction are not singular \cite{cww,bckr}. The situation is very
different from the approximate Fermi-Breit forms, which contain singular
potentials and necessitate the introduction of arbitrary short-distance
cut-off parameters. The spin dependence of the relativistic potentials in
the exact Schr\"{o}dinger-like equation arises naturally from the
relativistic reduction procedure and it incorporates detailed minimal
interaction and dynamical recoil effects, characteristic of field theory. We
shall also use the term \textquotedblleft quasipotential" to represent this
effective, non-singular interaction.

To obtain the interaction used in the TBDE formalism, we first determined
the relativistic quasipotential to the lowest order in $\alpha $ for the Schr
\"{o}dinger-like equation in Ref.\ \cite{bckr} by comparing the effective
interaction with the interaction derived from the Bethe-Salpeter equation.
This, in turn led to an invariant Coulomb-like potential $A(r)=-\alpha /r$,
where $\alpha$ is the coupling constant. Insertion of this information into
the minimal interaction structures of the two body Dirac equations then
completely determined all aspects (spin-dependent as well as
spin-independent) of the interaction. (In \cite{decay} we gave a procedure
to construct the full 16-component solution to our coupled first-order Dirac
equations from a solution of the second-order equation for the reduced wave
function.)

Next, we showed that both the quantum mechanical perturbative and the TBDE
non-perturbative treatments (i.e. analytic or numerical) yield the standard
spectral results for QED and related interactions through order $\alpha ^{4}$
. Such an agreement depends crucially on the inclusion of the coupling
between various components of our 16-component Dirac wave functions and on
the short-distance behavior of the relativistic quasipotential in the
associated Schr\"{o}dinger-like equation. We then examined the speculations 
\cite{Won86} whether the quasipotentials (including the angular momentum
barrier) for some states in the $e^{+}e^{-}$ system may become attractive
enough at short distances to yield a pure QED resonance corresponding to the
anomalous positron peaks in heavy-ion collisions \cite{Sch83}. For the $
^{3}P_{0}$ state we found that, even though the quasipotential becomes
attractive and overwhelms the centrifugal barrier at short distances, the
spatial extension of the attraction is not large enough to hold a resonance
at the energy of 1.579 MeV \cite{bckr}. This result contradicted predictions
of such states by other authors \cite{Spe91} based on numerical solutions of
three-dimensional truncations of the Bethe-Salpeter equation, for which the
entire QED bound state spectrum has been treated successfully through order $
\alpha ^{4}$ only by perturbation theory.

In this paper we return to this problem of the magnetic resonance and
magnetic states, not motivated so much by new experimental data as by a
discovery of an additional peculiar solution of the TBDE overlooked in the
earlier work in Ref.\ \cite{bckr}. Our examination of the two body Dirac
equations reveals that at short distance for both $^{1}S_{0}$ and $
{}^{3}P_{0}$ states, the magnetic interactions is indeed quite strong. As a
consequence, they counterbalance other repulsive interactions to result in a
quasipotential for these states that behaves as $-\alpha ^{2}/r^{2}$ at
short distances.

In standard quantum mechanics for central interactions including the angular
momentum barrier $L(L+1)/r^{2}$ for states with $L\neq 0$ at short
distances, one generally retains only one of the two solutions for the
radial part of the wave function, $u=r\psi $, the one that grows with
distance as ($\sim r^{L+1})$, dropping the other solution($\sim r^{-L})$ as
being too singular. If we likewise represent the quasipotential as $\lambda
(\lambda +1)/r^{2}$ with $\lambda =(-1+\sqrt{(1-4\alpha ^{2}})/2$, it leads
to a short-distance solution that behaves as $r^{\lambda +1}$, which we call
the usual solution, in addition to a solution, whose radial part grows as $
r^{-\lambda }$, which we call the peculiar solution. However, both usual and peculiar states have quantum-mechanically acceptable
behaviors at short distances, as the wave functions at short-distances are square-integrable.

In the case of the spin singlet $^{1}S_{0}$ states, the eigenstates and
eigenenergies can be obtained analytically and are found to encompass both
usual and peculiar states. We find usual bound states with attributes the
same as those one usually encounters in QED and QCD, explicitly agreeing
with the standard perturbative results through order $\alpha ^{4}$. In
contrast, the peculiar $^{1}S_{0}$ ground state of a fermion-antifermion
pair with a fermion rest mass $m$ has a root-mean-square radius
approximately $1/m$, a binding energies approximately $B_{p}$$\sim $$(2-
\sqrt{2})m$, and a rest mass approximately $\sqrt{2}m$. However, the $(n+1)$
th ${}^{1}S_{0}$ peculiar state is nearly degenerate in energy and
approximately equal in size with the $n$th usual ${}^{1}S_{0}$.

The existence of both usual and peculiar eigenstates in the same system
brings with them conceptual and mathematical problems of the
non-self-adjoint property of the mass operator and the over-completeness of
the set of eigenstates. We resolve these problems by the introduction of a
new quantum number, the peculiarity quantum number, that makes the mass
operator self-adjoint and the combined set of usual and peculiar states a
complete set in an enlarged Hilbert space.

In the case of the $^{3}P_{0}$ states, both of the usual and peculiar
solutions reflect the overwhelmed centrifugal barrier and so differ
substantially from the $r^{L+1}$ and $r^{-L}$ behaviors at short distances
respectively. 
As a peculiar state radial wave function $u$ rises from the zero value at the
origin as $r^{-\lambda }\sim r^{\alpha ^{2}}$, the strongly attractive
magnetic interaction has the tendency of bending the wave function in such a
way to allow for the possibility of a resonance. Furthermore, as the
quasipotential obtained through the relativistic reduction is sensitively
energy dependent, we can explore the behavior of the two-body system over a
larger domain of energies. We find that the usual solutions lead to no
resonant behavior, but the peculiar
solution can lead to a $^{3}P_{0}$ resonance whose phase shift changes by $
\pi $ at an appropriate energy, depending on the description of the internal
structure of the charges, the mass of the constituent, and the coupling
constant.

This paper is organized as follows. In Sec.\ II we give a review of the two
body Dirac equations of constraint dynamics. \ For those readers who are
already familiar with the constraint approach we refer them to the TBDE
given in Eq.\ (\ref{tbde}) and their Schr\"{o}dinger-like Pauli reduction
given in Eq.\ (\ref{57}). We specialize to electromagnetic-like interactions
only in this paper. We give in Sec.\ III the single-component radial forms
of Eq. (\ref{57}) relevant to this paper. In Sec.\ IV we examine both
solutions for the $^{1}S_{0}$ states. In addition to examining new bound
state solutions, we show how the $^{1}S_{0}$ wave functions for positive
energies (and their corresponding phase shifts) can be determined
analytically in terms of Coulomb wave functions for noninteger angular
momentum. This is done for both the usual and peculiar solutions. We explain
why and how we introduce of a new quantum number, which we call the
peculiarity quantum number, to solve the problems of the non-self-adjoint
property of the mass operator and the over-completeness of the set of
eigenstates.  In Sections V we examine the short distance behaviors for the $
^{3}P_{0}$ state for the usual \ and peculiar solutions. In Sec.\ VI we
discuss the variable phase shift formalism of Calogero \cite{cal} and
outline how we use it for our phase shift analysis. \ Since the short
distance behavior of the $^{3}P_{0}$ quasipotential is the same as that of $
^{1}S_{0}$ quasipotential, we can use those same $^{1}S_{0}$ Coulomb wave
functions as reference wave functions in that region to compute phase
shifts. \ There is, however, an additional term (proportional to $\delta ({
\bb r)}$) that does not appear in the extreme short distance region for the $
^{1}S_{0}$ quasipotential. Even though this term does not contribute in the
case of the phase shift for the usual solution, its contribution to the
phase shift calculations for the peculiar solution must be considered. In
Sec.\ VI we discuss our numerical results and in Sec.\ VII our conclusions.
Various technical results are presented in the appendices. In Appendix A we
give an outline of the details on the relation between the two-body Dirac
equations and their Pauli reduced Schr\"{o}dinger forms. In Appendix B we
present the radial forms of those Schr\"{o}dinger-like equations for a
general angular momentum coupling. In Appendix C we present details of the $
^{1}S_{0}$ usual and peculiar bound states. \ In Appendix D we review the
connections between the Coulomb wave functions for noninteger angular
momentum index. \ Appendix E presents a review of the variable phase method
of Calogero \cite{cal} for our problem.

\section{TWO BODY DIRAC EQUATIONS}

We briefly review the two body Dirac equations of constraint dynamics \cite
{cra82,cww,crater2,tmlk,unusual,saz86} providing a covariant three
dimensional truncation of the Bethe Salpeter equation for the two body
system. \ Sazdjian \cite{saz85,saz92,saz97} has shown that the
Bethe-Salpeter equation can be algebraically transformed into two
independent equations. The first yields a covariant three dimensional
eigenvalue equation which for spinless particles takes the form 
\begin{equation}
\biggl(\mathcal{H}_{10}+\mathcal{H}_{20}+2\Phi \biggr )\Psi (x_{1},x_{2})=0,
\label{eq:sum}
\end{equation}
where $\mathcal{H}_{i0}=p_{i}^{2}+m_{i}^{2}$ . The quasipotential $\Phi $ is
a modified geometric series in the Bethe-Salpeter kernel $K$ such that in
lowest order in $K$ 
\begin{equation}
\Phi =\pi i\delta (P\cdot p)K,  \label{bsen}
\end{equation}
where $P=p_{1}+p_{2}$ is the total momentum, $p=\mu _{2}p_{1}-\mu _{1}p_{2}$
is the relative momentum, $w$ is the invariant total center of momentum
(c.m.) energy with $P^{2}=-w^{2}$. $\ $The $\mu _{i}$ must be chosen so that
the relative coordinate $x=x_{1}-x_{2}$ and $p$ are canonically conjugate,
i.e. $\mu _{1}+\mu _{2}=1$. The second equation, Eq.\ (\ref{bsen}),
overcomes the difficulty of treating the relative time in the center of
momentum system by setting an invariant condition on the relative momentum $p
$, 
\begin{equation}
(\mathcal{H}_{10}-\mathcal{H}_{20})\Psi (x_{1},x_{2})=0=2P\cdot p\Psi
(x_{1},x_{2}).  \label{eq:dif}
\end{equation}
Note that this implies $p^{\mu }\Psi =p_{\perp }^{\mu }\Psi \equiv (\eta
^{\mu \nu }+\hat{P}^{\mu }\hat{P}^{\nu })p_{\nu }\Psi $ in which $\hat{P}
^{\mu }=P^{\mu }/w$ is a time like unit vector $(\hat{P}^{2}=-1)$ in the
direction of the total momentum\footnote{
We use the metric $\eta ^{11}=\eta ^{22}=\eta ^{33}=-\eta ^{00}=1$.}.

One can further combine the sum and the difference of Eqs. (\ref{eq:sum})
and (\ref{eq:dif}) to obtain a set of two relativistic equations one for
each particle with each equation specifying two generalized mass-shell
constraints 
\begin{equation}
\mathcal{H}_{i}\Psi (x_{1},x_{2})=(p_{i}^{2}+m_{i}^{2}+\Phi )\Psi
(x_{1},x_{2})=0,~i=1,2,  \label{dir}
\end{equation}
including the interaction with the other particle. These constraint
equations are just those of Dirac's Hamiltonian constraint dynamics for
spinless particles \cite{dirac,cnstr,cra84s}. In order for Eq. (\ref{dir})
to have consistent solutions, Dirac's constraint dynamics stipulate that
these two constraints must be compatible among themselves, $[\mathcal{H}_{1},
\mathcal{H}_{2}]\Psi =0$, that is, they must be first class constraints.
This requires that the quasipotential $\Phi $ satisfy $[p_{1}^{2}-p_{2}^{2},
\Phi ]\Psi =0$. Working out the commutator shows that for this to be true in
general, $\Phi $ must depend on the relative coordinate $x=x_{1}-x_{2}$ only
through its component, $x_{\perp },$ perpendicular to $P,$ 
\begin{equation}
x_{\perp }^{\mu }=(\eta ^{\mu \nu }+\hat{P}^{\mu }\hat{P}^{\nu
})(x_{1}-x_{2})_{\nu }.  \label{ti}
\end{equation}
The invariant $x_{\perp }^{2}\equiv r^{2}$ becomes $\mathbf{r}^{2}$ in the
c.m. frame. Since the total momentum is conserved, the single component wave
function $\Psi ~$in coordinate space is a product of a plane wave eigenstate
of $P$ and an internal part $\psi (x_{\perp })$ \cite{cra87}\footnote{
We use the same symbol $P$ for the eigenvalue so that the $w$ dependence of $
m_{w}$ and $\varepsilon _{w}$ in Eq. (\ref{em}) is regarded as an eigenvalue
dependence. The wave function $\Psi $ can be viewed either as a relativistic
2-body wave function (similar in interpretation to the Dirac wave function)
or, if a close connection to field theory is required, related directly to
the Bethe Salpeter wave function $\chi {\bb~}$by \cite{saz92} $\Psi =-\pi
i\delta (P\cdot p)\mathcal{\ H}_{10}\chi =-\pi i\delta (P\cdot p)\mathcal{H}
_{20}\chi $.}.

We find a plausible structure for the quasipotential $\Phi $ by observing
that the one-body Klein-Gordon equation $(p^{2}+m^{2})\psi =({\bb p}
^{2}-\varepsilon ^{2}+m^{2})\psi =0$ takes the form $({\bb  p}
^{2}-\varepsilon ^{2}+m^{2}+2mS+S^{2}+2\varepsilon A-A^{2})\psi =0~$when one
introduces a scalar interaction and time-like vector interaction via the
minimum substitutions $m\rightarrow m+S~$and $\varepsilon \rightarrow
\varepsilon -A$. In the two-body case, separate classical \cite{fw} and
quantum field theory \cite{saz97} arguments show that when one includes
world scalar\ and vector interactions then $\Phi $ depends on two underlying
invariant functions $S(r)$ and $A(r)$ ($r=\sqrt{x_{\perp }^{2}}$) through
the two body Klein-Gordon-like potential form with the same general
structure, that is 
\begin{equation}
\Phi =2m_{w}S+S^{2}+2\varepsilon _{w}A-A^{2}.  \label{em}
\end{equation}
Those field theories further yield the c.m.\ energy dependent forms 
\begin{equation}
m_{w}=m_{1}m_{2}/w,  \label{mw}
\end{equation}
and 
\begin{equation}
\varepsilon _{w}=(w^{2}-m_{1}^{2}-m_{2}^{2})/2w,  \label{ew}
\end{equation}
ones that Tododov \cite{cnstr,tod71} introduced as the relativistic reduced
mass and effective particle energy for the two-body system.\ Similar to what
happens in the nonrelativistic two-body problem, in the relativistic case\
we have the motion of this effective particle taking place as if it were in
an external field \ (here generated by $S$ and $A$). \ The two kinematical
variables (\ref{mw}) and (\ref{ew}) are related to one another by the
Einstein condition 
\begin{equation}
\varepsilon _{w}^{2}-m_{w}^{2}=b^{2}(w),
\end{equation}
where the invariant 
\begin{equation}
b^{2}(w)\equiv
(w^{4}-2w^{2}(m_{1}^{2}+m_{2}^{2})+(m_{1}^{2}-m_{2}^{2})^{2})/4w^{2},
\label{bb}
\end{equation}
is the c.m. value of the square of the relative momentum expressed as a
function of $w$. One also has 
\begin{equation}
b^{2}(w)=\varepsilon _{1}^{2}-m_{1}^{2}=\varepsilon _{2}^{2}-m_{2}^{2},
\end{equation}
in which $\varepsilon _{1}$ and $\varepsilon _{2}$ are the invariant c.m.
energies of the individual particles satisfying 
\begin{equation}
\ \varepsilon _{1}+\varepsilon _{2}=w,\ \varepsilon _{1}-\varepsilon
_{2}=(m_{1}^{2}-m_{2}^{2})/w.  \label{es}
\end{equation}
In terms of these invariants, the relative momentum appearing in Eq. (\ref
{bsen}) and (\ref{eq:dif}) is given by 
\begin{equation}
p^{\mu }=(\varepsilon _{2}p_{1}^{\mu }-\varepsilon _{1}p_{2}^{\mu })/w 
\mathrm{,}  \label{relm}
\end{equation}
so that $\mu _{1}+\mu _{2}=(\varepsilon _{1}+\varepsilon _{2})/w=1$. In \cite
{tod} the forms for these two-body effective kinematic variables are given
sound justifications based solely on relativistic kinematics, supplementing
the dynamical arguments of \cite{fw} and \cite{saz97}.

This covariant and useful three-dimensional truncation of the Bethe-Salpeter
equation has been extended to the case of a two-fermion system where the two
constraint equations become the two body Dirac equations (TBDE)\ \cite
{cra82,cra82,cww,crater2,tmlk,unusual} 
\begin{align}
\mathcal{S}_{1}\psi & \equiv \gamma _{51}(\gamma _{1}\cdot (p_{1}-\tilde{A}
_{1})+m_{1}+\tilde{S}_{1})\Psi =0,  \notag \\
\mathcal{S}_{2}\psi & \equiv \gamma _{52}(\gamma _{2}\cdot (p_{2}-\tilde{A}
_{2})+m_{2}+\tilde{S}_{2})\Psi =0.  \label{tbde}
\end{align}
Here $\Psi $ is a sixteen component wave function consisting of an external
plane wave part that is an eigenstate of $P$ and an internal part $\psi
=\psi (x_{\perp })$. The vector potential$\ \tilde{A}_{i}^{\mu }$ was taken
to be an electromagnetic-like four-vector potential with the time-like and
space-like portions both arising from a single invariant function $A(r)$.
\footnote{
In particular, in a perturbative context that would mean that these aspects
of $\tilde{A}_{i}^{\mu }$ were regarded as arising from a Feynman gauge
vertex coupling of a form proportional to $\gamma _{1}^{\mu }\gamma _{2\mu }A
$ .} The tilde on these four-vector potentials indicate that they are not
only position dependent but also spin dependent by way of the gamma
matrices. The operators $\mathcal{S}_{1}$ and $\mathcal{S}_{2}$ must commute
or at the very least $[\mathcal{S}_{1},\mathcal{S}_{2}]\psi =0$ since they
operate on the same wave function\footnote{
The $\gamma _{5}$ matrices for each of the two particles are designated by $
\gamma _{5i}$ $i=1,2$. \ The reason for putting these matrices out front of
the whole expression is that including them facilitates the proof of the
compatibility condition, see \cite{cra82,cra87}.}. This compatibility
condition gives restrictions on the spin dependencies of the vector and
scalar potentials, 
\begin{equation}
\tilde{A}_{i}^{\mu }=\tilde{A}_{i}^{\mu }(A(r),p_{\perp },\hat{P},w,\gamma
_{1},\gamma _{2}),  \label{paul1}
\end{equation}
in addition to requiring that they depend on the invariant separation $
r\equiv \sqrt{x_{\perp }^{2}}$ through the invariant $A(r)$. The covariant
constraint (\ref{eq:dif}) can also be shown to follow from Eq. (\ref{tbde}).
We give the explicit connections between $\tilde{A}_{i}^{\mu }$ and the
invariant $A(r)$ in Appendix A. (A similar dependence occurs for $\tilde{S}
_{i}$ on $S(r).$) \ The general structural dependence on $A(r)$ and $S(r)$
and the spin dependence of $\tilde{A}_{i}^{\mu },\tilde{S}_{i}$ is a
consequence of the compatibility condition $[\mathcal{S}_{1},\mathcal{S}
_{2}]\psi =0.$

The Pauli reduction of these coupled Dirac equations lead to a covariant Schr
\"{o}dinger-like equation for the relative motion with an explicit
spin-dependent potential $\Phi $, 
\begin{equation}
{\bigg(}p_{\perp }^{2}+\Phi (S(r),A(r),p_{\perp },\hat{P},w,\sigma
_{1},\sigma _{2}){\bigg)}\psi _{+}=b^{2}(w)\psi _{+},  \label{schlike}
\end{equation}
with $b^{2}(w)$ playing the role of the eigenvalue\footnote{
Due to the dependence of $\Phi $ on $w,$ this is a nonlinear eigenvalue
equation.}. This eigenvalue equation can then be solved for the
four-component effective particle spinor wave function $\psi _{+}$ related
to the sixteen component spinor $\psi (x_{\perp })$ in appendix A.

In Appendix A we outline the steps needed to obtain the explicit c.m. form
of Eq. (\ref{schlike}). That form is \cite{liu}, \cite{saz94}, \cite
{crater2,tmlk,unusual} 
\begin{eqnarray}
&\{&{\bb p}^{2}+\Phi ({\bb r,}m_{1},m_{2},w,{\bb\sigma }_{1},{\bb\sigma }
_{2})~\}\psi _{+}  \notag \\
&=&\{{\bb p}^{2}+2m_{w}S+S^{2}+2\varepsilon _{w}A-A^{2}+\Phi _{D}  \notag \\
&&+{\bb L\cdot (\bb\sigma }_{1}{+\bb\sigma }_{2}{)}\Phi _{SO}+{\bb\sigma }
_{1}{\bb\cdot \hat{r}\sigma }_{2}{\bb\cdot \hat{r}L\cdot (\sigma }_{1}{\bb
+\sigma }_{2}{\bb)}\Phi _{SOT}  \notag \\
&&+{\bb\sigma }_{1}{\bb\cdot \sigma }_{2}\Phi _{SS}+(3{\bb\sigma }_{1}{\bb
\cdot \hat{r}\sigma }_{2}{\bb\ \cdot \hat{r}-\bb\sigma }_{1}{\bb\cdot \sigma 
}_{2})\Phi _{T}  \notag \\
&&+{\bb L\cdot (\sigma }_{1}{\bb-\sigma }_{2}{\bb)\Phi }_{SOD}+i{\bb L\cdot
\sigma }_{1}{\bb\times \sigma }_{2}\Phi _{SOX}\}\psi _{+}  \notag \\
&=&b^{2}\psi _{+},  \label{57}
\end{eqnarray}
where the detailed forms of the separate quasipotentials $\Phi _{i}$ are
given in Appendix A. The subscripts of most of the quasipotentials are self
explanatory\footnote{
The subscript on quasipotential $\Phi _{D}$ refers to Darwin. It consist of
what are called Darwin terms, those that are the two-body analogue of terms
that accompany the spin-orbit term in the one-body Pauli reduction of the
ordinary one-body Dirac equation, and ones related by canonical
transformations to Darwin interactions \cite{fw,sch73}, momentum dependent
terms arising from retardation effects. The subscripts on the other
quasipotentials refer respectively to $SO$ (spin-orbit), $SOD$ (spin-orbit
difference), $SOX$ (spin-orbit cross terms), $SS$ (spin-spin), $T$ (tensor), 
$SOT$ (spin-orbit-tensor).}. \ After the eigenvalue $b^{2}$ of (\ref{57}) is
obtained, the invariant mass of the composite two-body system $w$ can then
be obtained by inverting Eq.\ (\ref{bb}). It is given explicitly by 
\begin{equation}
w=\sqrt{b^{2}+m_{1}^{2}}+\sqrt{b^{2}+m_{2}^{2}}.
\end{equation}
For this reason we call the operator that appears to the left of Eq. (\ref
{57}) the invariant mass operator. The structure of the linear and quadratic
terms in Eq. (\ref{57}) as well as the Darwin and spin-orbit terms, are
plausible in light of the discussion given above Eq. (\ref{em}), and in
light of the static limit Dirac structures that come about from the Pauli
reduction of the Dirac equation. Their appearance as well as that of the
remaining spin structures are direct outcomes of the Pauli reductions of the
simultaneous TBDE Eq. (\ref{tbde}). In this paper we take the scalar
interaction $S(r)=0$.

\section{TBDE SINGLE COMPONENT WAVE EQUATIONS.}

The 4 component two-body wave function $\psi _{+}$ of the above Pauli-form ( 
\ref{57}) of the TBDE can be conveniently represented by spin-singlet $S=0$
and spin-triplet $S=1$ components with quantum numbers $\{J,L,S\}$ and basis
wave functions 
\begin{equation}
\langle {\bb r}|wJLS\rangle \equiv \psi _{JLS}({\bb r})=\frac{u_{JLS}(r)}{r}
Y_{JM}(\mathbf{\hat{r})}.
\end{equation}
In general, the singlet and triplet states are coupled. However, we see from
Appendix B that for the case of equal masses and certain angular momentum
states, the spin singlet and spin triplet components decouple, and the TBDE
reduce to a single component equation.

Specifically, for the spin-singlet $S=0$ state with $J=L$, (the $^{1}J_{J}$
state), the TBDE is 
\begin{equation}
\{-\frac{d^{2}}{dr^{2}}+\frac{J(J+1)}{r^{2}}+2\varepsilon _{w}A-A^{2}+\Phi
_{D}{-}3\Phi _{SS}\}u_{JJ0}=b^{2}u_{JJ0},  \label{ss}
\end{equation}
where, using the results in Appendix B, the magnetic interaction $-3\Phi
_{SS}$ is 
\begin{eqnarray}
-3\Phi _{SS} &=&-3\Phi _{SS}(A,A^{\prime },{\nabla }^{2}A)=-3(\frac{1}{r^{2}}
-\frac{3}{2r}\left( \frac{A^{\prime }}{w-2A}\right) )((\frac{1}{\sqrt{1-2A/w}
}+\sqrt{1-2A/w})-2)  \notag \\
&&-\frac{3}{2r}\left( \frac{A^{\prime }}{w-2A}\right) (\frac{1}{\sqrt{1-2A/w}
}-\sqrt{1-2A/w})\mathcal{-}\frac{21}{2}\left( \frac{A^{\prime }}{w-2A}
\right) ^{2}-\frac{3{\bb\nabla }^{2}A}{w-2A}  \label{20} \\
&=&-\Phi _{D}(A,A^{\prime },{\nabla }^{2}A),  \notag
\end{eqnarray}
which is attractive and singular, as we discussed in the Introduction. At
large distances and for $A=-\alpha /r$ potential, ${\bb\nabla }^{2}A=4\pi
\alpha \delta ({\bb r})$ and the spin-spin interaction indeed becomes a
singular interaction as described in \cite{Jac62}. In addition to the
magnetic spin-spin interaction, there is also the repulsive Darwin
quasipotential $\Phi _{D}$. \ In the $^{1}J_{J}$ state, the attractive
magnetic spin-spin quasipotential in the spin-singlet configuration exactly
cancels the repulsive Darwin quasipotential, 
\begin{equation}
{-}3\Phi _{SS}+\Phi _{D}=0.  \label{sd}
\end{equation}
As a result of this remarkable cancellation, the eigenvalue equation for the 
$^{1}J_{J}$ state in Eq.\ (\ref{20}) becomes simply 
\begin{equation}
\{-\frac{d^{2}}{dr^{2}}+\frac{J(J+1)}{r^{2}}+2\varepsilon
_{w}A-A^{2}\}u_{JJ0}=b^{2}u_{JJ0}.
\end{equation}
Of all spin-singlet states, only in the $^{1}S_{0}$ states ($J=L=0)$ do the effects of the quasipotential and the absence of a  centrifugal barrier make the combined quasipotential strongly
attractive at short-distances.    This, of course, would not happen were it
not for the highly attractive spin-spin interaction discussed in the
Introduction and in Eq.\ (\ref{20}).  Among the spin-singlet states with different $J$ quantum
numbers, we shall therefore focus our attention only on the $^{1}S_{0}$
states.

For the spin-triplet $S=1$ states, there are two states with single
component radial equations. The first is the $^{3}J_{J}$ state whose radial
equation takes the form ($J\geq 1$) 
\begin{equation}
\{-\frac{d^{2}}{dr^{2}}+\frac{J(J+1)}{r^{2}}+2\varepsilon _{w}A-A^{2}-\frac{
2A^{\prime }}{r\left( w-2A\right) }+3\left( \frac{A^{\prime }}{w-2A}\right)
^{2}+\frac{{\bb\nabla }^{2}A}{w-2A}\}u_{J1J}=b^{2}u_{JJ1}.
\end{equation}
The second is the $^{3}P_{0}$ equation which takes the form 
\begin{equation}
\{-\frac{d^{2}}{dr^{2}}+\frac{2}{r^{2}}+2\varepsilon _{w}A-A^{2}-\frac{
8A^{\prime }}{r\left( w-2A\right) }+8\left( \frac{A^{\prime }}{w-2A}\right)
^{2}+\frac{2{\bb\nabla }^{2}A}{w-2A}\}u_{011}=b^{2}u_{011}.  \label{3p0}
\end{equation}
Of the two spin-triplet cases, only in the $^{3}P_{0}$ states ($J=0,$ $L=1)~$
do the combined effects of the quasipotentials become so strongly attractive
at short-distances that they overwhelm the presence of the centrifugal
barrier. As discussed in the Introduction, this is due to the highly
attractive spin-orbit interaction (\textquotedblleft magnetic" interaction)
when the total spin and the orbital angular momentum are oppositely aligned.
In that case, the competing effects of both the short-distance attraction
and the presence of the potential barrier raise the question whether the
attraction is strong enough to hold a resonance state in the continuum.
Among the spin-triplet states with different $J$ and $L$ quantum numbers, we
shall therefore focus our attention only on the $^{3}P_{0}$ states.

In the last term of the quasipotential in Eq.\ (\ref{3p0}), the quantity $
\nabla ^{2}A$ is related to the particle charge density, $\rho (\bb r)$,
seen by each of the two particles by 
\begin{equation}
\nabla ^{2}A(\bb r)=4\pi \alpha \rho (\bb r).  \label{3p0A}
\end{equation}
Therefore, the equation for the two-body relative wave function for the $
^{3}P_{0}$ state becomes 
\begin{equation}
\biggl \{-\frac{d^{2}}{dr^{2}}+\frac{2}{r^{2}}+2\varepsilon _{w}A-A^{2}- 
\frac{8A^{\prime }}{r\left( w-2A\right) }+8\left( \frac{A^{\prime }}{w-2A}
\right) ^{2}+\frac{8\pi \alpha \rho (\bb r)}{w-2A}\biggr \}
u_{011}=b^{2}u_{011}.  \label{25}
\end{equation}

As we shall see in this case, the attractive magnetic interaction overwhelms
the centrifugal barrier, allowing the wave function to reach the
short-distance region where the particle charge density $\rho (r)$, if any,
can be exposed for scrutiny. This is in contrast to the situation for states
in which the centrifugal barrier dominates the short-distance region. In
that case, the centrifugal barrier will prevent the wave function from
reaching the short-distance region and the particle charge density will not
make as a significant difference in observable quantities\footnote{
Such would also be the case for $^{1}S_{0}$ states in which, due to the
cancellation in Eq. (\ref{sd}), the dependence on $\rho (r)$ is only
indirect or implicit through the altered form for $A(r)$.}. We obtain the
important result that the ${}^{3}P_{0}$ quasipotential depends explicitly on
the particle charge density $\rho (\bb r)$ at short distances. As a
consequence, some observable quantities may depend more critically on the
nature of the particle charge distribution and the forces binding the charge
elements together.

For the $^3P_0$ state, it is convenient to separate out the centrifugal
barrier $2/r^{2}$ and the quasipotential $\Phi $ to write the above equation
as 
\begin{equation}
\biggl \{-\frac{d^{2}}{dr^{2}}+\frac{2}{r^{2}}+\Phi (\bb r)\biggr \}
u_{011}=b^{2}u_{011},  \label{3P0}
\end{equation}
where 
\begin{equation}
\Phi (\bb r)=2\varepsilon _{w}A-A^{2}-\frac{8A^{\prime }}{r\left(
w-2A\right) }+8\left( \frac{A^{\prime }}{w-2A}\right) ^{2}+\frac{8\pi \alpha
\rho (\bb r)}{w-2A}.  \label{27}
\end{equation}

In our early work \cite{bckr}, we limited our attention to energy regions
around 1.579 MeV for the $(e^{+}e^{-})$ system and searched for $^3 P_0$
resonances whose wave functions start from the origin in the usual way. We
found no resonance states. We return to this problem again including now an
additional (peculiar) solution of the TBDE that was overlooked in the
earlier work but has quantum-mechanically acceptable behaviors at short
distances.

In Eqs.\ (\ref{25})), both the gauge field $A(r)$ and the gauge field source 
$\rho (r)$ appear in the equation of motion for the wave function in the $
^{3}P_{0}$ state. The appearance of the fermion charge source distribution $
\rho (r)$ brings into focus the question whether it is sufficient to
describe the magnetic interaction in the $^{3}P_{0}$ state completely within
quantum electrodynamics or quantum chromodynamics. Electrons in QED and
quarks in QCD are taken to be point particles with no structure. It may be
necessary to go beyond these field theories, to include additional auxiliary
interactions that hold the charge elements together, in order to properly
describe the internal structure of these particles. If these auxiliary
interactions act on the charged elements of the fermion to hold them
together, they can also act on the charged elements of the antifermion
charge and will affect the $^{3}P_{0}$ wave function in the interior region
of the charge distribution $\rho (r)$.

The nature of these auxiliary forces holding the charged elements together
is completely unknown, although there have been many attempts to carry out
such an investigation. For example, in the Dirac's model of an electron, a
surface tension from an unknown axillary interaction is invoked to hold the
electric charged elements of an electron together \cite
{Dir48,Dir51,Dir52,Dir53,Dir62,Dir65}. However, our knowledge on the
internal structures of electrons and quarks remain very uncertain. We shall
return to examine how such a lack of knowledge of the internal structures of
these elementary quanta leads to uncertainties in the $^{3}P_{0}$ magnetic
resonance states in Sec.\ VC.

\section{SOLUTIONS OF THE TWO BODY DIRAC EQUATIONS FOR THE $^{1}S_{0}$ STATE}

\subsection{ The $^{1}S_{0}$ quasipotential}

We first consider the case of the $^{1}S_{0}$ state of a point
fermion-antifermion pair with electric or color charges interact through an
electromagnetic-type interaction arising from the exchange of a single
photon or gluon. The single photon annihilation diagram does not contribute
because the $^{1}S_{0}$ state is a charge parity even state. We thus have 
\begin{equation}
A=-\frac{\alpha }{r}.
\end{equation}
For brevity of notation in this subsection, we shall abbreviate the radial
wave function $u_{0JJ}$=$u_{000}$ as $u$. Equation \ (\ref{ss}) for $u$
becomes 
\begin{equation}
\left\{ -\frac{d^{2}}{dr^{2}}-\frac{2\varepsilon _{w}\alpha }{r}-\frac{
\alpha ^{2}}{r^{2}}\right\} u=b^{2}u.  \label{up}
\end{equation}
with a short distance ($r<<\alpha /2\varepsilon _{w})$ behavior given by 
\begin{equation}
\left\{ -\frac{d^{2}}{dr^{2}}-\frac{\alpha ^{2}}{r^{2}}\right\} u=0,
\end{equation}
with solutions 
\begin{eqnarray}
u_{+} &\sim &r^{\lambda +1},  \notag \\
u_{-} &\sim &r^{-\lambda },  \notag \\
\lambda &=&(-1+\sqrt{1-4\alpha ^{2}})/2,  \label{a1}
\end{eqnarray}
or 
\begin{equation}
u_{\pm }\sim r^{(1\pm \sqrt{1-4a^{2}})/2},  \label{39}
\end{equation}
both of which approach zero as $r$ approaches zero. With these behaviors,
the probability 
\begin{equation}
\psi _{\pm }^{2}d^{3}r=\frac{u_{\pm }^{2}}{r^{2}}r^{2}drd\Omega =u_{\pm
}^{2}drd\Omega =r^{(1\pm \sqrt{1-4a^{2}})}drd\Omega ,  \label{a3}
\end{equation}
is finite for both signs. We call $u_{+}$ the usual solution, and it behaves
as $r^{\lambda +1}\sim r^{1-\alpha ^{2}}$ for small $\alpha $. We call $
u_{-} $ the peculiar solution, and it behaves as $r^{-\lambda }\sim
r^{\alpha ^{2}} $for small $\alpha $. Both of these behaviors are physically
acceptable near the origin in the sense of (i) $u(0)$$\rightarrow 0$, and
(ii) being square-integrable in the neighborhood of $r\sim 0$. \ \ We note
that if the sign in front of $\alpha ^{2}$ were positive or if we had
non-zero angular momentum such that $L(L+1)-\alpha ^{2}>0$ then the second
or peculiar set of solutions are not physically admissible states.

In \cite{khel} one finds a thorough discussion on the proper boundary
condition for the radial wave function of the Schr\"{o}dinger equation at
the origin. They discuss several conditions that appear in the literature:
(1) Continuity of $R=u/r$ at $r=0,$ requiring $u(0)=0$. (2) A finite
differential probability in the spherical slice $(r,r+dr$), that is $
R^{2}r^{2}dr<\infty $ requiring $u(r)\rightarrow r^{s+1},s>-1$ and again $
u(0)=0$. (3) Requiring a finite total probability inside a sphere of small
radius $a$ which allows a more singular behavior, namely $u(r)\rightarrow
r^{-1/2+\varepsilon }$ where $\varepsilon >0$ is a small positive constant,
which would also include a finite behavior of the norm. (4) Requiring time
independence of the norm leading to $u(r)\rightarrow cr^{-s+1}$, $s<1$ which
again leads to $u(0)=0.$ Reference \cite{khel} furthermore shows that the
radial Schr\"{o}dinger equation [$
-d^{2}/dr^{2}+l(l+1)/r^{2}+2mV(r)]u(r)=2mEu(r)$ is compatible with the full
Schr\"{o}dinger equation ($-\nabla ^{2}+2mV(r))\frac{u(r)}{r}Y_{lm}=2mE\frac{
u(r)}{r}Y_{lm}$ if and only if the condition$~u(0)=0$ is satisfied. This $
u(0)=0$ condition is clearly satisfied for both solutions in Eq. (\ref{39}).

In Schiff's Quantum Mechanics \cite{schiff}, a solution similar to the
peculiar one discussed here is examined for the case of the Klein-Gordon
equation for the Coulomb system. He argues that what we call the peculiar
solution can be discarded since the source of the Coulomb attraction is a
finite sized nucleus of radius $r_{0}$. In particular, he states that for $r<r_{0}$ for which the potential is finite all the way to
the origin, matching at $r_{0}$ would rule out the peculiar solution. In our
case, with point particles, the potential does not satisfy this condition.

\subsubsection{$^{1}S_{0}$ Bound States}

The  solutions of the $^1S_0$ bound states can be obtained analytically.
In Appendix C we show how we can obtain the two sets of $^{1}S_{0}$ bound state solutions that
correspond to the usual and peculiar short distance behaviors. The
respective sets of eigenvalues and normalized eigenfunctions for the state
with total invariant c.m. energy (mass)  $w_{\pm n}$ and the principle quantum
number $n$ is 
\begin{eqnarray}
w_{\pm n} &=&m\sqrt{2+2/\sqrt{1+{\alpha ^{2}}/{(}n\pm \sqrt{1/4-\alpha ^{2}}
-1/2{)^{2}}}},  \notag \\
u_{\pm n}(r) &=&\left[ \left( \frac{4\varepsilon _{w_{\pm }}\alpha r}{n_{\pm
}^{\prime }}\right) ^{2}\frac{n_{r}!}{2n_{\pm }^{\prime }(n_{\pm }^{\prime
}+\lambda _{\pm })!}\right] ^{1/2}\exp (-\frac{\varepsilon _{w_{\pm }}\alpha
r}{n_{\pm }^{\prime }})\left( \frac{2\varepsilon _{w_{\pm }}\alpha r}{n_{\pm
}^{\prime }}\right) ^{\lambda _{\pm }+1}L_{n_{r}}^{2\lambda _{\pm }+1}(\frac{
2\varepsilon _{w_{\pm }}\alpha r}{n_{\pm }^{\prime }}),  \label{uw}
\end{eqnarray}
where $n_{\pm }^{\prime }=n_{r}+\lambda _{\pm }+1=n+\lambda _{\pm }$ and 
\begin{eqnarray}
\varepsilon _{w_{\pm }}=(w_{\pm }^{2}-2m^{2})/2w_{\pm }.
\end{eqnarray}
For the usual
states $u_{+n}$, the bound state eigenvalues $w_{+n}$ agree with standard
QED perturbative results through order $\alpha ^{4}$, 
\begin{equation}
w_{+n}=2m-m{\alpha ^{2}}/{4}n^{2}-m\alpha ^{4}/2n^{3}(1-11/32/n)+O(\alpha
^{6}),~n=1,2,3,...
\end{equation}
For the set of peculiar states $u_{-n}$, note that the peculiar ground state 
$u_{-1}$ with $n=1$ has eigenenergy (mass) 
\begin{equation}
w_{-1}=m\sqrt{2+2/\sqrt{1+{\alpha ^{2}}/({1/2}-\sqrt{1/4-\alpha ^{2}}{)^{2}}}
}\sim \sqrt{2}m\sqrt{1+\alpha },
\end{equation}
which represents very tight binding, with a binding energy on the order 300
keV for an $e^{+}e^{-}$ state and a root-mean-square radius on the order of
a Compton wave length instead of an angstrom. In particular we find (see
Appendix C) 
\begin{equation}
\sqrt{\langle r^{2}\rangle _{-1}}\rightarrow \frac{1}{m}.
\end{equation}
We note further the anti-intuitive behavior of the peculiar ground state
energy (mass), increasing with increasing coupling constant $\alpha $
instead of decreasing. The excited states are quite near to the usual bound
states. We find the following pattern for those excited peculiar states 
\begin{equation}
w_{-n}=2m-m{\alpha ^{2}}/{4(}n-1)^{2}+m\alpha ^{4}/2(n-1)^{3}\left(
1+11/32(n-1)\right) +O(\alpha ^{6});\text{ }n=2,3,4,...
\end{equation}
\ In the nonrelativistic limit, where terms of order $\alpha ^{4}$ are
ignored we find that the states are degenerate with the $n$th usual state
identical to the $(n+1)$th peculiar state. \ If \ we include the $\alpha
^{4} $ corrections then we find that
\begin{equation}
w_{+n}-w_{-(n+1)}=-m\alpha ^{4}/n^{3}.
\end{equation}
$\ $For all of the usual states and the remaining peculiar states we have
\begin{eqnarray}
\langle r^{2}\rangle _{+n} &=&\frac{(n+\lambda _{+})^{2}}{6\left(
\varepsilon _{w_{+n}}\alpha \right) ^{2}}[(n+\lambda _{+})^{2}+5\alpha ^{2}+3]
\text{,~}n=1,2,3...,  \notag \\
\langle r^{2}\rangle _{-n} &=&\frac{(n+\lambda _{-})^{2}}{6\left(
\varepsilon _{w_{-n}}\alpha \right) ^{2}}[(n+\lambda _{-})^{2}+5\alpha ^{2}+3]
\text{,~}n=2,3...,
\end{eqnarray}
so that the size in the $(n+1)$th peculiar state is nearly the same as with
the $n$th usual state.

As shown in the Appendix C, the two sets of solutions, are not orthogonal
with respect to one another. \ For example, the two $n=1$ wave functions
have the respective forms 
\begin{eqnarray}
u_{+}(r) &=&c_{+}r^{\lambda _{+}+1}\exp (-\kappa _{+}\varepsilon
_{w_{+}}\alpha r),  \notag \\
\kappa _{+} &=&\frac{2}{1+\sqrt{1-4\alpha ^{2}}}=\frac{1}{\lambda _{+}+1}, 
\notag \\
u_{-}(r) &=&c_{-}r^{\lambda _{-}+1}\exp (-\kappa _{-}\varepsilon
_{w_{-}}\alpha r),  \notag \\
\kappa _{-} &=&\frac{2}{1-\sqrt{1-4\alpha ^{2}}}=\frac{1}{\lambda _{-}+1},
\end{eqnarray}
where for brevity of notation, we have omitted the principal quantum number
designation in $u\pm $ for the of the ground state. Clearly since they are
both zero node solutions we have 
\begin{equation}
\langle u_{-}|u_{+}\rangle =\int_{0}^{\infty }dru_{+}(r)u_{-}(r)\neq 0.
\label{ny}
\end{equation}
How do we reconcile this with the expected orthogonality of the
eigenfunctions of a self-adjoint operator corresponding to different
eigenvalues. In the present context, the naive self-adjoint property
requires that 
\begin{equation}
\langle u_{+}|(-\frac{d^{2}}{dx^{2}}-\frac{2}{x}-\frac{\alpha ^{2}}{x^{2}}
)|u_{-}\rangle =\langle u_{-}|(-\frac{d^{2}}{dx^{2}}-\frac{2}{x}-\frac{
\alpha ^{2}}{x^{2}})|u_{+}\rangle .
\end{equation}
This boils down to 
\begin{equation}
\int_{0}^{\infty }dxu_{+}\frac{d^{2}u_{-}}{dx^{2}}=\int_{0}^{\infty }dxu_{-}
\frac{d^{2}u_{+}}{dx^{2}}.
\end{equation}
Let us integrate by parts. \ \ Then we have
\begin{eqnarray}
\int_{0}^{\infty }dxu_{+}\frac{d^{2}u_{-}}{dx^{2}} &=&\left( u_{+}\frac{
du_{-}}{dx}\right) \biggr |_{0}^{\infty }-\int_{0}^{\infty }dx\frac{du_{+}}{
dx}\frac{du_{-}}{dx} \\
&=&\left( u_{-}\frac{du_{+}}{dx}\right) \biggr |_{0}^{\infty
}-\int_{0}^{\infty }dx\frac{du_{+}}{dx}\frac{du_{-}}{dx}.  \notag
\end{eqnarray}
We thus have a self-adjoint operator if 
\begin{equation}
\left( u_{+}\frac{du_{-}}{dx}\right) \biggr |_{0}^{\infty }=\left( u_{-}
\frac{du_{+}}{dx}\right) \biggr |_{0}^{\infty }.  \label{lr}
\end{equation}
Now clearly these vanish at the upper end points. \ Since we have that
\begin{eqnarray}
\frac{du_{+}}{dx} &=&u_{+}(\frac{\lambda _{+}+1}{x}-\frac{1}{\lambda _{+}+1}
),  \notag \\
\frac{du_{-}}{dx} &=&u_{-}(\frac{\lambda _{-}+1}{x}-\frac{1}{\lambda _{-}+1}
).
\end{eqnarray}
at the lower end point the LHS of Eq. (\ref{lr}) is
\begin{eqnarray}
&&\underset{x\rightarrow 0}{\lim }x^{\lambda _{+}+1}\exp (-x/(\lambda
_{+}+1))x^{\lambda _{-}+1}\exp (-x/(\lambda _{-}+1))(\frac{\lambda _{-}+1}{x}
-\frac{1}{\lambda _{-}+1})  \notag \\
&=&\underset{x\rightarrow 0}{\lim }x(\frac{\lambda _{-}+1}{x}-\frac{1}{
\lambda _{-}+1})=\lambda _{-}+1
\end{eqnarray}
whereas at the lower end point the RHS is $\lambda _{+}+1\neq \lambda _{-}+1$.
 Thus,  the second derivative is not self-adjoint in this context!  This accounts for the non-orthogonality of the usual and peculiar ground
states in Eq. (\ref{ny}).

In general beginning with a set of usual and peculiar wave functions $
\{u_{+n},u_{-n}\}$ such that
\begin{eqnarray}
\langle u_{+n}|u_{+n^{\prime }}\rangle  &=&\delta _{nn^{\prime }},  \notag \\
\langle u_{-n}|u_{-n^{\prime }}\rangle  &=&\delta _{nn^{\prime }},  \notag \\
\langle u_{-n}|u_{+n}\rangle  &\equiv &b_{nn^{\prime }}=b_{n^{\prime }n},
\end{eqnarray}
we find with 
\begin{eqnarray}
H &\equiv &\frac{1}{\left( \varepsilon _{w}\alpha \right) ^{2}}\left[ -\frac{
d^{2}}{dr^{2}}-\frac{2\varepsilon _{w}\alpha }{r}-\frac{\alpha ^{2}}{r^{2}}
\right] =\left[ -\frac{d^{2}}{dx^{2}}-\frac{2}{x}-\frac{\alpha ^{2}}{x^{2}}
\right] ,  \notag \\
Hu_{\pm n} &=&h_{\pm n}u_{\pm n},  \notag \\
h_{\pm n} &\equiv &-\kappa _{\pm n}^{2}=-1/(\lambda _{\pm }+n)^{2},~n=1,2,...
\end{eqnarray}
where $x=\varepsilon _{w}\alpha r$, that
\begin{eqnarray}
\langle u_{+n}|H|u_{+n^{\prime }}\rangle  &=&\delta _{nn^{\prime }}h_{+n} 
\notag \\
\langle u_{-n}|H|u_{-n^{\prime }}\rangle  &=&\delta _{nn^{\prime }}h_{-n}, 
\notag \\
\langle u_{-n}|H|u_{+n^{\prime }}\rangle  &=&b_{nn^{\prime }}h_{+n^{\prime }}
\notag \\
\langle u_{+n}|H|u_{-n^{\prime }}\rangle  &=&b_{nn^{\prime }}h_{-n^{\prime
}}\neq \langle u_{-n^{\prime }}|H|u_{+n}\rangle .
\end{eqnarray}
In the first two terms it does not matter whether the $H$ operators operate
to the left or the right. \ In the last two cases we explicitly have $H$
operating to the right. \ To emphasize that we write them as
\begin{eqnarray}
\langle u_{-n}|(H|u_{+n^{\prime }}\rangle ) &=&b_{nn^{\prime }}h_{+n^{\prime
}},  \notag \\
\langle u_{+n}|(H|u_{-n^{\prime }}\rangle ) &=&b_{nn^{\prime }}h_{-n^{\prime
}}.
\end{eqnarray}
It is evident that with both sets of basis, $H$ is not self-adjoint since\ $
\langle u_{-n}|(H|u_{+n^{\prime }}\rangle )\neq (\langle
u_{-n}|H)|u_{+n}\rangle $ and \ $\langle u_{+n}|(H|u_{-n^{\prime }}\rangle
)\neq (\langle u_{+n}|H)|u_{-n^{\prime }}\rangle $.

Let us see where the non-orthogonality leads us if we treat both basis on an
equal footing. \ In that case a general wave function for the $^{1}S_{0}$
system would be expanded as\footnote{
Strictly speaking we should include the continuum states. \ See section
below \ \ for \ discussion of those states. \ For the purpose here the use
of \ discrete states is sufficient.} 
\begin{equation}
\Psi =\sum_{n_{+}}c_{+n}u_{+n}+\sum_{n_{-}}c_{-n}u_{-n},  \label{ph}
\end{equation}
and applying the variational principle to 
\begin{equation}
\langle H\rangle =\frac{\langle \Psi |H|\Psi \rangle }{\langle \Psi |\Psi
\rangle },
\end{equation}
and defining (we show just a finite $n\times n$ portion of the matrices) 
\begin{eqnarray}
\mathbf{B} &\mathbf{=}&
\begin{bmatrix}
b_{11} & b_{12} & ... & b_{1n} \\ 
b_{21} & b_{22} & ... & b_{2n} \\ 
... & ... & ... & ... \\ 
b_{n1} & b_{n2} & ... & b_{nn}
\end{bmatrix}
,  \notag \\
\mathbf{H}_{+} &=&
\begin{bmatrix}
h_{+1} & 0 & ... & 0 \\ 
0 & h_{+2} & ... & 0 \\ 
... & ... & ... & ... \\ 
0 & 0 & ... & h_{+n}
\end{bmatrix}
,  \notag \\
\mathbf{H}_{-} &=&
\begin{bmatrix}
h_{-1} & 0 & ... & 0 \\ 
0 & h_{-2} & ... & 0 \\ 
... & ... & ... & ... \\ 
0 & 0 & ... & h_{-n}
\end{bmatrix}
,
\end{eqnarray}
then in block form we would have the eigenvalues equation 
\begin{equation}
\begin{bmatrix}
\mathbf{H}_{+} & \mathbf{B\mathbf{H}}_{-} \\ 
\mathbf{B\mathbf{H}_{+}} & \mathbf{H}_{-}
\end{bmatrix}
\begin{bmatrix}
\mathbf{c}_{+} \\ 
\mathbf{c}_{-}
\end{bmatrix}
=-\kappa ^{2}
\begin{bmatrix}
\mathbf{1} & \mathbf{B} \\ 
\mathbf{B} & \mathbf{1}
\end{bmatrix}
\begin{bmatrix}
\mathbf{c}_{+} \\ 
\mathbf{c}_{-}
\end{bmatrix}
\end{equation}
(Note that the way this stands , the matrix on the left is not
self-adjoint.) Multiplying both sides on the right by 
\begin{equation}
\begin{bmatrix}
\mathbf{1} & \mathbf{B} \\ 
\mathbf{B} & \mathbf{1}
\end{bmatrix}
^{-1}=
\begin{bmatrix}
\mathbf{(1-B}^{2})^{-1} & -\mathbf{B(1-B}^{2})^{-1} \\ 
-\mathbf{B(1-B}^{2})^{-1} & \mathbf{(1-B}^{2})^{-1}
\end{bmatrix}
\end{equation}
we obtain
\begin{equation}
\begin{bmatrix}
\mathbf{H}_{+} & \mathbf{0} \\ 
\mathbf{0} & \mathbf{H}_{-}
\end{bmatrix}
\begin{bmatrix}
\mathbf{c}_{+} \\ 
\mathbf{c}_{-}
\end{bmatrix}
=-\kappa ^{2}
\begin{bmatrix}
\mathbf{c}_{+} \\ 
\mathbf{c}_{-}
\end{bmatrix}
.
\end{equation}
It is clear that the eigenvectors corresponding to the eigenvalue sets of $
-\kappa _{+}^{2}$ and $-\kappa _{-}^{2}$ are of the form 
\begin{equation}
\begin{bmatrix}
\mathbf{c}_{+} \\ 
\mathbf{0}
\end{bmatrix}
,
\begin{bmatrix}
\mathbf{0} \\ 
\mathbf{c}_{-}
\end{bmatrix}
.
\end{equation}
From Eq. (\ref{uw}). one recalls that the two sets of basis functions $
\{u_{+n},u_{-n}\}$ have distinctly different behaviors at \ the origin,
corresponding to the usual and peculiar solutions. \ In particular
\begin{eqnarray}
u_{+n}(x) &=&c_{+n}x^{\lambda _{+}+1}\exp (-x)L_{n_{r}}^{2\lambda _{+}+1}(x),
\notag \\
u_{-n}(x) &=&c_{+n}x^{\lambda _{-}+1}\exp (-x)L_{n_{r}}^{2\lambda _{-}+1}(x),
\end{eqnarray}
\ These generalized Laguerre polynomials are orthonormal with respect to
different weight functions $x^{\lambda _{\pm }+1}\exp (-x)$. They would each
correspond to a complete set. \ Together they would constitute an
over-complete set. However, that does not imply that Eq. (\ref{ph}) is
incorrect as it allows for the function $\Psi (x)$ to be a linear
combination of functions with two distinct behaviors at the origin. \
Nevertheless, the set up here is a bit clumsy with questions of completeness
and the non-self-adjoint property remaining.

It should be realized that for the given quasipotential of the type $
-\alpha^2/r^2$ at short distances that is at hand, both the set of usual
states and the peculiar states are physically admissible states. There does
not appear to be reasons to exclude one set as being unphysical, if one is
given the attractive interaction near the origin as it is. We note
however that the peculiar states with the $r^{-\lambda }$ behavior at the
origin are excluded from existence if $\ $coefficient $\lambda (\lambda +1)$
for the $1/r^{2}$ term is greater than zero since that would lead to a $u(r) 
$ that is singular at the origin. Only for interactions with sufficient
attraction at the origin (so that $-1/4\leq $ $\lambda (\lambda +1)<0)$ can
these states be pulled into existence and appear as eigenstates in the
physically acceptable sheet, with regular non-singular radial wave functions
at the origin. It is desirable to find ways to admit both types of physical
states into a larger Hilbert space to accommodate both sets of states with
the mass operator to be self-adjoint and the states to be part of a complete
set. It is reasonable to assign a quantum number which we call
``peculiarity" for a states emerging into the physical sheet in this way as
physically acceptable states. The introduction of the peculiarity quantum
number enlarges the Hilbert space, allows the mass operator to be
self-adjoint, and the set of physically allowed states become a complete
set, as we shall demonstrate.

We introduce a new peculiarity observable $\hat{\zeta}$ with the quantum
number peculiarity $\zeta $ such that 
\begin{eqnarray}
\hat{\zeta}\chi _{+} &=&\zeta \chi _{+}~~\mathrm{with~eigenvalue~}\zeta =+1,
\notag \\
\hat{\zeta}\chi _{-} &=&\zeta \chi _{-}~~\mathrm{with~eigenvalue~}\zeta =-1,
\end{eqnarray}
with the corresponding spinor wave function $\chi _{\zeta }$ assigned to the
states so that a usual state is represented by the peculiarity spinor $\chi
_{+}$, 
\begin{equation}
\chi _{+}= 
\begin{pmatrix}
1 \\ 
0
\end{pmatrix}
,
\end{equation}
and a peculiar state is represented by the peculiarity spinor $\chi _{-}$, 
\begin{equation}
\chi _{-}= 
\begin{pmatrix}
0 \\ 
1
\end{pmatrix}
.
\end{equation}
With this introduction, a general wave function can be expanded in terms of
the complete set of basis functions $\{u_{+n},u_{-n}\}$ as 
\begin{equation}
\Psi =\sum_{\zeta n}a_{\zeta n}u_{\zeta n}\chi _{\zeta },
\end{equation}
where $n$ represent all the spin and spatial quantum numbers of the state
and $\zeta $ the peculiarity quantum number. The variational principle
applied to 
\begin{equation}
\langle H\rangle =\frac{\langle \Psi |H|\Psi \rangle }{\langle \Psi |\Psi
\rangle },  \label{h}
\end{equation}
would lead to
\begin{eqnarray}
Hu_{+n}\chi _{+} &=&-\kappa _{+n}^{2}u_{+n}\chi _{+},  \notag \\
Hu_{-n}\chi _{-} &=&-\kappa _{-n}^{2}u_{-n}\chi _{-}.
\end{eqnarray}

It is clear that in this context the usual and peculiar wave functions are
orthogonal, $H$ is self-adjoint, and the basis states are complete. That
is, 
\begin{equation}
\langle i|j\rangle =\langle u_{\zeta _{i}n_{i}}|u_{\zeta _{j}n_{j}}\rangle
\equiv \int_{0}^{\infty }dru_{\zeta _{i}n_{i}}\chi _{\zeta _{i}}u_{\zeta
_{j}n_{j}}\chi _{\zeta _{j}}=\delta _{\zeta _{i}\zeta _{j}}\delta
_{n_{i}n_{j}}=\delta _{ij}
\end{equation}
and so the set of basis functions $\{u_{+n}\zeta _{+},u_{-n}\zeta _{-}\}$,
containing both the usual states and peculiar states in the enlarged Hilbert
space, form a complete set. We also have 
\begin{eqnarray}
\langle i|H|j\rangle  &=&\langle u_{\zeta _{i}n_{i}}|H|u_{\zeta
_{j}n_{j}}\rangle \equiv \int_{0}^{\infty }dru_{\zeta _{i}n_{i}}\chi _{\zeta
_{i}}Hu_{\zeta _{j}n_{j}}\chi _{\zeta _{j}}  \notag \\
&=&h_{\zeta _{i}}\delta _{\zeta _{i}\zeta _{j}}\delta _{n_{i}n_{j}}=\langle
u_{\zeta _{j}n_{j}}|H|u_{\zeta _{i}n_{i}}\rangle   \notag \\
&=&\langle j|H|i\rangle ,
\end{eqnarray}
so that the mass operator $H$ in this enlarged Hilbert space is
self-adjoint. We see that the introduction of the peculiarity quantum number
resolves the problem of over-completeness property of the basis states and
the non-self-adjoint property of the mass operator.

\subsubsection{$^{1}S_{0}$ Scattering States}

The $^{1}S_{0}$ state equation 
\begin{equation}
\left\{ -\frac{d^{2}}{dr^{2}}-\frac{2\varepsilon _{w(r)}\alpha }{r}-\frac{
\alpha ^{2}}{r^{2}}\right\} u=b^{2}u  \label{62}
\end{equation}
has the same form as the nonrelativistic Schr\"{o}dinger equation for
Coulomb interaction 
\begin{equation}
\left\{ -\frac{d^{2}}{dr^{2}}-\frac{2m\alpha }{r}+\frac{L(L+1)}{r^{2}}
\right\} u=2mE\bar{u}=k^{2}u,
\end{equation}
except that the standard angular momentum term with $L(L+1)$ now take on the
value of $(-\alpha ^{2})$. The two solutions the above equation are given by
the regular $F_{L}$ and irregular $G_{L}$ Coulomb wave functions, 
\begin{align}
\bar{u}& =aF_{L}(\eta ,kr)+cG_{L}(\eta ,kr),  \notag \\
\eta & =-\frac{m\alpha }{k},
\end{align}
with only the regular Coulomb wave function having an acceptable behavior at
the origin. \ The long distance behaviors of the regular and irregular
solutions are 
\begin{eqnarray}
F_{L}(\eta ,kr &\rightarrow &\infty )\rightarrow \mathrm{const}\times \sin
(kr-\eta \log 2kr+\sigma _{L}-L\pi /2),  \notag \\
G_{L}(\eta ,kr &\rightarrow &\infty )\rightarrow \mathrm{const}\times \cos
(kr-\eta \log 2kr+\sigma _{L}-L\pi /2),
\end{eqnarray}
in which $\sigma _{L}$ is the Coulomb phase shift given by 
\begin{equation}
\sigma _{L}=\arg (\Gamma (L+1+i\eta ).
\end{equation}

Now we can solve Eq. (\ref{62}) exactly for $b^{2}>0$ by analytically
continuing the above solutions to an arbitrary (non-integer) angular
momentum $\lambda $ and making a few obvious replacements by analogy, \ 
\begin{align}
\bar{u}& =aF_{\lambda }(\eta ,br)+cG_{\lambda }(\eta ,br),  \notag \\
\lambda (\lambda +1)& =-\alpha ^{2},  \notag \\
\eta & =-\frac{\varepsilon _{w}\alpha }{b}.  \label{ufg}
\end{align}
Using the expressions for the analytically continued Coulomb wave functions
to non-integer $\lambda ~$\cite{klein} we will presently see that we have
solutions given by the $F$ and $G$ functions in Eqs. (\ref{f}) and (\ref{g})
below. We emphasize that both solutions have an acceptable behavior at the
origin. Since $\lambda $ is not an integer, one can replace the irregular
solution $G_{\lambda }(\eta ,br)$ by $F_{-\lambda -1}(\eta ,br)$.\footnote{
The reason that $G_{L}$ is used in place of \ $F_{-L-1}$ for $L$ integer is
that the latter is not linearly independent of $F_{L}$ in that case. It is
melded together with $F_{L}$ to produce $G_{L}$ by a limited process
analogous to how the Neumann function is obtained from the Bessel functions.
For $\lambda \neq $ integer, $F_{\lambda }$ and $F_{-\lambda -1}$ are
linearly independent.} \ \ In particular, as shown in Appendix D, in terms
of the confluent hypergeometric function $M(a,b;z)$ 
\begin{equation}
F_{\lambda }(\rho )=C_{\lambda }(\eta )\rho ^{\lambda +1}\exp (-i\rho
)M(\lambda +1-i\eta ,2\lambda +2;2i\rho ),
\end{equation}
one has with 
\begin{equation}
x(\lambda ,\eta )\equiv (\lambda +\frac{1}{2})\pi +\sigma _{-\lambda
-1}(\eta )-\sigma _{\lambda }(\eta ),
\end{equation}
that 
\begin{equation}
G_{\lambda }(\rho )=\frac{F_{-\lambda -1}(\rho )-\cos x(\lambda ,\eta
)F_{\lambda }(\rho )}{\sin x(\lambda ,\eta )},  \label{fg}
\end{equation}
a linear combination of $F_{\lambda }(\rho )$ and $F_{-\lambda -1}(\rho )$.
In others words, \ Eq. (\ref{ufg}) can be written as 
\begin{equation}
\bar{u}=dF_{\lambda }(\eta ,br)+eF_{-\lambda -1}(\eta ,br),
\end{equation}
where 
\begin{eqnarray}
\lambda &=&\frac{1}{2}(-1+\sqrt{1-4\alpha ^{2}})\equiv \lambda _{+},
\label{pecu} \\
-\lambda -1 &=&\frac{1}{2}(-1-\sqrt{1-4\alpha ^{2}})\equiv \lambda _{-}, 
\notag
\end{eqnarray}
corresponding to the separate $\zeta =\pm 1$ sectors. As with the solutions
in Eq. (\ref{ufg}), $F_{\lambda }(\eta ,br)$ and $F_{-\lambda -1}(\eta ,br)$
have acceptable behaviors at the origin corresponding to Eq. (\ref{39}).
Their respective long distance behaviors are given by 
\begin{align}
F_{\lambda }(\eta ,br& \rightarrow \infty )\rightarrow \mathrm{const}\times
\sin (br-\eta \log 2br+\sigma _{\lambda _{+}}-\lambda _{+}\pi /2),  \notag \\
F_{-\lambda -1}(\eta ,br& \rightarrow \infty )\rightarrow \mathrm{const}
\times \sin (br-\eta \log 2br+\sigma _{\lambda _{-}}-\lambda _{-}\pi /2).
\label{f}
\end{align}
Alternatively we can use the related $\ G$ \ functions to determine the
behaviors 
\begin{eqnarray}
G_{\lambda }(\eta ,br &\rightarrow &\infty )\rightarrow \mathrm{const}\times
\cos (br-\eta \log 2br+\sigma _{\lambda _{+}}-\lambda _{+}\pi /2),  \notag \\
G_{-\lambda -1}(\eta ,br &\rightarrow &\infty )\rightarrow \mathrm{const}
\times \cos (br-\eta \log 2br+\sigma _{\lambda _{-}}-\lambda _{-}\pi /2).
\label{g}
\end{eqnarray}
The respective total Coulomb phase shifts for Eq. (\ref{62}) are the phase
shifts for the usual and peculiar solutions over and above those due to any
angular barrier part (absent here). They are given by 
\begin{align}
\delta _{\lambda _{\pm }}& =\sigma _{\lambda _{\pm }}-\lambda _{\pm }\pi /2,
\notag \\
\sigma _{\lambda _{\pm }}& =\arg (\Gamma (\lambda _{\pm }+1+i\eta ),
\end{align}
in which 
\begin{align}
\arg \Gamma (\lambda _{\pm }+1+i\eta )& =\eta \psi (\lambda _{\pm
}+1)+\sum_{n=0}^{\infty }\left( \frac{\eta }{\lambda _{\pm }+1+n}-\arctan (
\frac{\eta }{\lambda _{\pm }+1+n})\right) ,  \notag \\
&
\end{align}
with the digamma function given by 
\begin{equation}
\psi (\lambda _{\pm }+1)=-\gamma +\lambda _{\pm }\zeta (2)-\lambda _{\pm
}^{2}\sum_{n=1}^{\infty }\frac{1}{n^{2}(n+\lambda _{\pm })}.
\end{equation}
The (modified) Coulomb phase shift $\sigma _{\lambda _{\pm }}$$-\lambda
_{\pm }\pi /2$ is that for the Coulomb $2\varepsilon _{w}A$ plus $-A^{2}$
term alone. (Again, the $\pm $ sign corresponds to the two sectors $\zeta
=\pm 1$, with usual ($+)$ and peculiar ($-)~$boundary conditions given in
Eq. (\ref{39}).)\ Without the $-A^{2}$ term \ the phase shift would be
simply $\sigma _{0}$.

\bigskip

\section{SOLUTIONS OF THE TWO BODY DIRAC EQUATIONS FOR THE $^{3}P_{0}$ STATE}

\subsection{ The $^{3}P_{0}$ quasipotential}

We now consider the case of the $^{3}P_{0}$ state of a fermion-antifermion
pair with electric or color charges interacting through an
electromagnetic-type interaction arising from the exchange of a single
photon or gluon. As with the $^{1}S_{0}$ state, the single photon
annihilation diagram does not contribute because the $^{3}P_{0}$ state is a
charge parity even state. Then, the two terms in Eq.\ (\ref{3p0}) that
precede the $\nabla ^{2}A$ term precisely cancel the barrier term $2/r^{2}$
at very short distances to give the equation for the radial wave function 
\begin{equation}
\left\{ -\frac{d^{2}}{dr^{2}}+\frac{2}{(r+2\alpha /w)^{2}}-\frac{
2\varepsilon _{w}\alpha }{r}-\frac{\alpha ^{2}}{r^{2}}+\frac{8\pi \alpha
r\delta ({\bb r})}{wr+2\alpha }\right\} u=b^{2}u.  \label{2e}
\end{equation}
The cancellation of terms takes place in the following way. In Eq.\ (\ref
{3p0}), the three terms beyond $-A^{2}$ arise from a combination of
spin-orbit, spin-spin, tensor and spin-orbit tensor interactions. From a
detailed examination of Eq.\ (\ref{3res}) in the Appendix B, we can see that
the spin-orbit and tensor terms gives rise to the first \textquotedblleft
magnetic interaction" term on the right hand side of Eq. (\ref{3res}) that
has a strongly attractive $-8\alpha /wr^{3}$ attractive part down to
distances on the order of $2\alpha /w$ after which this magnetic interaction
approaches $-4/r^{2}$. The dominance of the attractive magnetic interaction
at short distances that can overwhelm the centrifugal barrier is in
agreement with the simple intuitive classical picture presented in the
Introduction. The second term on the right-hand side of Eq. (\ref{3res}),
arising from a combination of Darwin, spin-spin and tensor terms, has a
stronger repulsive $8\alpha ^{2}/w^{2}r^{4}~$ part down to distances on the
order of $2\alpha /w$ after which it approaches $+2/r^{2}$. Together they
tend to exactly cancel the angular momentum barrier term $+2/r^{2}$ at very
short distances. In addition to the repulsive interaction containing $\delta
(\bb r)$ arising from the assumption that the electron and positron are
point particles, the quasipotential behaves as $-\alpha ^{2}/r^{2}$ at
short-distances, separated from the outside long-distance region by a
barrier. The interaction containing the delta function comes from a
combination of Darwin, spin-spin, and tensor terms. Three fourths of the
repulsive term containing $\delta (\bb r)$ comes from the Darwin piece while
one fourth from the combination of the spin-spin and tensor parts. For
brevity of nomenclature we shall just call it the delta function term.

One of us (HWC) examined in a previous work \cite{atk} the effects on bound
state energies due to a repulsive $\delta (\bb r)$ interaction by itself,
without additional radial dependence. It was found that for wave functions $
\psi $ that do not vanish at the origin and for potentials that are less
singular than $1/r^{2}$, the exact effects on the eigenvalue of including a
repulsive delta function do not agree with the results of perturbation
theory in the limit of weak coupling, when the delta function potential is
modeled as the limit of a sequence of spherically symmetric square wells. In
particular it is shown that the repulsive delta function, viewed as the
limit of square well potentials, produces no effects at all on bound state
energies. In our case here the appearance of the $\delta ({\bb r})$
potential differs from this reference in two aspects however. \ First of all
the $\delta ({\bb r})$ appears in conjunction with $r/(wr+2\alpha )$,
softening its repulsive effects. \ Secondly, the wave function $\psi =u/r$
for the solution without the delta function term diverges at the origin both
for what we call the usual solution and what we call the peculiar solution.
If the null effects on bound state energies and phase shifts seen in \cite
{atk} should occur in our case as well, this, however, does not lead to a
problem with perturbative agreement \ with the spectral results.

In the case of weak potentials where the denominator $(wr+2\alpha)$ is
replaced by $wr$, we have shown previously in \cite{bckr} that the remaining
terms in Eq. (\ref{2e}) without the delta function term, when treated
nonperturbatively, would produce numerically the same spectral results for
the $^{3}P_{0}$ state as the inclusion of the repulsive $\delta (\bb r)$
interaction treated perturbatively. The agreement of the perturbative
treatment with the delta function term for weak coupling with the
nonperturbative treatment containing no delta function term justifies the
first approximate analysis of ignoring the delta function term and treating
the remainder of the equation nonperturbatively in the following subsection.

\subsection{ Usual and Peculiar Solutions for the $^{3}P_{0}$ State}

The wave equation (\ref{2e}) for the $^{3}P_{0}$ state without the delta
function term becomes 
\begin{equation}
\left\{ -\frac{d^{2}}{dr^{2}}+\frac{2}{(r+2\alpha /w)^{2}}-\frac{
2\varepsilon _{w}\alpha }{r}-\frac{\alpha ^{2}}{r^{2}}\right\} u=b^{2}u,
\label{31}
\end{equation}
with a short distance ($r<<2\alpha /w)$ form 
\begin{equation}
\left\{ -\frac{d^{2}}{dr^{2}}-\frac{\alpha ^{2}}{r^{2}}\right\} u=0,
\label{32}
\end{equation}
the same as with the $^{1}S_{0}$ states. Thus, the $^3P_0$ states also have
the same types of solutions as the $^1 S_0$ states, with radial wave
functions near the origin as given in Eqs. (\ref{a1})-(\ref{39}). \ Thus,
there are usual $^{3}P_{0}$ states with peculiarity $1$, and peculiar $
^{3}P_{0}$ states with peculiarity $-1$.

Note that both the usual and the peculiar solutions $u_{\pm }$ $\sim
r^{(1\pm \sqrt{1-4a^{2}})/2}$ arise from the strong magnetic interaction
that significantly modifies the qualitative behavior of the interaction at
short distances, when the total spin and the orbital angular momentum are
oppositely aligned in the $^{3}P_{0}$ state. If the strong magnetic
interaction is absent, the $2/(r+2\alpha /w)^{2}$ term in Eq.\ (\ref{2e})
would be $2/r^{2}$, and the wave function near the origin would be 
\begin{equation}
u_{\pm }=ar^{(1\pm \sqrt{3^{2}-4a^{2}})/2},
\end{equation}
with 
\begin{equation}
\psi _{\pm }^{2}d^{3}r=r^{[(1\pm \sqrt{3^{2}-4a^{2}})]}drd\Omega .
\label{pec}
\end{equation}
In that case, as stated below Eq. (\ref{a3}), only the usual $u_{+}$
solution is quantum-mechanically admissible, while the $u_{-}$ state becomes
singular at short distances. Such a comparison shows that the peculiar
solution $u_{-}$ is not present when there is no strongly attractive
magnetic interaction at short distances or more generally for $J\neq 0.$

\subsection{The $\protect\delta $ function term and the charge distribution}

The discussions in the above subsection pertain to the quasipotential
without the delta function term. We now examine the full Eq. (\ref{3p0}) for
both the usual and peculiar solutions with the $\delta ({\bb r})$ term
included. Consider first the perturbative treatment of taking the
interaction containing $\delta (\bb r)$ as a perturbation. We evaluate the
expectation value of the interaction term containing $\delta (\bb r)$. Even
though both usual and peculiar solutions have a diverging $\psi _{\pm }({\bb 
r})$ near the origin they each are allowed as a probability amplitude since
the probability $\int_{\Delta V}\left\vert \psi _{\pm }({\bb r})\right\vert
^{2}d^{3}r$ for an arbitrarily small volume $\Delta V$ about the origin
would be finite, in addition to the essential boundary condition $u_\pm
(0)=0 $. \ With $\left\vert \psi _{\pm }({\bb r})\right\vert ^{2}$ near the
origin having the behavior of $r^{(-1\pm \sqrt{1-4a^{2}})}$, the expectation
value of $\delta (\bb r)/(w-2A)$, after performing the angular integration,
is 
\begin{eqnarray}
\int d^{3}r\frac{r\delta ({\bb r})}{wr+2\alpha }\left\vert \psi _{\pm }({\bb 
r})\right\vert ^{2} &\rightarrow &\int d^{3}rr^{\pm \sqrt{1-4a^{2}}}\frac{
\delta ({\bb r})}{wr+2\alpha },  \notag \\
&\rightarrow &\int drr^{\pm \sqrt{1-4a^{2}}}\frac{\delta (r)}{(wr+2\alpha )},
\label{div}
\end{eqnarray}
which is zero for the plus sign for the usual solution but diverges for the
minus sign for the peculiar solution.

The results of Eq.\ (\ref{div}) for the usual solution explains our previous
agreement between (i) the perturbative treatment with the delta function
term for weak coupling and (ii) the nonperturbative treatment without the
delta function term \cite{bckr}. The agreement arises because in Ref.\ \cite
{bckr} we limited our attention only to the usual solution for which the
expectation value of the delta function term is zero.

The results of Eq.\ (\ref{div}) for the peculiar solution indicates that the
delta function term cannot be treated as a perturbation in the present
formulation, as such a treatment will lead to a diverging energy. The delta
function term arises from the charge distribution of the interacting
particles, as it is related to the Laplacian of the gauge field, $\nabla
^{2}A$, as given in Eq.\ (\ref{3p0A}). A proper non-perturbative treatment
of the problem of the peculiar solution states requires the knowledge of the
wave function at very short distances. Therefore, it will require not only
the knowledge of the structure of the charge distribution but also the
necessary auxiliary interactions at even shorter distances that are needed
to bind the charge elements of the distribution together. The auxiliary
interactions will affect the solutions of the two-body wave functions at
very short distances and the states of the peculiar solution. At the present
moment, we have little knowledge of the structure of elementary charges,
much less the auxiliary forces that would bind the charge distribution
together at very short distances.

The structure of the charge distribution of elementary particles at very
short distances is basically an experimental question. As the strong
magnetic interaction allows the two interacting particles to probe the
short-distance region, it is therefore useful to investigate quantities that
may reveal information on the structure of the charge distribution. While
many possibilities can be opened for examination, we shall examine the
following possibilities in the present manuscript:

(i) We shall first examine the case in which the (unknown) auxiliary
interaction that binds the charge elements of the elementary particles
together and the repulsive interaction arising from the charge density $\rho
(r)$ counteract in such a way that the total interaction at short distances
would still be dominated by the $-\alpha ^{2}/r^{2}$ term. Under such a
circumstance, the effects of the auxiliary interaction would cause the delta
function term term in Eq. (\ref{2e}) to make no contribution at short
distances. Keeping the dominant terms, the equation of motion for the wave
function becomes Eq.\ (\ref{31}) without the delta function term. It also
must be recognized that for the usual solution, the perturbative effect of
the delta function term (in which we ignore the effect or the potential in
the denominator $w-2A$) is accounted for by a nonperturbative (numerical)
treatment of the entire $\Phi $ without the delta function term. \ So, our
treatment of the delta function term in this case parallels that used in our
earlier spectroscopy calculations \cite{bckr}.

(ii) We examine subsequently the case when the auxiliary interaction that
holds the charge element together leaves the gauge field $A(r)$ unchanged
while the delta function term in Eq.\ (\ref{2e}) is modified by treating the
delta function as the limit of a set of Gaussian distributions with
different widths.

(iii) We examine two additional models completely within QED (or QCD) with
an assumed basic charge distribution that generates the gauge field also in
the region interior to the charge distribution. However, the auxiliary
interactions that hold the charge together and that can interact with the
other antifermion are altogether neglected. It should be recognized that
within pure QED (or QCD), with the neglect of the auxiliary interactions
that hold the charge elements together, the charge distribution cannot be a
stable configuration.

In the next section we describe the method we use to indicate the presence
or absence of a resonance in the $^{3}P_{0}$ system.

\section{PHASE SHIFT ANALYSIS}

In our study of the $^{3}P_{0}$ state for both the usual and peculiar
solutions, we wish to find out whether or not there is an energy that will
lead to a $\pi /2$ phase shift for a given $\alpha $ and constituent mass $m$
. Equation (\ref{3P0}) for the $^{3}P_{0}$ state is a Schr\"{o}dinger-like
equation of the form 
\begin{equation}
\left\{ -\frac{d^{2}}{dr^{2}}+\frac{L(L+1)}{r^{2}}+\Phi (r)\right\}
u(r)=b^{2}u(r).
\end{equation}

We calculate the phase shift for this problem by the variable phase method
of Calogero \cite{cal}. We first describe this method generally (see
Appendix E for a more detailed review) and then later in this section
describe its application to the $^{3}P_{0}$ state. We take $W(r)$ to include
not only the quasipotential $\Phi (r)$ but also the angular momentum
barrier. 
\begin{equation}
W(r)=\frac{L(L+1)}{r^{2}}+\Phi (r).  \label{39a}
\end{equation}
Thus our equation has the form 
\begin{equation}
\left\{ -\frac{d^{2}}{dr^{2}}+W(r)\right\} u(r)=b^{2}u(r).  \label{40}
\end{equation}
The Calogero method relies on introducing a reference potential $\bar{W}(r)$
that can be solved exactly, with two independent solutions $u_{1}$ and $
u_{2} $, 
\begin{equation}
\left\{ -\frac{d^{2}}{dr^{2}}+\bar{W}(r)\right\}
u_{i}(r)=b^{2}u_{i}(r),~i=1,2.  \label{4}
\end{equation}
There are many ways to choose the reference potential $\bar{W}(r)$. To
display the general idea, we consider the case in which $W(r)$ is short
range. In that case the phase shift $\delta _{L}$ is defined by 
\begin{equation}
u(r\rightarrow \infty )\rightarrow \sin (br-L\pi /2+\delta _{L}).  \label{ur}
\end{equation}
The Calogero method uses two different types of $\bar{W}(r)$. In the first, $
\bar{W}(r)\equiv \bar{W}_{I}(r)$, the reference potential has the same long
and short distance behavior as $W(r)$. \ In the second $\bar{W}(r)\equiv
W_{II}(r)$, the reference potential does not have the same long and short
distance behavior as $W(r)$ but is especially simple.

We consider first Type I reference potential, $\bar{W}_{I}(r)=L(L+1)/r^{2}$,
the angular momentum barrier potential, for which the reference wave
functions $\bar{u}_{1}(r)$ and $\bar{u} _{2}(r)$ are the well known
spherical Bessel functions $\hat{\jmath}_{L}(br)$ and $\hat{n}_{L}(br)$ \cite
{cal}. The solution $\bar{u}_{1}(r)$ is taken to be the regular solution,
having the same short distance behavior as $u(r), $ in particular, $\bar{u}
_{1}(r\rightarrow 0)=0.$ The solution $\bar{u} _{2}(r)$ is taken to be the
irregular solution, $\bar{u}_{2}(r\rightarrow 0)\neq 0$. \ Those functions
together with their long distance behaviors are given by 
\begin{align}
\bar{u}_{1}(r)& =\hat{\jmath}_{L}(br)\rightarrow \mathrm{const}\sin (br-L\pi
/2),  \notag \\
\bar{u}_{2}(r)& =-\hat{n}_{L}(br)\rightarrow \mathrm{const}\cos (br-L\pi /2).
\label{jn}
\end{align}

We introduce the amplitude function $\alpha (r)$ and phase shift function $
\delta _{L}(r)$ to represent the wave function solutions for the $W(r)$
potential, $u(r)$ and $u^{\prime }(r)$, as 
\begin{align}
u(r)& =\alpha (r)(\cos \delta _{L}(r)\bar{u}_{1}(r)+\sin \delta _{L}(r)\bar{u
}_{2}(r)),  \notag \\
u^{\prime }(r)& =\alpha (r)(\cos \delta _{L}(r)\bar{u}_{1}^{\prime }(r)+\sin
\delta _{L}(r)\bar{u}_{2}^{\prime }(r)).  \label{a}
\end{align}
This leads to the following equation for the phase shift function (see
Appendix E) 
\begin{equation}
\tan \delta _{L}(r)=\frac{\bar{u}_{1}^{\prime }(r)u(r)-\bar{u}
_{1}(r)u^{\prime }(r)}{(\bar{u}_{2}(r)u^{\prime }(r)-\bar{u}_{2}^{\prime
}(r)u(r))}.  \label{50}
\end{equation}
Further manipulations lead to the differential equation for $\delta _{L}(r)$
given by 
\begin{equation}
\delta _{L}^{\prime }(r)=-\frac{[W(r)-\bar{W}(r)]}{b}\biggl [\hat{\jmath}
_{L}(br)\cos \delta _{L}(r)-\hat{n}_{L}(br)\sin \delta _{L}(r)\biggr ]^{2},
\end{equation}
To find the connection to the phase shift $\delta _{L}$ note that from Eq.\ (
\ref{a}) and (\ref{jn}), 
\begin{align}
u(r& \rightarrow \infty )=\mathrm{const}\left\{ \cos \delta
_{L}(r\rightarrow \infty )\sin (br-L\pi /2)+\sin \delta _{L}(r\rightarrow
\infty )\cos (br-L\pi /2)\right\}  \notag \\
& =\mathrm{const}\times \sin (br-L\pi /2+\delta _{L}(\infty )),
\end{align}
and so comparison with (\ref{ur}) gives the solution of the phase shift $
\delta _{L}$ for the $W(r)$ potential as 
\begin{equation}
\delta _{L}=\delta _{L}(\infty ).
\end{equation}
Thus, the second order linear different equation becomes a first order
non-linear equation whose solution at $r\rightarrow \infty $ gives the phase
shifts of the scattering problem with the $W(r)$ effective potential.\ The
boundary condition of $\delta _{L}(0)=0$ follows from Eq. (\ref{50}) when
one chooses $\bar{u}_{1}(r)$ to have the same behavior as $u(r)$ as $
r\rightarrow 0.$

We consider next type II of the short-range reference potentials $\bar{W}
_{II}(r)$ which do not need to have the same long distance behavior as $W(r)$
as long as the Schr\"{o}dinger Eq.\ (\ref{4}) containing the reference
potential $\bar{W}_{II}(r)$ has an exact solution. For example we may choose 
$\bar{W}_{II}(r)=0.$ Then the two exact reference solutions of Eq. (\ref{4})
are simply 
\begin{align}
\bar{u}_{1}(r)& =\sin (br),  \notag \\
\bar{u}_{2}(r)& =\cos (br).
\end{align}
One defines a phase shift function $\gamma _{L}(r)$ as in equation (\ref{a})
so that 
\begin{align}
u(r& \rightarrow \infty )=\mathrm{const}\times \{\cos \gamma
_{L}(r\rightarrow \infty )\sin (br)+\sin \gamma _{L}(r\rightarrow \infty
)\cos (br)\}  \notag \\
& =\mathrm{const}\times \sin (br+\gamma _{L}(\infty )).
\end{align}
Comparison with (\ref{ur}) gives 
\begin{equation}
\delta _{L}=\gamma _{L}(\infty )+\frac{L\pi }{2}.  \label{dg}
\end{equation}
Since the angular momentum barrier is excluded from the equations for $\bar{u
}_{i}(r)$ one finds that the phase shift equation for integrating the phase
shift function $\gamma _{L}(r)$ includes the repulsive barrier term in $W$
[Eq.\ (\ref{39})], 
\begin{align}
\gamma _{L}^{\prime }(r)& =-\frac{W(r)}{b}\biggl [\cos \gamma _{L}(r)\sin
(br)+\sin \gamma _{L}(r)\cos (br)\biggr ]^{2}  \notag \\
& =-\frac{W(r)}{b}\sin ^{2}(br+\gamma _{L}(r)).  \label{3}
\end{align}
Note that because of the $L(L+1)/r^{2}$ behavior of $W(r)-\bar{W}
_{II}(r)(r)=W(r)$, which dominates at large distances, one will have to
integrate quite far to obtain convergence for $\gamma _{L}(r)$.\footnote{
Alternatively Calogero gives a formula for avoiding integrating to large
distances to build up a centrifugal phase shift. \ (See \cite{cal}, p 92).}
For this case of $\bar{W}_{II}(r)=0$, one has an equation similar to (\ref
{50}) with $\delta _{L}(r)$ replaced by $\gamma _{L}(r)$. Thus even though $
\bar{u}_{1}(r)$ has a different behavior than $u(r)$, we still have the
boundary condition $\gamma _{L}(0)=0.$ Eq. (\ref{dg}) compensates for the $
-L\pi /2$ effective phase shift due to the barrier term in $W(r)$ in Eq. (
\ref{3}).\ We also have the additional boundary condition of (see Appendix
E) 
\begin{equation}
\gamma _{L}^{\prime }(0)=-\frac{bL}{L+1}.  \label{glp}
\end{equation}
$\allowbreak $

We now turn our attention to the $^{3}P_{0}$ system, in particular Eq. (\ref
{3P0}) for a general $\Phi (r)$. \ In this application of the Calogero
method we choose a reference potential $\bar{W}(r)\equiv \bar{W}_{III}(r)$
that in a sense is a hybrid of the\ two types of reference potentials
considered above. \ Since Eq. (\ref{3P0}) contains a long range Coulomb
interaction $-2\varepsilon _{w}\alpha /r$ we must include that interaction
into our choice for $\bar{W}_{III}(r)$. \ If it did not have the same
behavior as $W(r)$ at large distances we would have to have a way of
subtracting an infinite Coulomb phase shift, $\log 2br$ . \ So, in this way
our application is similar to the first type $\bar{W}_{I}$ (r)above. \ We
also include the $-\alpha ^{2}/r^{2}$ term in $\bar{W}_{III}(r)$ because as
seen in Eqs. (\ref{31}), (\ref{32})$~$\ and (\ref{39}) the solution displays
the desired short distance peculiar as well as usual behaviors. We do not
include the angular momentum barrier term $2/r^{2}$ however, as this would
prohibit a treatment of the peculiar solution (see comments below Eq. (\ref
{pec})). Thus we choose 
\begin{eqnarray}
\bar{W}_{III}(r) &=&-\frac{2\varepsilon _{w}\alpha }{r}-\frac{\alpha ^{2}}{
r^{2}},  \notag \\
W(r) &=&\frac{2}{r^{2}}+\Phi (r),  \label{5}
\end{eqnarray}
where \ $\Phi (r)$ is given in Eq.\ (\ref{27}). In this way $\bar{W}_{III}(r)
$ has some similarities to the second type $\bar{W}_{II}(r)$ discussed
above. Our choice for $\bar{W}_{III}(r)$ permits the two exact solutions $
\bar{u}_{1}(r),\bar{u}_{2}(r)$ of Eq. (\ref{4}) which becomes that of the $
^{1}S_{0}$ state 
\begin{equation}
\left\{ -\frac{d^{2}}{dr^{2}}-\frac{2\varepsilon _{w(r)}\alpha }{r}-\frac{
\alpha ^{2}}{r^{2}}\right\} \bar{u}=b^{2}\bar{u}.  \label{rel}
\end{equation}
Now to determine the phase shift for the actual $^{3}P_{0}$ state we return
to the conditions defined in Eq. (\ref{5}). Then the full solution has the
asymptotic form 
\begin{equation}
u(r\rightarrow \infty )\rightarrow \mathrm{const}\times \sin (br-\eta \log
2br+\sigma _{1}-\pi /2+\delta _{1}).
\end{equation}
The appearance of $\sigma _{1}$ and $\delta _{1}$ includes the effects of
the angular momentum barrier term $2/r^{2}$ in the presence of the Coulomb
interaction. In Appendix E, using 
\begin{eqnarray}
\bar{u}_{1}(r) &=&F_{\lambda }(\eta ,br),  \notag \\
\bar{u}_{2}(r) &=&G_{\lambda }(\eta ,br),
\end{eqnarray}
we show that the full $^{3}P_{0}$ phase shift $\delta $ is given by 
\begin{equation}
\delta =\delta _{1}+\sigma _{1}=\gamma _{\pm }(\infty )+\sigma _{\lambda
_{\pm }}+(1-\lambda _{\pm })\pi /2,
\end{equation}
where (in analogy to the proof of Eq. (\ref{3}) with $\bar{W}\neq 0)~$ $
\gamma _{\pm }(r)$satisfies the nonlinear equation 
\begin{equation}
\gamma _{\pm }^{\prime }(r)=-\frac{W(r)-\bar{W}_{III}(r)}{b}\biggl [\cos
\gamma _{\pm }(r)F_{\lambda _{\pm }}+\sin \gamma _{\pm }(r)G_{\lambda _{\pm
}}\biggr ]^{2},  \label{ps}
\end{equation}
subject to the boundary condition that $\gamma _{\pm }(0)=0$ (see Appendix
E). The functions $F_{\lambda _{\pm }}$ and $G_{\lambda _{\pm }}$ are the
regular and irregular Coulomb wave functions corresponding to the negative
effective centrifugal barrier $-\alpha ^{2}/r^{2}$. \ Again, because of the $
2/r^{2}$ behavior of $W(r)-\bar{W}_{III}(r)$ which takes over at large
distances, one will have to integrate quite far to obtain convergence for $
\gamma _{\pm }(r).$

We consider numerical solutions for both the usual solution with $\lambda
_{+}=(-1+\sqrt{1-4\alpha ^{2}})/2$, and the peculiar solution, with $\lambda
_{-}=(-1-\sqrt{1-4\alpha ^{2}})/2.~$In the next section we discuss the
results obtained in the numerical integration of the phase shift equation ( 
\ref{ps}) for different behaviors at very short distances.

\section{NUMERICAL RESULTS FOR $^3 P_0$ RESONANCES}

\subsection{ The case without the delta function term}

With the above general formalism, we can begin to examine states in the
quasipotential of Eq. (\ref{2e}) first without the delta function term. The
Schr\"{o}dinger equation for the $^{3}P_{0}$ state becomes Eq. (\ref{31}).
In order to gain an idea on the attractive magnetic interaction at short
distances for this ${}^{3}P_{0}$ state, we plot in Fig.\ 2 the corresponding
quasipotential including the angular momentum barrier, 
\begin{equation*}
W(r)=\frac{2}{(r+2\alpha /w)^{2}}-\frac{2\epsilon _{w}\alpha }{r}-\frac{
\alpha ^{2}}{r^{2}},
\end{equation*}
for $w=27.85$ MeV, $\alpha =1/137$ and a constituent electron mass of 0.511
MeV. One observes that at short distances $W(r)$ becomes very attractive and
behaves as $-\alpha ^{2}/r^{2}$. There is a barrier in the region between 10$
^{-2}$ to 10$^{-1}$ GeV$^{-1}$. Such a potential becomes singular at $
r\rightarrow 0$ when $\alpha $ exceeds 1/2 \cite{Case}.

\begin{figure}[h]
\includegraphics[scale=0.40]{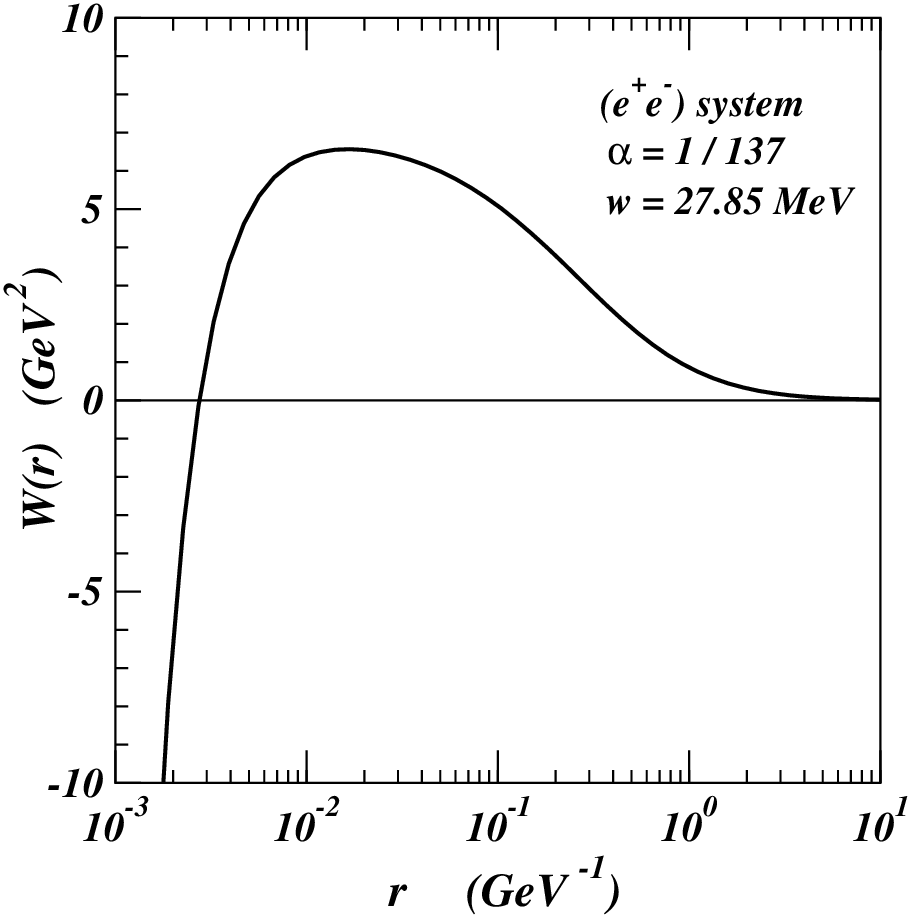} 
\caption{The effective potential $W(r)= 2/(r+2\protect\alpha/w)^{2}-2\protect\epsilon_w\protect\alpha/r-\protect\alpha^2/r^2 $ for the for the $(e^+e^-)$
system in the $^{3}P_0$ state with $\protect\alpha=1/137$ and $w=27.85 $
MeV. }
\end{figure}

We calculate the phase shift as a function of energy using the boundary
condition $\gamma _{\pm }(0)$=0, including the dependence of the potential
as a function of energy. For the usual solution ($\zeta =+1)$, our results
for the QED $e^{-}e^{+}$ system in the $^{3}P_{0}$ state with $\alpha =1/137$
and $m=0.511$ MeV show no evidence whatsoever for resonances for all c.m.
energies tested (from about 1 MeV to about 100 MeV). The magnitude of the
phase shifts are of the order of $\pi /100$.

\begin{table}[h]
\caption{ Variation of the resonant energy as a function of the quark mass
for a fixed $\protect\alpha_s=0.11$.}\vspace*{0.3cm} 
\begin{tabular}{lclcl}
\hline
quark & mass~~ & $w_{R}$ &  &  \\ \hline
up & ~~~3 MeV & 27 MeV &  &  \\ 
down & ~~~5 MeV & 45 MeV &  &  \\ 
strange~~ ~ & 135 MeV & 1220 MeV &  &  \\ 
charm & ~1.5 GeV & 13.6 GeV &  &  \\ 
bottom & ~4.5 GeV & 40.8 GeV &  &  \\ 
top & ~~~175 GeV~~~ & 1590 GeV &  &  \\ \hline
\end{tabular}
\end{table}

For the peculiar solution ($\zeta =-1)~$with the wave functions starting
with a less positive slope, the attraction at short distances is able to
bend the wave function downward to result in a very sharp resonance at about 
$27.85$ MeV. \ In Figure 3(a) we plot the phase shift $\delta =\delta
_{1}+\sigma _{1}$ as a function of the c.m.\ energy $w$ and $\sin ^{2}\delta 
$ versus $w$ in Fig.\ 3 (b). \ We start the integration at the origin and
extend to about 1 angstrom. As one observes, the phase shift undergoes a
transition from near zero to $\pi $. The resonance has a full width at half
maximum of 15 KeV. We also include a plot of the wave function in Fig. 4
from the origin up to about 1000 GeV$^{-1}$. The wave function rises as $
r^{(1-\sqrt{1-4\alpha ^{2}})/2}$ near the origin, and appears nearly flat at 
$r\sim 10^{-3}$ GeV$^{-1}$, and it slowly decreases near the barrier. It
oscillates when it emerges from the barrier at $r\sim 2\times 10^{-2}$ GeV$
^{-1}$.

\begin{figure}[h]
\includegraphics[scale=0.45]{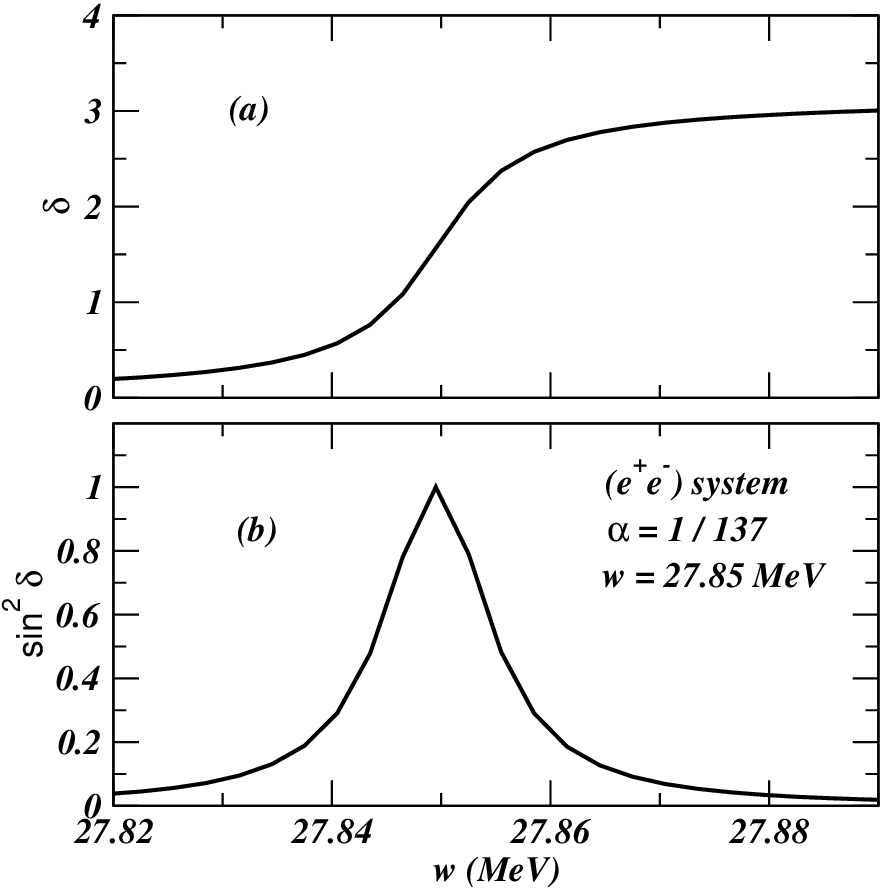} 
\caption{ The phase shift as a function of $w$ for the $(e^+e^-)$ system
with $\protect\alpha=1/137$ and $w=27.85$ MeV.}
\end{figure}

\begin{figure}[h]
\includegraphics[scale=0.45]{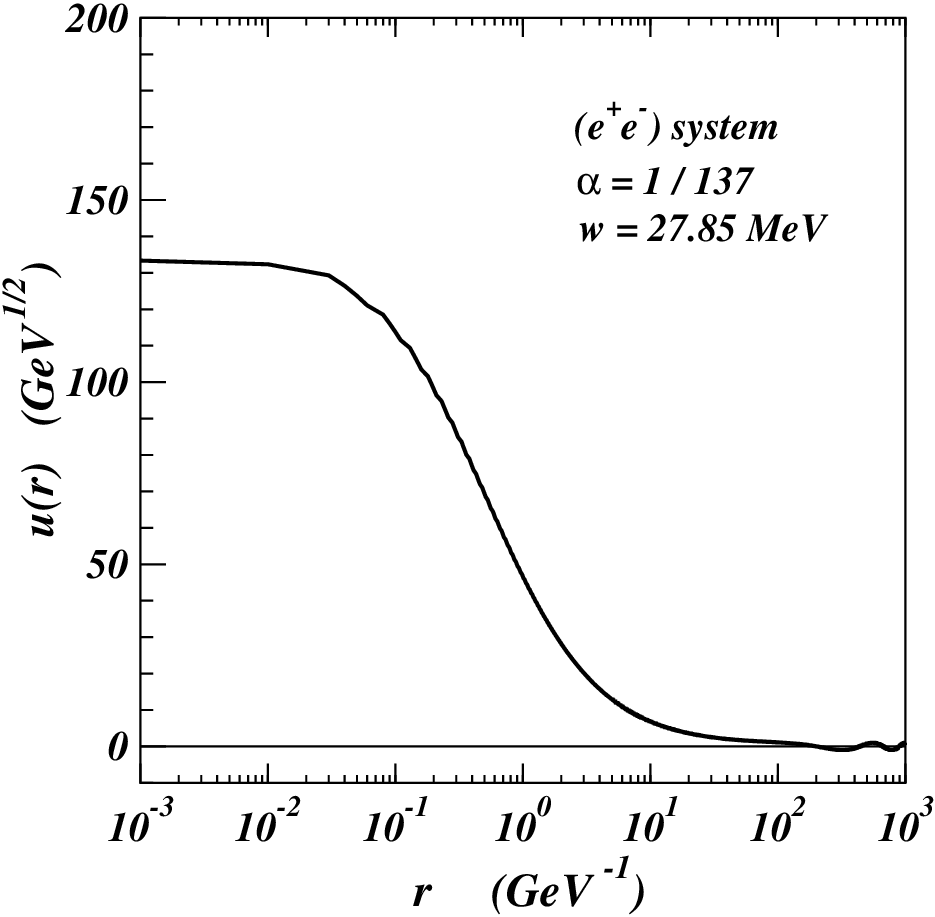} 
\caption{ The wave function $u$ of the peculiar resonance at $w=27.85$ MeV
for $\protect\alpha=1/137$ and $m=0.511$ MeV.}
\end{figure}

Having observed a resonance for the QED interaction with the $e^{+}$ and $
e^{-}$ constituents, we turn our attention to quarks and antiquarks
interacting with a color-coulomb type interaction with an effective coupling
constant $\alpha _{s}$. We focus here only on the $\zeta =-1$ sector. In the
color-singlet $(q\bar{q})$ states of interest, the effective interaction is
then $\alpha _{\mathrm{eff}}=4\alpha _{s}/3$. To get an idea of the order of
energy for these quark-antiquark two-body resonance states, we calculate the
resonance energies for the typical case of $\alpha _{s}=0.11$ For this
value, the resonance energy varies nearly linearly with quark mass. The
largest energy resonances occur with the largest quark masses. In Table I we
present the resonance energies $w_{R}$ for the families of quarks from the
up quark to the top quark. \ It should be pointed out that these resonance
values take into account only the Coulomb-like portion of $A(r)=-(4/3)\alpha
_{s}/r$ and ignores any affects on the resonance values of the confining
part of the potential.

To examine how the resonance energies varies with the coupling constant, we
have found that for fixed mass (e.g. 0.511 MeV) the resonance energy $w_{R}$ 
\textit{increases} as the coupling parameter \textit{\ decreases } until the
coupling constant gets to be on the order of $0.01$, when $w_{R}$ starts
decreasing again.

\subsection{ The case of representing the delta function by a Gaussian
function}

For the second case for the $^{3}P_{0}$ state given in Eqs.\ (\ref{2e}) and (
\ref{40}) using (\ref{ps}), we take 
\begin{equation}
W(r)=\frac{2}{(r+2\alpha /w)^{2}}-\frac{2\varepsilon _{w}\alpha }{r}-\frac{
\alpha ^{2}}{r^{2}}+\frac{8\pi \alpha r\delta ({\bb r})}{\left( wr+2\alpha
\right) },
\end{equation}
in which we model the three dimensional delta function by 
\begin{equation}
\delta ({\bb r})\rightarrow \delta _{\sigma }({\bb r})=\frac{\exp
(-r^{2}/2\sigma ^{2})}{(2\pi )^{3/2}\sigma ^{3}}.
\end{equation}

\begin{table}[h]
\caption{Variation of resonance energy with the width of the Gaussian
distribution}\vspace*{0.3cm} 
\begin{tabular}{ll}
\hline
$\sqrt{2}\sigma (\mathrm{fm})$ ~~~~~~~ & $w$(GeV) \\ \hline
$1000$ & $0.0279$ \\ 
$100$ & $0.0278$ \\ 
$10$ & $0.0279$ \\ 
$1$ & $0.0398$ \\ 
$0.1$ & $0.314$ \\ 
$0.01$ & $3.13$ \\ 
$0.001$ & $31.3$ \\ 
$0.0001$ & $313$ \\ 
$0.00001$ & $3130$ \\ 
$0.000001$ & $31300$ \\ 
$0.0000001$ & $313000$ \\ \hline
\end{tabular}
\end{table}

In this treatment, we keep the point charge source term for the $A(r)$ so
that $A(r)=-\alpha /r$. \ What we are attempting to do is just present a
mathematical representation of the delta function that will allow a
numerical solution. \ The reference potential $\bar{W}_{III}(r)$ is the same
as without the delta function. \ With this modeling of the delta function we
start off our Runge-Kutta integration of Eq. (\ref{ps}) with $\gamma
_{-}(0)=0$ since $W(r)-\bar{W}_{III}(r)$ is \ $w^{2}2\alpha ^{2}$ at the
origin just as without the delta function term. The function $\delta
_{\sigma }$ does not alter the extreme short distance\ behavior since it is
multiplied by $r$ and vanishes at the origin. We obtain the resonance energy
results as given in Table II. It is obvious that for small $r_{0}$ we obtain
a limiting behavior of $w=3.13$ GeV-fm/$\sqrt{2}\sigma $. There is
however, a difference between what we are doing here and what was done in 
\cite{atk}. There the delta function was just regarded as given, not related
to other parts of the potential. Here that is not the case. \ The delta
function arose from the Laplacian of $A(r).$ \ There may therefore be some
ambiguity of, in effect, modeling $\nabla ^{2}A$ in one part of the
quasipotential while leaving $A(r)$ unaffected in the other part. That leads
us then to the third case.

\subsection{ The case of representing the charge distribution by a
continuous function}

In Eq.\ (\ref{3p0}), if one replaces $A(r)$ by 
\begin{equation}
A=\left( \frac{\alpha }{r}-\frac{\alpha }{r_{0}}\right) \frac{1}{1+\exp
\{(r-r_{0})/\delta r_{0}\}}-\frac{\alpha }{r},
\end{equation}
or alternatively as 
\begin{equation}
A(r)=
\begin{cases}
\frac{\alpha r^{2}}{2r_{0}^{3}}-\frac{3\alpha }{2r_{0}} & \mathrm{~for~}
r\leq r_{0}\cr-\frac{\alpha }{r} & \mathrm{~for~}r\geq r_{0}.\cr
\end{cases}
.
\end{equation}
then our numerical solutions show that there is no ${}^{3}P_{0}$ resonance
for the peculiar solution for both cases, resulting in a phase shift of $\pi 
$ all the way down to threshold ($w=2m$). Both of these corresponds to
smeared charge distributions from $\nabla ^{2}A$ but neither have auxiliary
interactions at short distances that would bind the elements of the charge
distribution together. The reason no resonance is produced in this case is
that in the interior of the charge distribution ($r<r_{0}$), the angular
momentum barrier in Eq.\ (\ref{3p0}) comes out from under the dominance of
the magnetic interaction terms as $A(r)$ tends to a finite constant. By
following steps similar to those used to determine $\gamma _{-}(0)$ in
Appendix E for the point charge one can show this results in an initial
value for $\gamma _{-}(0)$ defined by 
\begin{equation}
\tan \gamma _{-}(0)=\tan x(\lambda ,\eta ).
\end{equation}
This is positive and even though small ($\sim 0.007$) is large enough to
prevent the formation of a resonance. \ Note that this differs from the
previous section in that here we are giving a physical connection $\nabla
^{2}A=4\pi \alpha \rho (\bb r)$ between the smeared delta function and the
invariant potential $A(r)$, whereas in the previous section we simply
mathematically modeled the delta function in isolation.

\section{DISCUSSION AND CONCLUSION}

Magnetic interactions in the $^{1}S_{0}$ and $^{3}P_{0}$ states are very
attractive and singular at short distances. In the two-body Dirac equation
formulated in constraint dynamics, the magnetic interactions lead to
quasipotentials that behave as $-\alpha ^{2}/r^{2}$ near the origin and
admit two different types of states. At short distances, the radial wave
functions $u(r)$ of the usual states, grow as $r^{\lambda +1}$, while the
radial wave functions of the peculiar states grow as $r^{-\lambda }$, where $
\lambda =(-1+\sqrt{1-4\alpha ^{2}})/2$. They have drastically different
properties.

The existence of usual and peculiar states for the same fermion-antifermion
system poses conceptual and mathematical problems. If we keep both sets of
states in the same Hilbert space, then each set is complete by itself, but
the two sets of states are not orthogonal to each other. Our system is thus
over-complete. Furthermore, the matrix element of $H$ (the scaled invariant
mass operator for these states) between states of one type and state of the
other type are not symmetric and the $H$ operator is not self-adjoint.

Given our quasipotential of the type $-\alpha ^2/r^2$ at short distances for
the $^{1}S_{0}$ and $^{3}P_{0}$ states, both the usual and peculiar states are
physically admissible. There do not appear to be compelling reasons to
exclude one of the two sets as being unphysical, if one is given the
attractive interaction $-\alpha ^{2}/r^{2}$ near the origin as it is. It is
desirable to find ways to admit both types of states as physical states
while maintaining the self-adjoint property of the mass operator and the
completeness property of the set of basis states.

We are therefore motivated to introduce a quantum number $\zeta$, which we
call ``peculiarity", to specify the usual or peculiar properties of a state.
The peculiarity quantum number $\zeta$ is 1 for usual states which have
properties the same as those one usually encounters in QED and QCD. The
peculiarity quantum number is $-1$ for peculiar states which intrudes into
the physical region, when the interaction near the origin becomes very
attractive, such as the $\lambda(\lambda+1)/r^2$ interaction with $-1/4
\leq\lambda(\lambda+1)< 0$. The introduction of the peculiarity quantum
number enlarges the Hilbert space, makes the mass operator self-adjoint, and
the enlarged physical basis states containing both usual and peculiar states
in a complete set. It is also clear from our discussions that to maintain the self-adjoint property of the mass operator and to have a single complete set, the presence
of the peculiarity quantum number will be a general phenomenon, when the
 mass operator contains very attractive interactions at short
distances such that there are more than one set of eigenstates satisfying
the boundary conditions at the origin.

It should be emphasized that the quasipotential $-\alpha ^{2}/r^{2}$ has
been obtained under the assumption of a point fermion and a point
antifermion for which the gauge field potential between them is $
A(r)=-\alpha /r$. The point nature of an electron may be a good experimental
concept as the lower limits on the QED cut-off parameter $\Lambda _{\mathrm{
cut}}$ with the present day high energy accelerators exceeds the value of 250
GeV, suggesting that the electron, muon, and tauon, behave as point
particles down to 10$^{-3}$ fm. The asymptotic freedom is a good description
for the interaction of quarks at short distances. It may appear that point
charge particles may be a reasonable description. On the other hand, a
finite structure of the electron or quarks may modify significantly the
short-distance attractive interactions so substantially that the peculiar
states may be pushed out of existence. The experimental search of the
peculiar states, which follows from the point charge potential, can provide
a probe of the point nature of these particles and the interaction at short
distances.

Our first focus on the attractive magnetic interaction is for the $
{}^{1}S_{0}$ states, where the spins of the fermion and antifermion of
opposite electric or color charges are oppositely aligned. The usual bound $
^{1}S_{0}$ states possess attributes the same as those one usually
encounters in QED and QCD, with bound state energies explicitly agreeing
with the standard perturbative results through order $\alpha ^{4}$. In
contrast, the peculiar bound $^{1}S_{0}$ states, yet to be observed, not
only have different behaviors at the origin, but also distinctly different
bound state properties (and scattering phase shifts). For the peculiar $
^{1}S_{0}$ ground state of a fermion-antifermion pair with fermion rest mass 
$m$, the root-mean-square radius is approximately $1/m$, binding energies
approximately $(2-\sqrt{2})m$, and a rest mass approximately $\sqrt{2}m$. On
the other hand, the $(n+1)$${}^{1}S_{0}$ peculiar state with principal
quantum number $(n+1)$ is nearly degenerate in energy and approximately
equal in size with the $n$$^{1}S_{0}$ usual states.

Our second focus is for the $ {}^{3}P_{0}$ state where the total spin and
the orbital angular momentum are oppositely aligned. The magnetic
interaction overwhelms the centrifugal repulsion at short distances and the
wave function admits a peculiar solution that grows with radial distances as 
$u\sim r^{(1- \sqrt{1-4\alpha ^{2}})/2}$. The particle charge density $\rho
(r)$ and auxiliary interactions that bind the charge elements together can
be exposed for scrutiny. As the structures of elementary particles are
basically experimental questions, it is useful to utilize the magnetic
interaction to probe such charge distributions at very short distances.
While many possibilities can be opened for examination, we have investigated
only a few possibilities in the present manuscript.

The $^{3}P_{0}$ quasipotential contains a term proportional to $\delta (\bb 
r)$ . As the delta function term does not contribute to the usual QED $
{}^{3}P_{0}$ bound state energies, it was plausible to ignore it as one of
our explored possibilities. In that case, we find that there is a magnetic $
{}^{3}P_{0}$ resonance at 27.85 MeV for the peculiar solution of the $
(e^{+}e^{-})$ system. For various $(q\bar{q})$ systems of different flavors,
we find magnetic ${}^{3}P_{0}$ resonances at energies of the peculiar
solution ranging from many tens of MeV to thousands of GeV. It is
interesting to note that these ${}^{3}P_{0}^{++}$ resonances have the same
quantum number as the vacuum.

In another one of our explored possibilities, if we mathematically model the
delta function at short distances by a sequence of Gaussians of different
widths without changing the gauge field $A(r)=-\alpha /r,$ then a completely
different behavior for the resonance energies ensues as they occur at
different energies, depending on the width of the Gaussian. In the third of
our explored possibilities, if we replace the delta function by a charge
distribution that also alters the gauge field $A(r)$, we obtain no resonance
at all.

Because of 1) the limited knowledge of the unknown auxiliary interactions
and charge distributions at very short distances, not to mention possible
alterations on the angular momentum barrier itself, and 2) the ambiguity of
treating the delta function in isolation nonperturbatively, and 3) the fact
that the delta function term does not contribute to the $^{3}P_{0}$ usual
bound state solution, we speculate that the first case may provided a more
reliable representation of the physics. It furthermore makes a clear
prediction of a QED resonance in a region that has not been investigated.

While we have studied the resonance $^3 P_0$ states, future work calls for
the investigation of possible $^3 P_0$ peculiar bound states where the
attractive interaction near the origin may allow the formation of bound
states. The presence of a delta function repulsion at the origin will also
lead to difficulties and problems similar to the ones we encounter here with
the $^3 P_0$ peculiar resonances.

Fermion-antifermion states as we know them experimentally belong to the
usual states. Peculiar states have not been observed. Can the peculiar
states be observed? How do the usual and peculiar states interplay between
them? Will there be transitions between the usual states and peculiar
states? Clearly, the stability of peculiar states first and foremost depends
on the strong attraction near the origin, which in turn depends on the
point-like nature of the elementary particles. As we discussed earlier,
substantial modification of the attractive interaction at the origin may
push the peculiar states out of existence. Only for interactions with
sufficient attraction at the origin can the peculiar states be pulled into
existence and appear as eigenstates in the physically acceptable sheet, with
non-singular radial wave functions at the origin. This is true for both $
^{1}S_{0}$ and $^{3}P_{0}$ states. From such a perspective, we expect that
interactions at short distances have important bearings on the existence or
non-existence of the peculiar states, and presumably also on the transition
between the usual and peculiar states. However, the interactions at short
distances that may allow the peculiar states to be stable and may effect
transitions between states with different peculiarity quantum numbers
(flipping the peculiarity spinor) are not yet known. They can only be obtained
by careful experimental investigations. The first task of such
investigations should be to locate these peculiar states in high-energy
experiments where interactions at short-distance may be involved and these
strong interactions at short distances may lead us to probe short-distance
transition from the usual to the peculiar states. These new $^{1}S_{0}$
peculiar bound states correspond to a very tightly bound state and a set of $
(n+1$)th excited states nearly degenerate with the $n$th usual states. It
will also be of interest to search for these states as a result of some
tunneling process between the usual and peculiar states, relying on the
small probability of the usual states to explore short-distance regions
where the interaction at short distances may induce a transition from a
usual state to a peculiar state. The fact that peculiar states of $(n+1)$th$
^{1}S_{0}$ state is nearly degenerate with the usual $n$th $^{1}S_{0}$ state
may facilitate such a tunneling transition. Whether or not these
quantum-mechanically acceptable resonances correspond to physical states
remains to be further investigated. Future experimental as well as
theoretical work on this interesting topic will be of great interest in
shedding light on the question whether magnetic bound states and resonances play any role in
the states of fermion-antifermion systems.

Future work should include the effects of the weak interactions, in
particular the exchange of the $Z^{0}$ boson. Since the mass of the $
Z^{0}$ is about 92.5 GeV the range is on the order of $10^{-2}$ GeV$^{-1}$.
The exchange of this particle corresponds to not only a vector interaction
but also a pseudovector interaction.  The coupling corresponding to the
vector portion is \cite{wein} 
\begin{eqnarray}
e^{\ast } &\equiv &+\frac{g^{2}-g^{\prime 2}}{4\sqrt{g^{2}+g^{\prime 2}}}+
\frac{g^{\prime }}{2},  \notag \\
g &=&-\frac{e}{\sin \theta },  \notag \\
g^{\prime } &=&-\frac{e}{\cos \theta },
\end{eqnarray}
and so 
\begin{eqnarray}
e^{\ast } &=&e[\frac{\frac{1}{\sin ^{2}\theta }-\frac{1}{\cos ^{2}\theta }}{4
\sqrt{\frac{1}{\sin ^{2}\theta }+\frac{1}{\cos ^{2}\theta }}}-\frac{1}{2\cos
\theta }]  \notag \\
&=&e[\frac{\cos 2\theta }{2\sin 2\theta }-\frac{1}{2\cos \theta }].
\end{eqnarray}
With $\sin ^{2}\theta \sim 0.23$ we find that 
\begin{equation}
e^{\ast }\sim -0.25e
\end{equation}
so, its coupling appears with the same sign as that of the photon. Since $
\alpha ^{\ast }=e^{\ast 2}\sim 0.063$\ Its effect should be small but not
negligible. \ There is also the question of the effects of the pseudovector
interaction, not discussed in this appendix but in \cite{jmath},\cite{long}.

Finally, there are however important mathematical and conceptual issues associated with these two-body fermion-antifermion system at short distances that require future careful considerations. In standard QED theory, the charge and mass of a single charged object due to vacuum polarization and self energy corrections need to be renormalized or regularized to render them finite for comparison with observables. For the case with two-body magnetic bound and resonance states,  for example, how are the two-body Green's functions  regularized,  with internal lines off mass shell in a way that reflects the Dirac constraints?  How do such regularizations modify the short distance two-body interaction? Can the regularization affects the magnetic interaction at short distances so substantially that the peculiar states no longer survive to intrude into the physical states? Are these peculiar states stable against fluctuation of the vacuum in quantum field theory. These are some of the many interesting questions associated with the two-body problem raised by the possibility of magnetic states under consideration.

\appendix

\section{Details of the equivalent Relativistic Schr\"{o}dinger Equation}

\subsection{\protect\bigskip\ Connections between TBDE and the equivalent
Relativistic Schr\"odinger equation [Eq. (\protect\ref{57})]}

Here we present an outline of some details of Eq. (\ref{tbde}) and its
Pauli-Schr\"{o}dinger reduction given in full elsewhere (see \cite
{cra87,jmath,long,liu}).This appendix and the one following it are
specializations of Appendices A and B given in \cite{tmlk}. Each of the two
Dirac equations in (\ref{tbde}) has a form similar to a single particle
Dirac equation in an external\ four-vector and scalar potential but here
acting on sixteen component wave function $\Psi $ which is the product of an
external part being a plane wave eigenstate of $P~$multiplying the internal
wave function $\psi $ 
\begin{equation}
\psi = 
\begin{bmatrix}
\psi _{1} \\ 
\psi _{2} \\ 
\psi _{3} \\ 
\psi _{4}
\end{bmatrix}
.
\end{equation}
The four $\psi _{i}$ are each four-component spinor wave functions. To
obtain the actual general spin dependent forms of those $\tilde{A}_{i}^{\mu
} $ potentials (including scalar interactions in general) which were
required by the compatibility condition $[\mathcal{S}_{1},\mathcal{S}
_{2}]\psi =0$ was a most perplexing problem, involving the discovery of
underlying supersymmetries in the case of scalar and time-like vector
interactions \cite{cra82},\cite{cra87}. \ Extending those external potential
forms to more general covariant interactions necessitated an entirely
different approach leading to what is called the hyperbolic form of the
TBDE. \ Their most general form for compatible TBDE is 
\begin{align}
\mathcal{S}_{1}\psi & =(\cosh (\Delta ){\bb S}_{1}+\sinh (\Delta ){\bb S}
_{2})\psi =0\mathrm{,}  \notag \\
\mathcal{S}_{2}\psi & =(\cosh (\Delta ){\bb S}_{2}+\sinh (\Delta ){\bb S}
_{1})\psi =0,  \label{hyp1}
\end{align}
where $\Delta $ represents any invariant interaction singly or in
combination. \ It has a matrix structure in addition to coordinate
dependence. Depending on that matrix structure we have either covariant
vector, scalar or more general covariant tensor interactions \cite{jmath}.
The operators ${\bb S}_{1}$ and ${\bb S}_{2}$ are auxiliary constraints
satisfying 
\begin{align}
{\bb S}_{1}\psi & \equiv (\mathcal{S}_{10}\cosh (\Delta )+\mathcal{S}
_{20}\sinh (\Delta )~)\psi =0,  \notag \\
{\bb S}_{2}\psi & \equiv (\mathcal{S}_{20}\cosh (\Delta )+\mathcal{S}
_{10}\sinh (\Delta )~)\psi =0,  \label{hyp2}
\end{align}
in which the\ $\mathcal{S}_{i0}$ are the free Dirac operators 
\begin{equation}
\mathcal{S}_{i0}=\frac{i}{\sqrt{2}}\gamma _{5i}(\gamma _{i}\cdot
p_{i}+m_{i}).  \label{es0}
\end{equation}
This, in turn leads to the two compatibility conditions \cite
{cww,jmath,saz86} 
\begin{equation}
\lbrack \mathcal{S}_{1},\mathcal{S}_{2}]\psi =0,
\end{equation}
and 
\begin{equation}
\lbrack {\bb S}_{1},{\bb S}_{2}]\psi =0,
\end{equation}
provided that $\ \Delta (x)=\Delta (x_{\perp }).$ These compatibility
conditions do not restrict the gamma matrix structure of $\Delta $. \ That
matrix structure is determined by the type of vertex-vertex structure we
wish to incorporate in the interaction. \ \ The three types of invariant
interactions $\Delta $ that was used in the relativistic quark model based
on this approach (as most recently discussed in \cite{unusual},\cite{tmlk})
are 
\begin{align}
\Delta _{\mathcal{L}}(x_{\perp })& =-1_{1}1_{2}\frac{\mathcal{L}(x_{\perp }) 
}{2}\mathcal{O}_{1},\ \mathcal{O}_{1}=-\gamma _{51}\gamma _{52},~~~\text{
scalar}\mathrm{,}  \notag \\
\Delta _{\mathcal{J}}(x_{\perp })& =\beta _{1}\beta _{2}\frac{\mathcal{J}
(x_{\perp })}{2}\mathcal{O}_{1},~~~\text{time-like\ vector}\mathrm{,}  \notag
\\
\Delta _{\mathcal{G}}(x_{\perp })& =\gamma _{1\perp }\cdot \gamma _{2\perp } 
\frac{\mathcal{G}(x_{\perp })}{2}\mathcal{O}_{1},~~\text{space-like\ vector}{
,}  \label{hyp3}
\end{align}
where 
\begin{align}
\gamma _{5i}& =\gamma _{i}^{0}\gamma _{i}^{1}\gamma _{i}^{2}\gamma _{i}^{3},
\notag \\
\beta _{i}& =-\gamma _{i}\cdot \hat{P}.  \label{beta}
\end{align}
$~$ For general independent scalar, time-like vector, and space-like vector
interactions we have 
\begin{equation}
\Delta (x_{\perp })=\Delta _{\mathcal{L}}+\Delta _{\mathcal{J}}+\Delta _{ 
\mathcal{G}}.  \label{Delta}
\end{equation}
The special case of an electromagnetic-like interaction (in the Feynman
gauge) applied in this paper and in \cite{bckr} corresponds to $\mathcal{J}
=- \mathcal{G}$ or 
\begin{align}
\Delta _{\mathcal{J}}+\Delta _{\mathcal{G}}& \equiv \Delta _{\mathcal{EM}
}=(-\gamma _{1}\cdot \hat{P}\gamma _{2}\cdot \hat{P}+\gamma _{1\perp }\cdot
\gamma _{2\perp })\frac{\mathcal{G}(x_{\perp })}{2}\mathcal{O}_{1}  \notag \\
& =\gamma _{1}\cdot \gamma _{2}\frac{\mathcal{G}(x_{\perp })}{2}\mathcal{O}
_{1}.  \label{vecc}
\end{align}
and for scalar and electromagnetic interaction, 
\begin{equation}
\Delta (x_{\perp })=\Delta _{\mathcal{L}}+\Delta _{\mathcal{EM}}.
\label{A10}
\end{equation}
This leads to\footnote{
In short, \ one inserts Eq. (\ref{hyp2}) into (\ref{hyp1}) and brings the
free Dirac operator (\ref{es0}) to the right of the matrix hyperbolic
functions. \ Using commutators and $\cosh ^{2}\Delta -\sinh ^{2}\Delta =1$
one arrives at Eq. (\ref{extd}). \ The structure of these equations are very
much the same as that of a Dirac equation for each of the two particles,
with $M_{i}$ and $E_{i}$ playing the roles that $m+S$ and $\varepsilon -A$
do in the single particle Dirac equation. Over and above the usual kinetic
part, the spin-dependent modifications involving $G\mathcal{P}_{i}$ and the
last set of derivative terms are two-body recoil effects essential for the
compatibility (consistency) of the two equations} \cite{jmath,long} 
\begin{align}
\mathcal{S}_{1}\psi & =\big(-G\beta _{1}\Sigma _{1}\cdot \mathcal{P}
_{2}+E_{1}\beta _{1}\gamma _{51}+M_{1}\gamma _{51}-G\frac{i}{2}\Sigma
_{2}\cdot \partial (\mathcal{L}\beta _{2}\mathcal{-J}\beta _{1})\gamma
_{51}\gamma _{52}\big)\psi =0,  \notag \\
\mathcal{S}_{2}\psi & =\big(G\beta _{2}\Sigma _{2}\cdot \mathcal{P}
_{1}+E_{2}\beta _{2}\gamma _{52}+M_{2}\gamma _{52}+G\frac{i}{2}\Sigma
_{1}\cdot \partial (\mathcal{L}\beta _{1}\mathcal{-J}\beta _{2})\gamma
_{51}\gamma _{52}\big)\psi =0,  \label{extd}
\end{align}
in which $\partial _{\mu }=\partial /\partial x^{\mu }.$ \ \ With 
\begin{align}
G& =\exp \mathcal{G},  \notag \\
\mathcal{P}_{i}& \equiv p_{\perp }-\frac{i}{2}\Sigma _{i}\cdot \partial 
\mathcal{G}\Sigma _{i}.  \label{d2}
\end{align}
The connections between what we call the vertex invariants $\mathcal{L}, 
\mathcal{J},\mathcal{G}$ and the mass and energy potentials $M_{i},E_{i}$
are 
\begin{align}
M_{1}& =m_{1}\ \cosh \mathcal{L}\ +m_{2}\sinh \mathcal{L},  \notag \\
M_{2}& =m_{2}\ \cosh \mathcal{L}\ +m_{1}\ \sinh \mathcal{L},  \notag \\
E_{1}& =\varepsilon _{1}\ \cosh \mathcal{J}\ +\varepsilon _{2}\sinh \mathcal{
J},  \notag \\
E_{2}& =\varepsilon _{2}\ \cosh \mathcal{J}+\varepsilon _{1}\sinh \mathcal{J}
.  \label{d2b}
\end{align}
Eq. (\ref{extd}) depends on standard Pauli-Dirac representation of gamma
matrices in block forms (see Eq.\ (2.28) in \cite{crater2} for their
explicit forms) and where\footnote{
\ Just as $x^{\mu }$ is a four vector, so is $P^{\mu }.$ \ Thus, the
time-like and space-like interactions in Eq. (\ref{hyp3}) become $\gamma
_{1}^{0}\gamma _{2}^{0}$ and ${\bb\gamma } _{1}\cdot {\bb\gamma }_{2}$ only
in the c.m. system due to the fact that from Eq. (\ref{beta}), $\beta
_{i}=\gamma _{i}^{0}$ only in the c.m. frame. \ Likewise, $\Sigma _{i}^{\mu
}=(0,{\bb\Sigma )}$ only in the c.m. frame just as is $x_{\perp }^{\mu }=(0,{
\bb r)}$ in that frame only.} 
\begin{equation}
\Sigma _{i}=\gamma _{5i}\beta _{i}\gamma _{\perp i}.  \label{d3}
\end{equation}

\subsection{\protect\bigskip\ Vector potentials $\tilde{A}_{i}^{\protect\mu 
} $ in terms of the invariant $A(r)$}

Comparing Eq. (\ref{extd}) with Eq. (\ref{tbde}) we find that the
spin-dependent electromagnetic-like vector interactions of Eq. (\ref{tbde})
are \cite{cra87,bckr} 
\begin{align}
\tilde{A}_{1}^{\mu }& =\big((\varepsilon _{1}-E_{1})\big )\hat{P}^{\mu
}+(1-G)p_{\perp }^{\mu }-\frac{i}{2}\partial G\cdot \gamma _{2}\gamma
_{2}^{\mu },  \notag \\
A_{2}^{\mu }& =\big((\varepsilon _{2}-E_{2})\big )\hat{P}^{\mu
}-(1-G)p_{\perp }^{\mu }+\frac{i}{2}\partial G\cdot \gamma _{1}\gamma
_{1}^{\mu },
\end{align}
Note that the first portion of the vector potentials is time-like (parallel
to $\hat{P}^{\mu })$ while the next two portions are space-like (transverse
to $\hat{P}^{\mu })$. The spin-dependent scalar potentials $\tilde{S}_{i}$
are 
\begin{align}
\tilde{S}_{1}& =M_{1}-m_{1}-\frac{i}{2}G\gamma _{2}\cdot \partial \mathcal{L}
,  \notag \\
\tilde{S}_{2}& =M_{2}-m_{2}+\frac{i}{2}G\gamma _{1}\cdot {\partial }\mathcal{
\ L}{.}  \label{scalp}
\end{align}

We have chosen a parametrization for the vertex invariants $\mathcal{L},~ 
\mathcal{J}=-\mathcal{G}$ that takes advantage of the Todorov effective
external potential forms and at the same time will display the correct
static limit form for the Pauli reduction. \ The logic of the choice for
these parametrizations is strengthened by the fact that for classical \cite
{fw} or quantum field theories \cite{saz97} for separate scalar and
time-like vector interactions one can show that the spin independent part of
the quasipotential $\Phi ~$ involves the difference of squares of the
invariant mass and energy potentials 
\begin{equation}
M_{i}^{2}=m_{i}^{2}+2m_{w}S+S^{2};\ E_{i}^{2}=\varepsilon
_{i}^{2}-2\varepsilon _{w}A+A^{2},  \label{kg1}
\end{equation}
so that \ 
\begin{equation}
M_{i}^{2}-E_{i}^{2}=2m_{w}S+S^{2}+2\varepsilon _{w}A-A^{2}-b^{2}(w).
\label{kg}
\end{equation}

Eqs. (\ref{tbde}) and (\ref{extd}) involve combined scalar and
electromagnetic-like vector interactions (without the separate time-like
interactions this amounts to working in the Feynman gauge with the simplest
relation between space- and time-like parts, see Eqs.\ (\ref{vecc}), (\ref
{A10}), and \cite{cra88,crater2}). \ In that case the\ mass and energy
potentials in place of Eq.\ (\ref{kg1}) are respectively 
\begin{align}
M_{i}^{2}& =m_{i}^{2}+\exp (2\mathcal{G)(}2m_{w}S\mathcal{+}S^{2}),  \notag
\\
E_{i}^{2}& =\exp (2\mathcal{G(A))(}\left( \varepsilon _{i}-A)^{2}\right) , 
\notag \\
M_{i}^{2}-E_{i}^{2}& =\exp (2\mathcal{G(A))[}2m_{w}S+S^{2}+2\varepsilon
_{w}A-A^{2}-b^{2}(w]  \label{tree}
\end{align}
so that from Eq. (\ref{d2b}), 
\begin{align}
\exp (\mathcal{L})& =\exp (\mathcal{L}(S,A))=\frac{\sqrt{m_{1}^{2}+\exp (2 
\mathcal{G)(}2m_{w}S\mathcal{+}S^{2})}+\sqrt{m_{2}^{2}+\exp (2\mathcal{G)(}
2m_{w}S\mathcal{+}S^{2})}}{m_{1}+m_{2}},\   \label{1.5} \\
\exp (\mathcal{J})& =\exp (-\mathcal{G)}  \notag
\end{align}
with 
\begin{equation}
\exp (2\mathcal{G(}A\mathcal{))=}\frac{1}{(1-2A/w)}\equiv G^{2},
\label{three}
\end{equation}
or 
\begin{align}
-\mathcal{G}& \mathcal{=}\frac{1}{2}\log (1-2A/w)=\log \frac{E_{1}+E_{2}}{w},
\end{align}
and the spin-independent minimal coupling appears like 
\begin{equation}
\Phi _{SI}=2m_{w}S+S^{2}+2\varepsilon _{w}A-A^{2}.
\end{equation}

\subsection{Interaction terms in the equivalent Relativistic Schr\"odinger
Equation [Eq. (\protect\ref{57})]}

The Klein-Gordon like potential energy terms appearing in the Pauli form ( 
\ref{57}) arise from (see Eq. (\ref{tree})) 
\begin{equation}
M_{i}^{2}-E_{i}^{2}=\exp (2\mathcal{G)[}2m_{w}S+S^{2}+2\varepsilon
_{w}A-A^{2}-b^{2}(w)].
\end{equation}
To obtain the simple Pauli form of Eq.\ (\ref{schlike}) and the subsequent
detailed form in Eq. (\ref{57}) involves steps similar to those used in the
Pauli reduction of the single particle Dirac equation \cite{unusual} but
with the combinations $\phi _{\pm }=\psi _{1}\pm \psi _{4}$ and $\chi _{\pm
}=\psi _{2}\pm \psi _{3}$ instead of the upper and lower components of the
single particle wave function. This reduces the Pauli forms to 4 uncoupled 4
component relativistic Schr\"{o}dinger equations \cite
{saz94,long,crater2,liu}. \ We work in the c.m. frame in which $\hat{P}=(1,{
\ \bb0)}$ and $\hat{r}=(0,{\bb {\hat r}).}$ We also define four component
wave functions $\psi _{\pm },\eta _{\pm }$ \ by \cite{liu} 
\begin{align}
\phi _{\pm }& =\exp (\mathcal{F}+\mathcal{K}\boldsymbol{\sigma }_{1}{\bb 
\cdot \hat{r}}\boldsymbol{\sigma }_{2}{\bb\cdot \hat{r}})\psi _{\pm }=(\exp 
\mathcal{F})(\cosh \mathcal{K}+\boldsymbol{\sigma }_{1}{\bb\cdot \hat{r}} 
\boldsymbol{\sigma }_{2}{\bb\cdot \hat{r}}\sinh \mathcal{K})\psi _{\pm }, 
\notag \\
\chi _{\pm }& =\exp (\mathcal{F}+\mathcal{K}\boldsymbol{\sigma }_{1}{\bb 
\cdot \hat{r}}\boldsymbol{\sigma }_{2}{\bb\cdot \hat{r}})\eta _{\pm }=(\exp 
\mathcal{F})(\cosh \mathcal{K}+\boldsymbol{\sigma }_{1}{\bb\cdot \hat{r}} 
\boldsymbol{\sigma }_{2}{\bb\cdot \hat{r}}\sinh \mathcal{K})\eta _{\pm },
\label{fk}
\end{align}
in which 
\begin{align}
\mathcal{F}& =\frac{1}{2}\log \frac{\mathcal{D}}{\varepsilon
_{2}m_{1}+\varepsilon _{1}m_{2}}-\mathcal{G},  \notag \\
\mathcal{D}& \mathcal{=}E_{2}M_{1}+E_{1}M_{2},  \notag \\
\mathcal{K}& =\frac{(\mathcal{L}+\mathcal{G})}{2}.  \label{kf}
\end{align}
The substitution (\ref{fk}) has the convenient property that in the
resultant bound state equation, the coefficients of the first order relative
momentum terms vanish.

\ Using the results in \cite{liu} and \cite{unusual} we obtain for the
general case of unequal masses the relativistic Schr\"{o}dinger equation ( 
\ref{57}) that is a detailed c.m.\ form of Eq.\ (\ref{schlike}). In that
equations we have introduced the abbreviations\footnote{
Minor misprints of the equations below have appeared in appendices in \cite
{unusual} and \cite{tmlk}. \ The ones presented here are corrected.}

\begin{align}
\Phi _{D}& =-\frac{2(\mathcal{F}^{\prime }+1/r)(\cosh 2\mathcal{K}-1)}{r}+ 
\mathcal{F}^{\prime 2}+\mathcal{K}^{\prime 2}+\frac{2\mathcal{K}^{\prime
}\sinh 2\mathcal{K}}{r}-{\bb\nabla }^{2}\mathcal{F}+m(r),  \notag \\
\Phi _{SO}& =-\frac{\mathcal{F}^{\prime }}{r}-\frac{(\mathcal{F}^{\prime
}+1/r)(\cosh 2\mathcal{K}-1)}{r}+\frac{\mathcal{K}^{\prime }\sinh 2\mathcal{
\ K }}{r},  \notag \\
\Phi _{SOD}& =(l^{\prime }\cosh 2\mathcal{K}-q^{\prime }\sinh 2\mathcal{K}),
\notag \\
\Phi _{SOX}& =(q^{\prime }\cosh 2\mathcal{K}+l^{\prime }\sinh 2\mathcal{K}),
\notag \\
\Phi _{SS}& =\kappa (r)+\frac{2\mathcal{K}^{\prime }\sinh 2\mathcal{K}}{3r}- 
\frac{2(\mathcal{F}^{\prime }+1/r)(\cosh 2\mathcal{K}-1)}{3r}+\frac{2 
\mathcal{F}^{\prime }\mathcal{K}^{\prime }}{3}-\frac{{\bb\nabla }^{2} 
\mathcal{K}}{3},  \notag \\
\Phi _{T}& =\frac{1}{3}[n(r)+\frac{(3\mathcal{F}^{\prime }-\mathcal{K}
^{\prime }+3/r)\sinh 2\mathcal{K}}{r}+\frac{(\mathcal{F}^{\prime }-3\mathcal{
\ \ \ K}^{\prime }+1/r)(\cosh 2\mathcal{K}-1)}{r}+2\mathcal{F}^{\prime } 
\mathcal{\ K}^{\prime }-{\bb\nabla }^{2}\mathcal{K}],  \notag \\
\Phi _{SOT}& =-\mathcal{K}^{\prime }\frac{\cosh 2\mathcal{K}-1}{r}-\frac{ 
\mathcal{K}^{\prime }}{r}+\frac{(\mathcal{F}^{\prime }+1/r)\sinh 2\mathcal{K}
}{r},  \label{54}
\end{align}
where 
\begin{align}
n(r)& =\nabla ^{2}\mathcal{K}-\frac{1}{2}\nabla ^{2}\mathcal{G}+\frac{3( 
\mathcal{G}-2\mathcal{K})^{\prime }}{2r}+\mathcal{F}^{\prime }\mathcal{G}
^{\prime }-2\mathcal{F}^{\prime }\mathcal{K}^{\prime },  \notag \\
\kappa (r)& =\frac{1}{3}\nabla ^{2}(\mathcal{G}+\mathcal{K})-\frac{1}{2} 
\mathcal{G}^{\prime 2}-\frac{2\mathcal{F}^{\prime }(\mathcal{G}+\mathcal{K}
)^{\prime }}{3},  \notag \\
m(r)& =-\frac{1}{2}\nabla ^{2}\mathcal{G+}\frac{3}{4}\mathcal{G}^{\prime 2}- 
\mathcal{K}^{\prime 2}+\mathcal{G}^{\prime }\mathcal{F}^{\prime },  \notag \\
l^{\prime }& =-\frac{\mathcal{(L-G)}^{\prime }}{2r}\frac{
E_{2}M_{2}-E_{1}M_{1}}{E_{2}M_{1}+E_{1}M_{2}},  \notag \\
q^{\prime }& =\frac{\mathcal{(L-G)}^{\prime }}{2r}\frac{
E_{1}M_{2}-E_{2}M_{1} }{E_{2}M_{1}+E_{1}M_{2}}.  \label{knm}
\end{align}
(The prime symbol stands for $d/dr,$ and the explicit forms of the
derivatives are given in Eq. (\ref{der})$\,$). For $L=J$ states, the
hyperbolic terms cancel and the spin-orbit difference terms in general
produce spin mixing except for equal masses or $J=0$. For ease of use we
have listed below the explicit forms that appear in the above $\Phi $s in
Eqs. (\ref{54}) -(\ref{knm}) in terms of the general invariant potentials $
A(r)$ and $S(r).~$\ The radial components of Eq. (\ref{57}) are given in
Appendix B.

\bigskip

\subsection{\protect\bigskip Explicit expressions for terms in the
relativistic Schr\"{o}dinger Equation (\protect\ref{57}) from $A(r)$ and $
S(r)$}

Given the functions $A(r)$ and $S(r)$ for the interaction, users of the
relativistic Schr\"{o}dinger equation (\ref{57}) will find it convenient to
have an explicit expression in an order that would be useful for programing
the terms in the associated equation (\ref{54}). We use the definitions
above given in Eqs.\ (\ref{tree} )-(\ref{three}), and (\ref{kf}). In order
that the terms in Eq.\ (\ref{54}) be reduced to expressions involving just $
A(r),~$and $S(r)$ and their derivatives, we list the following formulae 
\begin{align}
\mathcal{F}^{\prime }& =\frac{(\mathcal{L}^{\prime }-\mathcal{G}^{\prime
})(E_{2}M_{2}+E_{1}M_{1})}{2(E_{2}M_{1}+E_{1}M_{2})}-\mathcal{G}^{\prime }, 
\notag \\
\mathcal{G}^{\prime }& =\frac{A^{\prime }}{w-2A},  \notag \\
\mathcal{L}^{\prime }& =\frac{M_{1}^{\prime }}{M_{2}}=\frac{M_{2}^{\prime }}{
M_{1}}=\frac{w}{M_{1}M_{2}}\left( \frac{S^{\prime }(m_{w}+S)}{w-2A}+\frac{
(2m_{w}S+S^{2})A^{\prime }}{(w-2A)^{2}}\right) ,  \notag \\
.\mathcal{K}^{\prime }& =\frac{(\mathcal{L}^{\prime }+\mathcal{G}^{\prime }) 
}{2}.  \label{der}
\end{align}
Also needed are 
\begin{align}
\cosh 2\mathcal{K}& =\frac{1}{2}\left( \frac{(\varepsilon _{1}+\varepsilon
_{2})(M_{1}+M_{2})}{(m_{1}+m_{2})(E_{1}+E_{2})}+\frac{
(m_{1}+m_{2})(E_{1}+E_{2})}{(\varepsilon _{1}+\varepsilon _{2})(M_{1}+M_{2})}
\right) ,  \notag \\
\sinh 2\mathcal{K}& =\frac{1}{2}\left( \frac{(\varepsilon _{1}+\varepsilon
_{2})(M_{1}+M_{2})}{(m_{1}+m_{2})(E_{1}+E_{2})}-\frac{
(m_{1}+m_{2})(E_{1}+E_{2})}{(\varepsilon _{1}+\varepsilon _{2})(M_{1}+M_{2})}
\right) ,
\end{align}
and 
\begin{align}
{\bb\nabla }^{2}\mathcal{F}& =\frac{({\bb\nabla }^{2}\mathcal{L}-{\bb\nabla }
^{2}\mathcal{G})(E_{2}M_{2}+E_{1}M_{1})}{2(E_{2}M_{1}+E_{1}M_{2})}-(\mathcal{
\ \ L}^{\prime }-\mathcal{G}^{\prime })^{2}\frac{(m_{1}^{2}-m_{2}^{2})^{2}}{
2\left( E_{2}M_{1}+E_{1}M_{2}\right) ^{2}}-{\bb\nabla }^{2}\mathcal{G}, 
\notag \\
{\bb\nabla }^{2}\mathcal{L}& =-\frac{\mathcal{L}^{\prime
2}(M_{1}^{2}+M_{2}^{2})}{M_{1}M_{2}}  \notag \\
& +\frac{w}{M_{1}M_{2}}\left( \frac{{\bb\nabla }^{2}S(m_{w}+S)+S^{\prime 2}}{
w-2A}+\frac{4S^{\prime }(m_{w}+S)A^{\prime }+(2m_{w}S+S^{2}){\bb\nabla }
^{2}A }{(w-2A)^{2}}+\frac{4(2m_{w}S+S^{2})A^{\prime 2}}{(w-2A)^{3}}\right) ,
\notag \\
{\bb\nabla }^{2}\mathcal{G}& =\frac{{\bb\nabla }^{2}A}{w-2A}+2\mathcal{G}
^{\prime 2}.
\end{align}
The expressions for $\kappa (r),m(r),$ $\ $and $n(r)$ that appear in Eqs.\ ( 
\ref{54})) are given in Eqs.\ (\ref{knm}). They can be evaluated using the
above expressions plus 
\begin{equation}
{\bb\nabla }^{2}\mathcal{K}=\frac{{\bb\nabla }^{2}\mathcal{L}+{\bb\nabla }
^{2}\mathcal{G}}{2}.  \label{blw}
\end{equation}
The only remaining parts of Eq.\ (\ref{54}) that need expressing are those
for $l^{\prime }$ and $q^{\prime }.$ Using Eq.\ (\ref{kf}) they can be
obtained in terms of the above formulae.\vspace*{0.3cm}

\section{Radial Equations}

The following are radial eigenvalue equations \cite{liu,unusual}
corresponding to Eq.\ (\ref{57}) . For a general singlet $^{1}J_{J}$ wave
function $u_{LSJ}=u_{J0J}\equiv u_{0}$ coupled to a general triplet $
^{3}J_{J}$ wave function $u_{J1J}\equiv u_{1}$, the wave equation

\begin{align}
& \{-\frac{d^{2}}{dr^{2}}+\frac{J(J+1)}{r^{2}}+2m_{w}S+S^{2}+2\varepsilon
_{w}A-A^{2}+\Phi _{D}{-}3\Phi _{SS}\}u_{0}  \notag \\
& +2\sqrt{J(J+1)}(\Phi _{SOD}-\Phi _{SOX})u_{1}  \notag \\
& =b^{2}u_{0},
\end{align}
is coupled to

\begin{align}
& \{-\frac{d^{2}}{dr^{2}}+\frac{J(J+1)}{r^{2}}+2m_{w}S+S^{2}+2\varepsilon
_{w}A-A^{2}+\Phi_{D}  \notag \\
& -2\Phi_{SO}+\Phi_{SS}+2\Phi_{T}-2\Phi_{SOT}\}u_{1}+2\sqrt{J(J+1)}
(\Phi_{SOD}+\Phi_{SOX})u_{0}  \notag \\
& =b^{2}u_{1}.
\end{align}
For a general $S=1,$ $J=L+1$ \ wave function $u_{J-11J}\equiv u_{+}~$coupled
to a general $S=1,$ $J=L-1~$wave function $u_{J+11J}\equiv u_{-}$ \ the
equation

\begin{align}
& \{-\frac{d^{2}}{dr^{2}}+\frac{J(J-1)}{r^{2}}+2m_{w}S+S^{2}+2\varepsilon
_{w}A-A^{2}+\Phi_{D}  \notag \\
& +2(J-1)\Phi_{SO}+\Phi_{SS}+\frac{2(J-1)}{2J+1}(\Phi_{SOT}-\Phi_{T})\}u_{+}
\notag \\
& +\frac{2\sqrt{J(J+1)}}{2J+1}\{3\Phi_{T}-2(J+2)\Phi_{SOT}\}u_{-}  \notag \\
& =b^{2}u_{+},  \label{pl}
\end{align}
$\allowbreak\allowbreak\allowbreak$is coupled to 
\begin{align}
& \{-\frac{d^{2}}{dr^{2}}+\frac{(J+1)(J+2)}{r^{2}}+2m_{w}S+S^{2}+2
\varepsilon_{w}A-A^{2}+\Phi_{D}  \notag \\
& -2(J+2)\Phi_{SO}+\Phi_{SS}+\frac{2(J+2)}{2J+1}(\Phi_{SOT}-\Phi_{T})\}u_{-}
\notag \\
& +\frac{2\sqrt{J(J+1)}}{2J+1}\{3\Phi_{T}+2(J-1)\Phi_{SOT}\}u_{+}  \notag \\
& =b^{2}u_{-}.  \label{mi}
\end{align}
For the uncoupled $^{3}P_{0}$ states the single equation is 
\begin{align}
& \{-\frac{d^{2}}{dr^{2}}+\frac{2}{r^{2}}+2m_{w}S+S^{2}+2\varepsilon
_{w}A-A^{2}+\Phi_{D}  \notag \\
& -4\Phi_{SO}+\Phi_{SS}+4(\Phi_{SOT}-\Phi_{T})\}u_{-}=b^{2}u_{-}.
\end{align}

\subsection{Specialization to vector interactions, equal masses and $J=0$.}

In this case we need only consider the $^{1}S_{0}$ and $^{3}P_{0}$ states. \
The corresponding equations are 
\begin{equation}
\{-\frac{d^{2}}{dr^{2}}+2\varepsilon _{w}A-A^{2}+\Phi _{D}{-}3\Phi
_{SS}\}u_{0}=b^{2}u_{0},
\end{equation}
and 
\begin{align}
& \{-\frac{d^{2}}{dr^{2}}+\frac{2}{r^{2}}+2\varepsilon _{w}A-A^{2}+\Phi _{D}
\notag \\
& -4\Phi _{SO}+\Phi _{SS}+4(\Phi _{SOT}-\Phi _{T})\}u_{-}=b^{2}u_{-}.
\end{align}
We consider the explicit forms for the quasipotentials given above that
appear in these equations for the case of vector interactions only, for $J=0$
and equal masses. \ In that case we have 
\begin{align}
\mathcal{F}^{\prime }& =-\frac{3\mathcal{G}^{\prime }}{2},  \notag \\
\mathcal{G}^{\prime }& =\frac{A^{\prime }}{w-2A},  \notag \\
\mathcal{L}^{\prime }& =0,  \notag \\
\mathcal{J}^{\prime }& =-\mathcal{G}^{\prime }=-\frac{A^{\prime }}{w-2A}, 
\notag \\
.\mathcal{K}^{\prime }& =\frac{(\mathcal{L}^{\prime }-\mathcal{J}^{\prime }) 
}{2}=\frac{\mathcal{G}^{\prime }}{2}.
\end{align}
Also needed are 
\begin{align}
\cosh 2\mathcal{K}& =\cosh \mathcal{G}=\frac{1}{2}(\frac{1}{\sqrt{1-2A/w}}+ 
\sqrt{1-2A/w}),  \notag \\
\sinh 2\mathcal{K}& =-\sinh \mathcal{G=-}\frac{1}{2}(\frac{1}{\sqrt{1-2A/w}}
- \sqrt{1-2A/w}),
\end{align}
and 
\begin{align}
{\bb\nabla }^{2}\mathcal{F}& =-\frac{3}{2}{\bb\nabla }^{2}\mathcal{G}, 
\notag \\
{\bb\nabla }^{2}\mathcal{L}& =0 ,  \notag \\
{\bb\nabla }^{2}\mathcal{J}& \mathcal{=}\mathcal{-}{\bb\nabla }^{2}\mathcal{
\ G }=-\frac{{\bb\nabla }^{2}A}{w-2A}-2\mathcal{G}^{\prime 2}.
\end{align}
In that case we have that the combination for the $^{1}S_{0}$ equation is 
\begin{align}
& \Phi _{D}-3\Phi _{SS}  \notag \\
& =-\frac{2(\mathcal{F}^{\prime }+1/r)(\cosh 2\mathcal{K}-1)}{r}+\mathcal{F}
^{\prime 2}+\mathcal{K}^{\prime 2}+\frac{2\mathcal{K}^{\prime }\sinh 2 
\mathcal{K}}{r}-{\bb\nabla }^{2}\mathcal{F}+m(r)  \notag \\
& -3\kappa (r)-\frac{2\mathcal{K}^{\prime }\sinh 2\mathcal{K}}{r}+\frac{2( 
\mathcal{F}^{\prime }+1/r)(\cosh 2\mathcal{K}-1)}{r}-2\mathcal{F}^{\prime } 
\mathcal{K}^{\prime }+{\bb\nabla }^{2}\mathcal{K}  \notag \\
& =\nabla ^{2}(-\mathcal{F+K-}\frac{\mathcal{G}}{2}-\mathcal{G-K)+F}^{\prime
2}+\frac{9}{4}\mathcal{G}^{\prime 2}+3\mathcal{F}^{\prime }\mathcal{G}
^{\prime }  \notag \\
& =0,  \label{d0}
\end{align}
while the combination that appears in the $^{3}P_{0}$ equation is 
\begin{align}
& \Phi _{D}-4\Phi _{SO}+\Phi _{SS}+4(\Phi _{SOT}-\Phi _{T})  \notag \\
& =-\frac{2(\mathcal{F}^{\prime }+1/r)(\cosh 2\mathcal{K}-1)}{r}+\mathcal{F}
^{\prime 2}+\mathcal{K}^{\prime 2}+\frac{2\mathcal{K}^{\prime }\sinh 2 
\mathcal{K}}{r}-{\bb\nabla }^{2}\mathcal{F}+m(r)  \notag \\
& +\frac{4\mathcal{F}^{\prime }}{r}+\frac{4(\mathcal{F}^{\prime }+1/r)(\cosh
2\mathcal{K}-1)}{r}-\frac{4\mathcal{K}^{\prime }\sinh 2\mathcal{K}}{r} 
\notag \\
& +\kappa (r)+\frac{2\mathcal{K}^{\prime }\sinh 2\mathcal{K}}{3r}-\frac{2( 
\mathcal{F}^{\prime }+1/r)(\cosh 2\mathcal{K}-1)}{3r}+\frac{2\mathcal{F}
^{\prime }\mathcal{K}^{\prime }}{3}-\frac{{\bb\nabla }^{2}\mathcal{K}}{3} 
\notag \\
& -4\mathcal{K}^{\prime }\frac{\cosh 2\mathcal{K}-1}{r}-\frac{4\mathcal{K}
^{\prime }}{r}+\frac{4(\mathcal{F}^{\prime }+1/r)\sinh 2\mathcal{K}}{r} 
\notag \\
& -\frac{4}{3}[n(r)+\frac{(3\mathcal{F}^{\prime }-\mathcal{K}^{\prime
}+3/r)\sinh 2\mathcal{K}}{r}+\frac{(\mathcal{F}^{\prime }-3\mathcal{K}
^{\prime }+1/r)(\cosh 2\mathcal{K}-1)}{r}+2\mathcal{F}^{\prime }\mathcal{K}
^{\prime }-{\bb\nabla }^{2}\mathcal{K}]  \notag \\
& =-\frac{8A^{\prime }}{r\left( w-2A\right) }+8\left( \frac{A^{\prime }}{
w-2A }\right) ^{2}+\frac{2{\bb\nabla }^{2}A}{w-2A} .  \label{3res}
\end{align}

\bigskip Thus we have the two $J=0$ single component equations reducing to 
\begin{equation}
\{-\frac{d^{2}}{dr^{2}}+2\varepsilon _{w}A-A^{2}\}u_{0}=b^{2}u_{0},
\end{equation}
and 
\begin{equation}
\{-\frac{d^{2}}{dr^{2}}+\frac{2}{r^{2}}+2\varepsilon _{w}A-A^{2}-\frac{
8A^{\prime }}{r\left( w-2A\right) }+8\left( \frac{A^{\prime }}{w-2A}\right)
^{2}+\frac{2{\bb\nabla }^{2}A}{w-2A}\}=b^{2}u_{-}.
\end{equation}

We consider the case in which 
\begin{eqnarray}
A &=&-\frac{\alpha }{r},  \notag \\
A^{\prime } &=&\frac{\alpha }{r^{2}},  \notag \\
\nabla ^{2}A &=&4\pi \delta (\mathbf{r).}
\end{eqnarray}
In that case 
\begin{eqnarray}
-\frac{8A^{\prime }}{r\left( w-2A\right) } &=&-\frac{8\alpha }{r^{2}\left(
wr+2\alpha \right) }\underset{r\rightarrow 0}{\rightarrow }-\frac{4}{r^{2}},
\notag \\
+8\left( \frac{A^{\prime }}{w-2A}\right) ^{2} &=&\frac{8}{r^{2}}\left( \frac{
\alpha }{wr+2\alpha }\right) ^{2}\underset{r\rightarrow 0}{\rightarrow }+ 
\frac{2}{r^{2}}.
\end{eqnarray}
This displays explicitly how the spin-orbit and other effects completely
overwhelm the angular momentum barrier leaving a nonsingular potential at
the origin \ In particular, combining with $2/r^{2}$ we obtain 
\begin{equation}
\frac{2}{r^{2}}-\frac{8A^{\prime }}{r\left( w-2A\right) }+8\left( \frac{
A^{\prime }}{w-2A}\right) ^{2}=\frac{2}{(r+2\alpha /w)^{2}}.
\end{equation}
From this we obtain Eq. (\ref{2e}).

\bigskip

\subsection{\protect\bigskip Specialization to vector interactions, equal
masses, and $J=L>0$.}

In this case we need only consider the $^{1}J_{J}$ and $^{3}J_{J}$ states. \
The corresponding equations are

\begin{equation}
\{-\frac{d^{2}}{dr^{2}}+\frac{J(J+1)}{r^{2}}+2\varepsilon _{w}A-A^{2}+\Phi
_{D}{-}3\Phi _{SS}\}u_{0}=b^{2}u_{0},
\end{equation}
and

\begin{align}
& \{-\frac{d^{2}}{dr^{2}}+\frac{J(J+1)}{r^{2}}+2\varepsilon _{w}A-A^{2}+\Phi
_{D}-2\Phi _{SO}+\Phi _{SS}+2\Phi _{T}-2\Phi _{SOT}\}u_{1}  \notag \\
& =b^{2}u_{1}.
\end{align}
The first equation simplifies as before $\Phi _{D}{=}3\Phi _{SS}$ while for
the second equation we have 
\begin{align}
& \Phi _{D}-2\Phi _{SO}+\Phi _{SS}+2\Phi _{T}-2\Phi _{SOT}  \notag \\
& =\frac{2\mathcal{G}^{\prime }}{r}+\nabla ^{2}\mathcal{G-G}^{\prime 2} 
\notag \\
& =-\frac{2}{r}\frac{A^{\prime }}{w-2A}+3\left( \frac{A^{\prime }}{w-2A}
\right) ^{2}+\frac{{\bb\nabla }^{2}A}{w-2A}.
\end{align}
Hence, our two $J=1$ uncoupled equations become 
\begin{equation}
\{-\frac{d^{2}}{dr^{2}}+\frac{2}{r^{2}}+2\varepsilon
_{w}A-A^{2}\}u_{0}=b^{2}u_{0},
\end{equation}
and 
\begin{align}
& \{-\frac{d^{2}}{dr^{2}}+\frac{2}{r^{2}}+2\varepsilon _{w}A-A^{2}-\frac{1}{
r }\frac{A^{\prime }}{w-2A}+\frac{3}{2}\left( \frac{A^{\prime }}{w-2A}
\right) ^{2}+\frac{1}{2}\frac{{\bb\nabla }^{2}A}{w-2A}\}u_{1}  \notag \\
& =b^{2}u_{1}.
\end{align}

\section{\protect\bigskip Solutions of Eq. (\ref{up}) for Usual and Peculiar $^1 S_0$ Bound States }

Let us use the Coulomb variable $r=x/\varepsilon _{w}\alpha $ so that our $
^{1}S_{0}$ equation becomes
\begin{eqnarray}
Hu &\equiv &(-\frac{d^{2}}{dx^{2}}-\frac{2}{x}-\frac{\alpha ^{2}}{x^{2}})u=
\frac{(\varepsilon _{w}^{2}-m_{w}^{2})}{\varepsilon _{w}^{2}\alpha ^{2}}
u\equiv -\kappa ^{2}u,  \notag \\
u &=&x^{\lambda +1}v(x)\exp (-\kappa x),
\end{eqnarray}
in which the two solutions for $\lambda $ are
\begin{eqnarray}
\lambda _{+} &=&\frac{1}{2}(-1+\sqrt{1-4\alpha ^{2}}),  \notag \\
\lambda _{-} &=&\frac{1}{2}(-1-\sqrt{1-4\alpha ^{2}}).
\end{eqnarray}
corresponding to the usual and peculiar solutions respectively. Then our
equation becomes
\begin{equation}
-v^{\prime \prime }+2v^{\prime }\kappa -\frac{2(\lambda +1)v^{\prime }}{x}+
\frac{2\kappa (\lambda +1)v}{x}-\frac{2}{x}v=0,  \label{la}
\end{equation}
Let
\begin{equation}
v=\sum_{n_{r}=0}^{\infty }v_{n_{r}}x^{n_{r}},
\end{equation}
and we obtain
\begin{equation}
v_{n_{r}+1}=\frac{(2\kappa n_{r}-2+2\kappa (\lambda +1))}{
(n_{r}+1)(n_{r}+2(\lambda +1))}v_{n_{r}},  \label{rec}
\end{equation}
For bound states we have
\begin{equation}
\kappa =\frac{1}{n_{r}+\lambda +1},~n_{r}=0,1,2,..
\end{equation}
We let 
\begin{equation}
n^{\prime }=n_{r}+\lambda +1.
\end{equation}
If $\lambda $ were an integer then this would be the principle quantum
number $n$. \ \ We write
\begin{equation}
(-\frac{d^{2}}{dx^{2}}-\frac{2}{x}-\frac{\alpha ^{2}}{x^{2}}+\kappa ^{2})u=0,
\end{equation}
as
\begin{equation}
(\frac{d^{2}}{dy^{2}}+\frac{1}{y\kappa }+\frac{\alpha ^{2}}{y^{2}}-\frac{1}{4
})u=0,
\end{equation}
where $x=y/\left( 2\kappa \right) ,$ so that \cite{arf}
\begin{equation}
u(y)=\exp (-y/2)y^{\lambda +1}L_{n_{r}}^{2\lambda +1}(y)
\end{equation}
Let 
\begin{equation}
r=\frac{x}{\varepsilon _{w}\alpha }=\frac{y}{2\kappa \varepsilon _{w}\alpha }
,
\end{equation}
and so our radial wave function is
\begin{equation}
u(r)=k\exp (-\frac{\varepsilon _{w}\alpha r}{n^{\prime }})\left( \frac{
2\varepsilon _{w}\alpha r}{n^{\prime }}\right) ^{\lambda
+1}L_{n_{r}}^{2\lambda +1}(\frac{2\varepsilon _{w}\alpha r}{n^{\prime }}).
\end{equation}
The corresponding hydrogenic radial wave function is
\begin{equation}
u(r)=k\exp (-\frac{r}{na_{0}})\left( \frac{2}{na_{0}}\right)
^{L+1}L_{n_{r}}^{2L+1}(\frac{2r}{na_{0}}).
\end{equation}
Using the result \cite{arf} for the hydrogenic wave function
\begin{equation}
\langle r^{2}\rangle =\frac{a_{0}^{2}n^{2}}{6}[n^{2}-5L(L+1)+3]
\end{equation}
and identifying $L(L+1)\rightarrow -\alpha ^{2}$, $n\rightarrow n^{\prime }$
,$a_{0}\rightarrow 1/(\varepsilon _{w}\alpha )$ we see that for our states
\begin{equation}
\langle r^{2}\rangle =\frac{n^{\prime 2}}{6\left( \varepsilon _{w}\alpha
\right) ^{2}}[n^{\prime 2}+5\alpha ^{2}+3].
\end{equation}

\bigskip Our total c.m. energy eigenvalues come from
\begin{eqnarray}
\frac{(\varepsilon _{w}^{2}-m_{w}^{2})}{\varepsilon _{w}^{2}\alpha ^{2}}
&=&-\kappa ^{2}=-\frac{1}{n^{\prime 2}}  \notag \\
\varepsilon _{w}^{2}(1+\frac{\alpha ^{2}}{n^{\prime 2}}) &=&m_{w}^{2}, 
\notag \\
\varepsilon _{w} &=&\pm \frac{m_{w}}{\sqrt{(1+\frac{\alpha ^{2}}{n^{\prime 2}
})}}.  \notag \\
n^{\prime } &=&n_{r}+\lambda +1,~n_{r}=0,1,...
\end{eqnarray}

In the static limit case for which $m_{2}>>m_{1}$ we use $
w=m_{2}+\varepsilon $ in which $\varepsilon <<m_{2}$ includes the rest mass
and binding energy of particle 1. Then
\begin{eqnarray}
m_{w} &=&\frac{m_{1}m_{2}}{m_{2}+\varepsilon }\rightarrow m_{1},  \notag \\
\varepsilon _{w} &=&\frac{m_{2}^{2}+2\varepsilon m_{2}+\varepsilon
^{2}-m_{1}^{2}-m_{2}^{2}}{2m_{1}m_{2}}  \notag \\
&\rightarrow &\frac{2\varepsilon m_{2}+\varepsilon ^{2}-m_{1}^{2}}{
2m_{1}m_{2}}\rightarrow \varepsilon .
\end{eqnarray}
In that case the above solution would be for the binding energy
\begin{equation}
\varepsilon =\pm \frac{m}{\sqrt{(1+\frac{\alpha ^{2}}{n^{\prime 2}})}}.
\end{equation}
Since we do not include negative energies we dispense with the lower sign.

Let us solve for the total c.m. energy in the case of equal masses $
m_{1}=m_{2}\equiv m$,
\begin{eqnarray}
\frac{\varepsilon _{w}}{m_{w}} &=&\frac{w^{2}-2m^{2}}{2m^{2}}=f(\alpha
)\equiv \frac{1}{\sqrt{(1+\frac{\alpha ^{2}}{n^{\prime 2}})}},  \notag \\
w^{2} &=&2m^{2}(1+f(\alpha )).
\end{eqnarray}
Thus the solutions are 
\begin{eqnarray}
w_{\pm } &=&\sqrt{2}m\sqrt{1+\frac{1}{\sqrt{(1+\frac{\alpha ^{2}}{\left(
n_{r}+\lambda _{\pm }+1\right) ^{2}})}}}  \notag \\
&&
\end{eqnarray}

Since $L=0$ we take our principle quantum number to be $n=n_{r}+1$. \ This
leads to the results in the text for the spectrum. \ The value of $\langle
r^{2}\rangle $ for the peculiar ground state is
\begin{eqnarray}
\langle r^{2}\rangle _{-} &=&\frac{n^{\prime 2}}{6\left( \varepsilon
_{w}\alpha \right) ^{2}}[n^{\prime 2}+5\alpha ^{2}+3]  \notag \\
&=&\frac{(1-\sqrt{1-4\alpha ^{2}})^{2}}{8\left( \varepsilon _{w}\alpha
\right) ^{2}}[\frac{1}{4}(1-\sqrt{1-4\alpha ^{2}})^{2}+5\alpha ^{2}+3] 
\notag \\
&\rightarrow &\frac{\alpha ^{2}}{2\varepsilon _{w}^{2}}\rightarrow \frac{
\alpha ^{2}}{2(m\alpha /\sqrt{2})^{2}}=\frac{1}{m^{2}}
\end{eqnarray}
so that $\sqrt{\langle r^{2}\rangle _{-}}$ is the electron Compton radius. $
\ $For all of the usual states and the remaining peculiar states they have
the following forms
\begin{eqnarray}
\langle r^{2}\rangle _{+} &=&\frac{n_{+}^{\prime 2}}{6\left( \varepsilon
_{w_{+}}\alpha \right) ^{2}}[n_{+}^{\prime 2}+5\alpha ^{2}+3]=\frac{
(n+\lambda _{+})^{2}}{6\left( \varepsilon _{w_{+}}\alpha \right) ^{2}}
[(n+\lambda _{+})^{2}+5\alpha ^{2}+3]\text{,~}n=1,2,3...,  \notag \\
\langle r^{2}\rangle _{-} &=&\frac{n_{-}^{\prime 2}}{6\left( \varepsilon
_{w_{-}}\alpha \right) ^{2}}[n_{-}^{\prime 2}+5\alpha ^{2}+3]=\frac{
(n+\lambda _{-})^{2}}{6\left( \varepsilon _{w_{+}}\alpha \right) ^{2}}
[(n+\lambda _{-})^{2}+5\alpha ^{2}+3]\text{,~}n=2,3...,
\end{eqnarray}
and we see that the size of the $nth$ usual state is very nearly the same as
the size of the $n+1st$ peculiar state.

In light of this one might wonder how the excited peculiar states (which
have the size of angtroms) can be orthogonal to the peculiar ground state,
that has size of a Compton wave length. \ As an example, as seen from Eq. (
\ref{rec}) the first node of the first excited state occurs at
\begin{eqnarray}
x &=&(\lambda _{-}+1)(\lambda _{-}+2)\sim \alpha ^{2},  \notag \\
r &\sim &\frac{\alpha }{\varepsilon _{w}}\sim \frac{\sqrt{2}}{m}.
\end{eqnarray}
which is on the order of 560 fermis.

\section{\protect\bigskip The Connection between $F_{\protect\lambda }(
\protect\eta ,br)$ and $G_{\protect\lambda }(\protect\eta ,br)$}

We begin with \cite{whit,hum,abram} 
\begin{equation}
F_{\lambda }(\rho )=C_{\lambda }(\eta )\rho ^{\lambda +1}\exp (-i\rho
)M(\lambda +1-i\eta ,2\lambda +2;2i\rho ),
\end{equation}
and

\begin{eqnarray}
G_{\lambda }(\rho ) &=&\frac{1}{2}\left\vert \Gamma (\lambda +1+i\eta
)\right\vert \exp (\pi \eta /2)[\frac{\exp (i\pi \lambda /2)}{\Gamma
(\lambda +1+i\eta )}W_{i\eta },_{\lambda +1/2}(2i\rho )  \notag \\
&&+\frac{\exp (-i\pi \lambda /2)}{\Gamma (\lambda +1-i\eta )}W_{-i\eta
},_{\lambda +1/2}(-2i\rho )].
\end{eqnarray}
We introduce the Coulomb phase shift 
\begin{eqnarray}
\sigma _{\lambda }(\eta ) &=&\frac{1}{2i}[\log (\Gamma (\lambda +1+i\eta
)-\log (\Gamma (\lambda +1-i\eta )],  \notag \\
\Gamma (\lambda +1+i\eta ) &=&\left\vert \Gamma (\lambda +1+i\eta
)\right\vert \exp (i\sigma _{\lambda }(\eta )),
\end{eqnarray}
and so 
\begin{eqnarray}
G_{\lambda }(\rho ) &=&\frac{1}{2}[\exp (-i[\sigma _{\lambda }(\eta
)-(\lambda -i\eta )\pi /2)W_{i\eta },_{\lambda +1/2}(2i\rho )+\exp (i[\sigma
_{\lambda }(\eta )-(\lambda -i\eta )\pi /2)W_{-i\eta },_{\lambda
+1/2}(-2i\rho )]  \notag \\
&\equiv &\frac{1}{2}[\psi _{-}(\lambda ,\eta ,\rho )+\psi _{+}(\lambda ,\eta
,\rho )],
\end{eqnarray}
where 
\begin{eqnarray}
\psi _{+}(\lambda ,\eta ,\rho ) &=&\exp (i[\sigma _{\lambda }(\eta
)-(\lambda -i\eta )\pi /2])W_{-i\eta },_{\lambda +1/2}(-2i\rho ),  \notag \\
\psi _{-}(\lambda ,\eta ,\rho ) &=&\exp (-i[\sigma _{\lambda }(\eta
)-(\lambda -i\eta )\pi /2])W_{i\eta },_{\lambda +1/2}(2i\rho ),
\end{eqnarray}
and since $\lambda ,\eta ,\rho $ are all real 
\begin{equation}
\psi _{-}(\lambda ,\eta ,\rho )=\psi _{+}^{\ast }(\lambda ,\eta ,\rho ).
\end{equation}
Also we have 
\begin{eqnarray}
F_{\lambda }(\rho ) &=&\frac{1}{2i}[\psi _{+}(\lambda ,\eta ,\rho )-\psi
_{-}(\lambda ,\eta ,\rho )], \\
\psi _{\pm }(\lambda ,\eta ,\rho ) &=&G_{\lambda }(\rho )\pm iF_{\lambda
}(\rho ).  \notag
\end{eqnarray}
Note that since the Whittaker function $W_{\kappa ,\upsilon }(z)$ is an even
function of $\mu $ we have that 
\begin{eqnarray}
\psi _{+}(-\lambda -1,\eta ,\rho ) &=&\exp (i[\sigma _{-\lambda -1}(\eta
)-(-\lambda -1-i\eta )\pi /2])W_{-i\eta },_{-\lambda -1/2}(-2i\rho )  \notag
\\
&=&\exp (ix(\lambda ,\eta ))\exp (i[\sigma _{\lambda }(\eta )-(\lambda
-i\eta )\pi /2])W_{-i\eta },_{\lambda +1/2}(-2i\rho )  \notag \\
&=&\exp (ix(\lambda ,\eta ))\psi _{+}(\lambda ,\eta ,\rho ),
\end{eqnarray}
where 
\begin{equation}
x(\lambda ,\eta )=(\lambda +\frac{1}{2})\pi +\sigma _{-\lambda -1}(\eta
)-\sigma _{\lambda }(\eta ).
\end{equation}
Similarly 
\begin{equation}
\psi _{-}(-\lambda -1,\eta ,\rho )=\exp (-ix(\lambda ,\eta ))\psi
_{-}(\lambda ,\eta ,\rho ).
\end{equation}
As a result of this we have 
\begin{eqnarray}
F_{-\lambda -1}(\rho ) &=&\frac{1}{2i}[\psi _{+}(-\lambda -1,\eta ,\rho
)-\psi _{-}(-\lambda -1,\eta ,\rho )]  \notag \\
&=&\frac{1}{2i}[\exp (ix(\lambda ,\eta ))\psi _{+}(\lambda ,\eta ,\rho
)-\exp (-ix(\lambda ,\eta ))\psi _{-}(\lambda ,\eta ,\rho )]  \notag \\
&=&\frac{1}{2i}[\exp (ix(\lambda ,\eta ))\left[ G_{\lambda }(\rho
)+iF_{\lambda }(\rho )\right] -\exp (-ix(\lambda ,\eta ))\left[ G_{\lambda
}(\rho )-iF_{\lambda }(\rho )\right]  \notag \\
&=&\cos x(\lambda ,\eta )F_{\lambda }(\rho )+\sin x(\lambda ,\eta
)G_{\lambda }(\rho ),
\end{eqnarray}
and thus 
\begin{equation}
G_{\lambda }(\rho )=\frac{F_{-\lambda -1}(\rho )-\cos x(\lambda ,\eta
)F_{\lambda }(\rho )}{\sin x(\lambda ,\eta )}.
\end{equation}

\section{ The Variable Phase Method of Calogero}

Here we outline the variable phase method, first applied to short range
potentials and then to long range potentials. \ We begin with the short
range potentials. \ \ We consider the following two sets of differential
equations 
\begin{align}
u^{\prime \prime }+(b^{2}-W)u& =0,  \notag \\
\bar{u}_{i}^{\prime \prime }+(b^{2}-\bar{W}_{I})\bar{u}_{i}& =0,~i=1,2, 
\notag \\
\bar{u}_{1}(0)& =u(0)=0,  \notag \\
\bar{W}_{I}& =\frac{L(L+1)}{r^{2}},
\end{align}
where $W(r)$ is a short range potential less singular at the origin than $
const/r^{2}$\ and 
\begin{align}
\bar{u}_{1}(r)& =\hat{\jmath}_{L}(br)\rightarrow const\sin (br-L\pi /2), 
\notag \\
\bar{u}_{2}(r)& =-\hat{n}_{L}(br)\rightarrow const\cos (br-L\pi /2).
\end{align}

Let 
\begin{align}
u(r)& =\alpha (r)(\cos \delta _{L}(r)\bar{u}_{1}(r)+\sin \delta _{L}(r)\bar{
u }_{2}(r))  \notag \\
u(r& \rightarrow \infty )=const(\cos \delta _{L}(r\rightarrow \infty )\sin
(br-L\pi /2)  \notag \\
+\sin \delta _{L}(r& \rightarrow \infty )\cos (br-L\pi /2)  \notag \\
& =const\sin (br-L\pi /2+\delta _{L}(\infty ))\rightarrow \sin (br-L\pi
/2+\delta _{L}),
\end{align}
and so 
\begin{equation}
\delta _{L}=\delta _{L}(\infty ).
\end{equation}

To find the differential equation that $\delta _{L}(r)$ satisfies, define 
\begin{equation}
u^{\prime }(r)=\alpha (r)(\cos \delta _{L}(r)\bar{u}_{1}^{\prime }(r)+\sin
\delta _{L}(r)\bar{u}_{2}^{\prime }(r)),  \label{ab}
\end{equation}
and so 
\begin{align}
\frac{u^{\prime }(r)}{u(r)}& =\frac{(\cos \delta _{L}(r)\bar{u}_{1}^{\prime
}(r)+\sin \delta _{L}(r)\bar{u}_{2}^{\prime }(r))}{(\cos \delta _{L}(r)\bar{
u }_{1}(r)+\sin \delta _{L}(r)\bar{u}_{2}(r))}=\frac{(\bar{u}_{1}^{\prime
}(r)+\tan \delta _{L}(r)\bar{u}_{2}^{\prime }(r))}{(\bar{u}_{1}(r)+\tan
\delta _{L}(r)\bar{u}_{2}(r))},  \notag \\
\tan \delta _{L}(r)& =\frac{\bar{u}_{1}^{\prime }(r)u(r)-\bar{u}
_{1}(r)u^{\prime }(r)}{(\bar{u}_{2}(r)u^{\prime }(r)-\bar{u}_{2}^{\prime
}(r)u(r))}.  \label{ba}
\end{align}
Then 
\begin{align}
\delta _{L}^{\prime }(r)\sec ^{2}\delta _{L}(r)& =\delta _{L}^{\prime
}(r)(1+\tan ^{2}\delta _{L}(r))  \notag \\
& =\frac{\left( \bar{u}_{2}u^{\prime }-\bar{u}_{2}^{\prime }u\right) (\bar{u}
_{1}^{\prime \prime }u-\bar{u}_{1}u^{\prime \prime })-\left( \bar{u}
_{1}^{\prime }u-\bar{u}_{1}u^{\prime }\right) (\bar{u}_{2}u^{\prime \prime
}- \bar{u}_{2}^{\prime \prime }u)}{(\bar{u}_{2}(r)u^{\prime }(r)-\bar{u}
_{2}^{\prime }(r)u(r))^{2}}  \notag \\
& =-\frac{(W-\bar{W}_{I})u^{2}b}{(\bar{u}_{2}u^{\prime }-\bar{u}_{2}^{\prime
}u)^{2}},  \label{c}
\end{align}
where we have used the Wronskian relation 
\begin{equation}
\bar{u}_{2}\bar{u}_{1}^{\prime }-\bar{u}_{2}^{\prime }\bar{u}_{1}=const=b,
\end{equation}
and so 
\begin{equation}
\delta _{L}^{\prime }(r)=-\frac{(W-\bar{W}_{I})b}{\sec ^{2}\delta _{L}(\bar{
u }_{2}u^{\prime }/u-\bar{u}_{2}^{\prime })^{2}}.
\end{equation}
Further manipulations lead to 
\begin{align}
\delta _{L}^{\prime }(r)& =-\frac{(W-\bar{W}_{I})(\hat{\jmath}_{L}(br)\cos
\delta _{L}(r)-\hat{n}_{L}(br)\sin \delta _{L}(r))^{2}}{b}.  \notag \\
&
\end{align}

Note that in case of type two reference potentials $(\bar{W}=\bar{W}
_{II}(r)=0)$ we would obtain 
\begin{equation}
\tan \gamma _{L}(r)=\frac{\bar{u}_{1}^{\prime }(r)u(r)-\bar{u}
_{1}(r)u^{\prime }(r)}{(\bar{u}_{2}(r)u^{\prime }(r)-\bar{u}_{2}^{\prime
}(r)u(r))},
\end{equation}
and with 
\begin{eqnarray}
\bar{u}_{1}(r &\rightarrow &0)\rightarrow br,  \notag \\
\bar{u}_{2}(r &\rightarrow &0)\rightarrow 1  \notag \\
u(r &\rightarrow &0)=c(br)^{L+1},
\end{eqnarray}
we obtain 
\begin{equation}
\tan \gamma _{L}(r\rightarrow 0)\rightarrow \frac{
bc(br)^{L+1}-brcb(L+1)(br)^{L}}{cb(L+1)(br)^{L}}\rightarrow 0,
\end{equation}
and so we obtain the same boundary condition as with the type I reference
potentials. \ From Eq. (\ref{3}) 
\begin{equation}
\gamma _{L}^{\prime }(r)=-\frac{W}{b}\sin ^{2}(br+\gamma _{L}(r)),
\end{equation}
at short distances becomes 
\begin{equation}
\gamma _{L}^{\prime }(0)=-\frac{L(L+1)}{b}\sin ^{2}(b+\gamma _{L}^{\prime
}(0)),
\end{equation}
with the solution given in Eq. (\ref{glp}).

Next we sketch an analogous derivation for the phase shift equation which
involves long range potentials corresponding to Eq. (\ref{2e}) in which the
Coulomb potential appears. \ As discussed in the text we begin with the
following two sets of differential equations 
\begin{align}
u^{\prime \prime }+(b^{2}-W)u& =0,  \notag \\
\bar{u}_{i}^{\prime \prime }+(b^{2}-\bar{W}_{III})\bar{u}_{i}& =0,~i=1,2, 
\notag \\
\bar{W}_{III}& =-\frac{2\varepsilon _{w}\alpha }{r}-\frac{\alpha ^{2}}{r^{2}}
,  \notag \\
W& =\frac{2}{(r+2\alpha /w)^{2}}-\frac{2\varepsilon _{w}\alpha }{r}-\frac{
\alpha ^{2}}{r^{2}}.
\end{align}
Note that the total potential plus barrier term $W~$appears in the equation
for $u$. We are not including the angular momentum barrier in the
definitions of $\bar{u}_{i}(r).$

The solutions $\bar{u}_{1},\bar{u}_{2}$ to 
\begin{equation}
\bar{u}_{i}^{\prime \prime }+(b^{2}+\frac{2\varepsilon _{w}\alpha }{r}+\frac{
\alpha ^{2}}{r^{2}})\bar{u}_{i}=0,~i=1,2,
\end{equation}
are Coulomb wave functions 
\begin{align}
\bar{u}_{1}& =aF_{\lambda _{\pm }}+cG_{\lambda _{\pm }}  \notag \\
\bar{u}_{2}& =dF_{\lambda _{\pm }}+fG_{\lambda _{\pm }}.  \label{gf}
\end{align}
We choose the constants so that $\bar{u}_{1}$ has the same behavior at the
origin that $u$ does.

Even though four functions are listed here, only two are linearly
independent (see Eq. (\ref{fg}\.{)}). \ To determine the phase shift
equation let us write down first the wave function $u(r)$ in terms of $\bar{
u 
}_{1},\bar{u}_{2}$ \ 
\begin{equation}
u(r)=\alpha (r)(\cos \gamma _{\pm }(r)\bar{u}_{1}(r)+\sin \gamma _{\pm }(r) 
\bar{u}_{2}(r)).
\end{equation}
In that case 
\begin{align}
u(r& \rightarrow \infty )\rightarrow (\cos \gamma _{\pm }(r\rightarrow
\infty )\sin (br-\eta \log 2br+\sigma _{\lambda _{\pm }}-\lambda _{\pm }\pi
/2)  \notag \\
+\sin \gamma _{\pm }(r& \rightarrow \infty )\cos (br-\eta \log 2br+\sigma
_{\lambda _{\pm }}-\lambda _{\pm }\pi /2)  \notag \\
& =\sin (br-\eta \log 2br+\sigma _{\lambda _{\pm }}-\lambda _{\pm }\pi
/2+\gamma _{\pm }(\infty )).
\end{align}
This defines the phase shift function $\gamma _{\pm }(r)$ \ and its relation
to the asymptotic behavior of $u(r)$. \ On the other hand since $u(r)$ is
the wave function for a potential that includes at $r>>2\alpha /w$ the
modified angular momentum barrier $(2-\alpha ^{2})/r^{2}\equiv \kappa
(\kappa +1)/r^{2}$ in addition to the Coulomb term, we must have 
\begin{equation}
u(r\rightarrow \infty )\rightarrow \sin (br-\eta \log 2br+\sigma _{\kappa
}-\kappa \pi /2+\delta _{\kappa }),
\end{equation}
and so comparison gives 
\begin{equation}
\sigma _{\lambda _{\pm }}-\lambda _{\pm }\pi /2+\gamma _{\pm }(\infty
)=\sigma _{\kappa }-\kappa \pi /2+\delta _{\kappa }.
\end{equation}
Thus with 
\begin{eqnarray}
\kappa (\kappa +1) &=&2-\alpha ^{2}, \\
\kappa &=&\frac{-1+\sqrt{9-4\alpha ^{2}}}{2},  \notag
\end{eqnarray}
the full phase shift is 
\begin{align}
\delta _{\kappa }+\sigma _{\kappa }& =\sigma _{\lambda _{\pm }}+(\kappa
-\lambda _{\pm })\pi /2+\gamma _{\pm }(\infty )  \notag \\
& =\arg \Gamma (\lambda _{\pm }+1+i\eta )+(\kappa -\lambda _{\pm })\pi
/2+\gamma _{\pm }(\infty )  \notag \\
& \sim \arg \Gamma (\lambda _{\pm }+1+i\eta )+(1-\lambda _{\pm })\pi
/2+\gamma _{\pm }(\infty ).
\end{align}

To find the differential equation that $\gamma _{\pm }(r)$ satisfies, define 
\begin{align}
u^{\prime }(r)& =\alpha (r)(\cos \gamma _{\pm }(r)\bar{u}_{1}^{\prime
}(r)+\sin \gamma _{\pm }(r)\bar{u}_{2}^{\prime }(r)),  \notag \\
u(r)& =\alpha (r)(\cos \gamma _{\pm }(r)\bar{u}_{1}(r)+\sin \gamma _{\pm
}(r) \bar{u}_{2}(r)).
\end{align}
Then following a procedure similar that given in Eqs. (\ref{ab}) we obtain 
\begin{equation}
\tan \gamma _{\pm }(r)=\frac{\bar{u}_{1}^{\prime }(r)u(r)-\bar{u}
_{1}(r)u^{\prime }(r)}{(\bar{u}_{2}(r)u^{\prime }(r)-\bar{u}_{2}^{\prime
}(r)u(r))}.  \label{tn}
\end{equation}
Also 
\begin{equation}
\gamma _{\pm }^{\prime }(r)\sec ^{2}\gamma _{\pm }(r)=-\frac{Wu^{2}b}{(\bar{
u }_{2}u^{\prime }-\bar{u}_{2}^{\prime }u)^{2}},
\end{equation}
where we have used the Wronskian relation 
\begin{align}
\bar{u}_{2}\bar{u}_{1}^{\prime }-\bar{u}_{2}^{\prime }\bar{u}_{1}& =const 
\notag \\
& =\cos ()\cos ()(b-\frac{\eta }{r})+\sin ()\sin ()(b-\frac{\eta }{r}) 
\notag \\
& \rightarrow b
\end{align}
and so 
\begin{equation}
\gamma _{\pm }^{\prime }(r)=-\frac{(W-\bar{W}_{III})b}{\sec ^{2}\gamma _{\pm
}(\bar{u}_{2}u^{\prime }/u-\bar{u}_{2}^{\prime })^{2}}.
\end{equation}
Now use 
\begin{equation}
\frac{u^{\prime }(r)}{u(r)}=\frac{(\cos \gamma _{\pm }(r)\bar{u}_{1}^{\prime
}(r)+\sin \gamma _{\pm }(r)\bar{u}_{2}^{\prime }(r))}{(\cos \gamma _{\pm
}(r) \bar{u}_{1}+\sin \gamma _{\pm }(r)\bar{u}_{2})},
\end{equation}
and hence, with $\bar{u}_{1}=F_{\lambda _{\pm }},\bar{u}_{2}=G_{\lambda
_{\pm }}$ we have 
\begin{equation}
\gamma _{\pm }^{\prime }(r)=-\frac{(W-\bar{W}_{III})(\cos \gamma _{\pm
}(r)F_{\lambda _{\pm }}(r)+\sin \gamma _{\pm }(r)G_{\lambda _{\pm }}(r))^{2} 
}{b}.  \label{gm}
\end{equation}
Because of the $2/r^{2}$ behavior of $W$ for large $r$ one will have to
integrate quite far to obtain a convergence for $\gamma _{\pm }(r)$ and
after that one must subtract the phase shift $-\pi /2$ due to the $2/r^{2}$
angular momentum barrier. An alternative form of this equation is 
\begin{equation}
\tan ^{\prime }\gamma _{\pm }(r)=-\frac{(W-\bar{W}_{III})(F_{\lambda _{\pm
}}(r)+\tan \gamma _{\pm }(r)G_{\lambda _{\pm }}(r))^{2}}{b}.  \label{or}
\end{equation}

The question now arises about the boundary condition at the origin for $
\gamma _{\pm }(r).$ \ We focus on Eq. (\ref{tn}) to determine the boundary
condition at the origin for $\gamma _{\pm }(r)$, 
\begin{equation}
\tan \gamma _{\pm }(r)=\frac{\bar{u}_{1}^{\prime }(r)u(r)-\bar{u}
_{1}(r)u^{\prime }(r)}{(\bar{u}_{2}(r)u^{\prime }(r)-\bar{u}_{2}^{\prime
}(r)u(r))}.  \label{tnm}
\end{equation}
We determine the behavior at the origin by evaluating the right hand side
for very small $r$. \ The dominant term for the quasipotential for both case
is $-\alpha ^{2}/r^{2}$. \ Thus it is sufficient to focus on the first case

We use Eq. (\ref{tnm}) with 
\begin{equation}
\left\{ -\frac{d^{2}}{dr^{2}}+\frac{2}{(r+2\alpha /w)^{2}}-\frac{
2\varepsilon _{w}\alpha }{r}-\frac{\alpha ^{2}}{r^{2}}\right\} u=b^{2}u,
\end{equation}
and 
\begin{eqnarray}
\left\{ -\frac{d^{2}}{dr^{2}}-\frac{2\varepsilon _{w}\alpha }{r}-\frac{
\alpha ^{2}}{r^{2}}\right\} \bar{u}_{1,2} &=&b^{2}\bar{u}_{1,2},  \notag \\
\bar{u}_{1} &=&F_{\lambda }(br),  \notag \\
\bar{u}_{2} &=&G_{\lambda }(br).
\end{eqnarray}
At short distance, the potential energy for $u$ is the same as that for $ 
\bar{u}_{1,2}$. \ \ At very short distance, we choose 
\begin{eqnarray*}
u,\bar{u}_{1} &\rightarrow &const\rightarrow r^{\lambda +1}, \\
\bar{u}_{2} &\rightarrow &ar^{\lambda +1}+br^{-\lambda }.
\end{eqnarray*}
Clearly then $\tan \gamma _{\pm }(0)=0$ as the numerator vanishes in both
cases where as the denominator is proportional to the Wronskian of $\bar{u}
_{1}$ and $\bar{u}_{2}$ which is $b^{2}$. \ \ This case allows us to
integrate either Eq. (\ref{gm}) or (\ref{or}) with the boundary condition of 
$\tan \gamma _{\pm }(0)=\gamma _{\pm }(0)=0$.

To find the total wave function we need to find the additional differential
equation for the amplitude of the wave function. \ We use 
\begin{align}
u^{\prime }(r)& =\alpha (r)(\cos \gamma _{\pm }(r)\bar{u}_{1}^{\prime
}(r)+\sin \gamma _{\pm }(r)\bar{u}_{2}^{\prime }(r))  \notag \\
& =\alpha ^{\prime }(r)(\cos \gamma _{\pm }(r)\bar{u}_{1}(r)+\sin \gamma
_{\pm }(r)\bar{u}_{2}(r))  \notag \\
& +\alpha (r)(\cos \gamma _{\pm }(r)\bar{u}_{1}^{\prime }(r)+\sin \gamma
_{\pm }(r)\bar{u}_{2}^{\prime }(r))  \notag \\
& +\gamma _{\pm }^{\prime }\alpha (r)(-\sin \gamma _{\pm }(r)\bar{u}
_{1}(r)+\cos \gamma _{\pm }(r)\bar{u}_{2}(r)),
\end{align}
and thus, using Eq. (\ref{gm}) 
\begin{align}
\frac{\alpha ^{\prime }(r)}{\alpha (r)}& =-\frac{W(r)-W_{III}}{b}[\frac{
\left( \bar{u}_{1}^{2}-\bar{u}_{2}^{2}\right) \sin 2\gamma _{\pm }(r)}{2}-
\bar{u} _{1}\bar{u}_{2}\cos 2\gamma _{\pm }],  \notag \\
\alpha (r)& =\alpha (r_{0})\exp \{-\int_{0}^{r}\frac{W(r)-W_{III}(r)}{b}[
\frac{ \left( \bar{u}_{1}^{2}-\bar{u}_{2}^{2}\right) \sin 2\gamma _{\pm }(r)
}{2}- \bar{u}_{1}\bar{u}_{2}\cos 2\gamma _{\pm }(r)]\}.
\end{align}
So, the total wave function is 
\begin{align}
u(r)& =\alpha (r_{0})\exp \{-\int_{0}^{r}\frac{W(r)-W_{III}(r)}{b}[\frac{
\left( \bar{u}_{1}^{2}-\bar{u}_{2}^{2}\right) \sin 2\gamma _{\pm }(r)}{2}-
\bar{u} _{1}\bar{u}_{2}\cos 2\gamma _{\pm }(r)]\}  \notag \\
& \times (\cos \gamma _{\pm }(r)\bar{u}_{1}(r)+\sin \gamma _{\pm }(r)\bar{u}
_{2}(r)).
\end{align}

\bigskip

\vspace{0.5cm} \centerline{\bf Acknowledgment} \vskip.5cm The authors would
like to thank Profs.\ Jin-Hee Yoon, R. L. Becker, and L. Hulett for helpful discussions. The research was
sponsored in part by the Office of Nuclear Physics, U.S. Department of
Energy.

\end{document}